\documentclass[%
 reprint,
includegraphics,
bibnotes,
 amsmath,amssymb,
aps,
rmp,
eqsecnum,
floatfix,
]{revtex4-2}
\usepackage{graphicx}
\usepackage{subfiles} 
\usepackage{dcolumn}
\usepackage{bm}
\usepackage{hyperref}
\hypersetup{
     colorlinks   = true,
     citecolor    = blue,
     linkcolor    = blue,
     urlcolor     = blue
}
\usepackage{float}

\renewcommand{\selectlanguage}[1]{}   
\newcommand{\BDR}{$R_{BD}$}
\usepackage[normalem]{ulem} 
\usepackage{xcolor}
\usepackage{miller}  
\usepackage{siunitx}
\usepackage{tabularx}
\usepackage{float}

\newcommand{\SC}[1]{\textcolor{olive}{#1}}

\begin{document}

\preprint{RC10129}

\title{Fundamentals of Vacuum Breakdown in High-Field Systems}

\author{Walter Wuensch}
 \email{Walter.Wuensch@cern.ch}
\affiliation{CERN, European Organization for Nuclear Research, 1211 Geneva, Switzerland}

\author{Sergio Calatroni}
\email{}
\affiliation{CERN, European Organization for Nuclear Research, 1211 Geneva, Switzerland}

\author{Flyura Djurabekova}
\email{}
\affiliation{Helsinki Institute of Physics and Department of Physics, University of Helsinki, P.O. Box 43 (Pietari Kalmin katu 2), 00014 Helsingin yliopisto, Finland}

\author{Andreas Kyritsakis}
\email{}
\affiliation{Institute of Technology, University of Tartu,  Nooruse 1, 50411 Tartu, Estonia}

\author{Yinon Ashkenazy}
\email{yinon.ash@mail.huji.ac.il}
\affiliation{Racah Institute of Physics and the Center for Nanoscience and Nanotechnology, Hebrew University of Jerusalem, Jerusalem 9190401, Israel}

\date{\today}

\begin{abstract}
This review consolidates experimental, theoretical, and simulation work examining the behavior of high-field devices and the fundamental process of vacuum arc initiation, commonly referred to as breakdown.
Detailed experimental observations and results relating to a wide range of aspects of high-field devices, including conditioning, field and temperature dependence of breakdown rate, and the ability to sustain high electric fields as a function of device geometry and materials, are presented.
The different observations are then addressed theoretically, and with simulation, capturing the sequence of processes that lead to vacuum breakdown and explaining the major observed experimental dependencies.
The core of the work described in this review was carried out by a broad multi-disciplinary collaboration in an over a decade-long program to develop high-gradient, 100 MV/m-range, accelerating structures for the CLIC project, a possible future linear-collider high-energy physics facility. 
Connections are made to the broader linear collider, high-field, and breakdown communities.  

\end{abstract}

\maketitle

\tableofcontents


\section{\label{s:intro}Introduction}
\subsection{\label{ss:intro-history} General introduction to breakdown and context for this work}

A vacuum arc, also known as a vacuum breakdown, is the sudden formation of plasma above a vacuum-facing metal surface exposed to a sufficiently high electric field. The plasma allows electron currents to flow with little resistance in a volume that was previously insulated by vacuum, resulting in the collapse of the applied fields.  In many applications, vacuum breakdowns are a failure mode, for example in the radio-frequency systems of particle accelerators, fusion reactors, X-ray sources, vacuum interrupters, and solar cell systems in satellites. In other applications the resulting vacuum arc is desired, for example in plasma coating systems and plasma thrusters for satellites.

Although vacuum breakdowns play a critical role in a wide variety of applications, the fundamentals of their initiation have remained an open question. However, the knowledge of these mechanisms may dramatically improve the controllability of 
both desired and failure mode-type systems. In general terms, a device starts with an applied high electrostatic or radio-frequency electromagnetic field, possibly producing low-level currents caused by field electron emission, i.e. emission of electrons due to extremely high surface electric fields, frequently referred to also as field emission. Field emitted currents are typically a micro-ampere or less. This metastable high-field state can be sustained over long periods before the device undergoes an abrupt, nanosecond timescale, transition to a breakdown. Systems for which breakdown is a failure mode are designed to operate in the metastable insulating state. When a breakdown occurs at some position on a high-field surface, the insulating vacuum becomes conductive through the formation of a plasma that carries high current, 100s of amperes and more, between electrodes or through the volume of radio frequency structures. The accelerated high current draws power from the power source that sets up the high field and causes it to drop to a small fraction of that in the insulating state. This field collapse constitutes a failure in some systems, while the high current and lower field is the operating mode in others. This report focuses on systems of the former type. 

As we will see in this review, breakdowns occur randomly in both space and time. Breakdown nucleation sites are very small, some tens of nanometer-scale or less, the upper limit coming from observations, while device surfaces typically have dimensions from millimeters even up to meters. The onset of breakdown occurs in nanoseconds while fields, sometimes continuous and sometimes pulsed, can be applied for hours, days, or even longer depending on the application. The combination of randomness and wide span of time and length scales make it very challenging to measure and analyse the processes underlying the initiation and early-stage evolution of breakdowns. The work described in this report addresses this challenge. 

Vacuum breakdown has been the focus of dedicated studies by different fields affected by it, or using it, for over a century. Several books are devoted to summarizing the experimental evidence, the early modelling work and the technological consequences and uses of vacuum arcs  \cite{lafferty_vacuum_1980, latham_high_1981, mesyats_pulsed_1989, latham_high_1995, boxman_handbook_1996, anders_cathodic_2008}. The arc initiation dynamics in particular has been the focus of several reviews \cite{davies_initiation_1973, juttner_vacuum_1988} drawing from  seminal works dedicated to the study and consequences of field-emitted electrons on the initiation of vacuum breakdowns \cite{kartsev_investigation_1970, litvinov_field_1983, vibrans_vacuum_1964, chatterton_theoretical_1966, williams_field-emitted_1972}. The elusiveness of breakdown initiation has hindered a full understanding of the underlying mechanisms, particularly the physical origins of the observed breakdown rates.


In recent years there has been a major new effort to understand the fundamentals of vacuum arcs, driven this time by a push to develop TeV-range energy electron-positron linear colliders for high-energy physics. There have been several different projects to design and develop key technology for such a collider including CLIC and NLC/JLC \cite{aicheler_multi-tev_2012, larsen_next_2001}. To limit length, and consequently cost, linear colliders depend on high-gradient accelerating structures. The rf (radio frequency) accelerating structures for CLIC target an accelerating gradient of 100 MV/m, resulting in surface electric fields over 200 MV/m. This high field makes vacuum breakdown a major limitation. 

The limitation of gradient by breakdown gives a clear scientific goal and has motivated a long-term effort to deepen the understanding of breakdown in order to achieve as high an accelerating gradient as possible, while minimizing the breakdown rate, i.e. the frequency at which breakdowns occur. This review will summarize the latest experimental and corresponding theoretical work carried out towards understanding vacuum breakdown, with a particular focus on the processes that lead to its initiation. The review draws mainly on work carried out by a multidisciplinary collaboration loosely centered on the CLIC study but also includes important related work carried out by other groups, within and outside of linear colliders studies.  

In addition, the progress described in this report has been made possible due to advances in development of simulation tools and methodologies, which had not been earlier either available or applied to address the physics underlying vacuum breakdown. Generally, to develop a consistent theoretical concept explaining the onset of vacuum arcs, the construction of a single multiscale and multiphysics simulation model where all relevant processes are interconnected is necessary. The processes to include span (i) the redistribution of electron densities on the metal surface as a result of applied electric fields, (ii) the consequent non-uniform distribution of field-induced charges especially on surface irregularities, (iii) the modification of atomic dynamics due to the resulting field-induced stress, especially on surface and under-surface defects, and (iv) the complex interaction of field emission currents with metal tips resulting in their thermal runaway, evaporation, and an eventual plasma buildup. The simulation of plasma initiation can enable the link between the theory and experimental observations of post-breakdown surface features.

Until recently the theoretical efforts in this direction were limited. The effects of electric field on materials, for example, were considered either within analytical theories developed for field emission applications \cite{forbes_new_2001,forbes_physics_2008} or within static quantum-mechanical calculations, see e.g. \cite{lozovoi_reconstruction_2003, bengtsson_dipole_1999,scivetti_electrostatic_2013,scivetti_electrostatics_2023}. The dynamic effects in materials in the field emission regime were addressed in the studies of thermal response of intensively field emitting microtips, however, within a static tip approximation \cite{fursey_field_2005}. Lately, developments of atom probe tomography \cite{miller_atom_1996, vurpillot_simulation_2018} based on field-evaporation phenomena \cite{forbes_field_1995} triggered a number of density functional theory calculations, which provided interesting insights on electron density distributions near metal surfaces in the presence of high electric fields \cite{sanchez__field-evaporation_2004}. All these approaches were developed to address specific technological problems but have not provided a consistent theoretical picture of physics triggered in metals exposed to high electric fields. Neither the atom-level insights on the modification of surface morphology or defect dynamics within the bulk of the materials exposed to high electric fields were available.

First attempts to apply molecular dynamic methods to the vacuum breakdown phenomenon date back to the beginning of 2000s, where the surface atoms were modeled as charged ions to which the pulling force due to an applied electric field (a fixed maximal value) was applied \cite{norem_triggers_2005, hassanein_effects_2006}. The lack of physical background of the proposed model limited the application of this approach to explain the vast majority of experimental evidence collected in many laboratories. Later, more rigorous approaches \cite{djurabekova_atomistic_2011,djurabekova_local_2013,veske_dynamic_2018,veske_dynamic_2020} were proposed that were developed to follow the changes within the metal exposed to a high electric field. These simulations have enabled many non-intuitive insights into the processes governing the onset of vacuum breakdowns. 

In this review, we outline the multiscale and multiphysics model developed in the collaborative effort within the linear colliders community, which we believe will be insightful for other fields dealing with this complex multiphysics phenomenon.

Linear collider development programs have successfully achieved their target, 100 MV/m-range, accelerating gradient objective. In addition to the advances in the understanding of vacuum breakdown initiation that occurred in the development programs, the success has led to the adoption of the underlying technology by an ever-increasing number of applications. These include small compact accelerator applications such as photon sources, XFELs (X-ray Free Electron Lasers) and ICSs (Inverse Compton Sources), industrial and medical applications. These applications, among the others, now carry forward the work described in this review. 

\subsection{\label{ss:intro-concept} An overview of vacuum breakdown}

This review will present a body of experimental and theoretical results that show that vacuum breakdown involves a combination of processes that occur inside the electrode material, on its surface, in the initially insulating vacuum, and in the powering circuit. The different mechanisms occur to some extent sequentially but there are numerous, and key, feedback interactions between them. An initial, simplified description of the different mechanisms is presented here, along with an identification of the most important interactions among the mechanisms in order to guide the reader through the in-depth discussions of the rest of the review.

\begin{figure}
    \centering
    \includegraphics[width=1\linewidth]{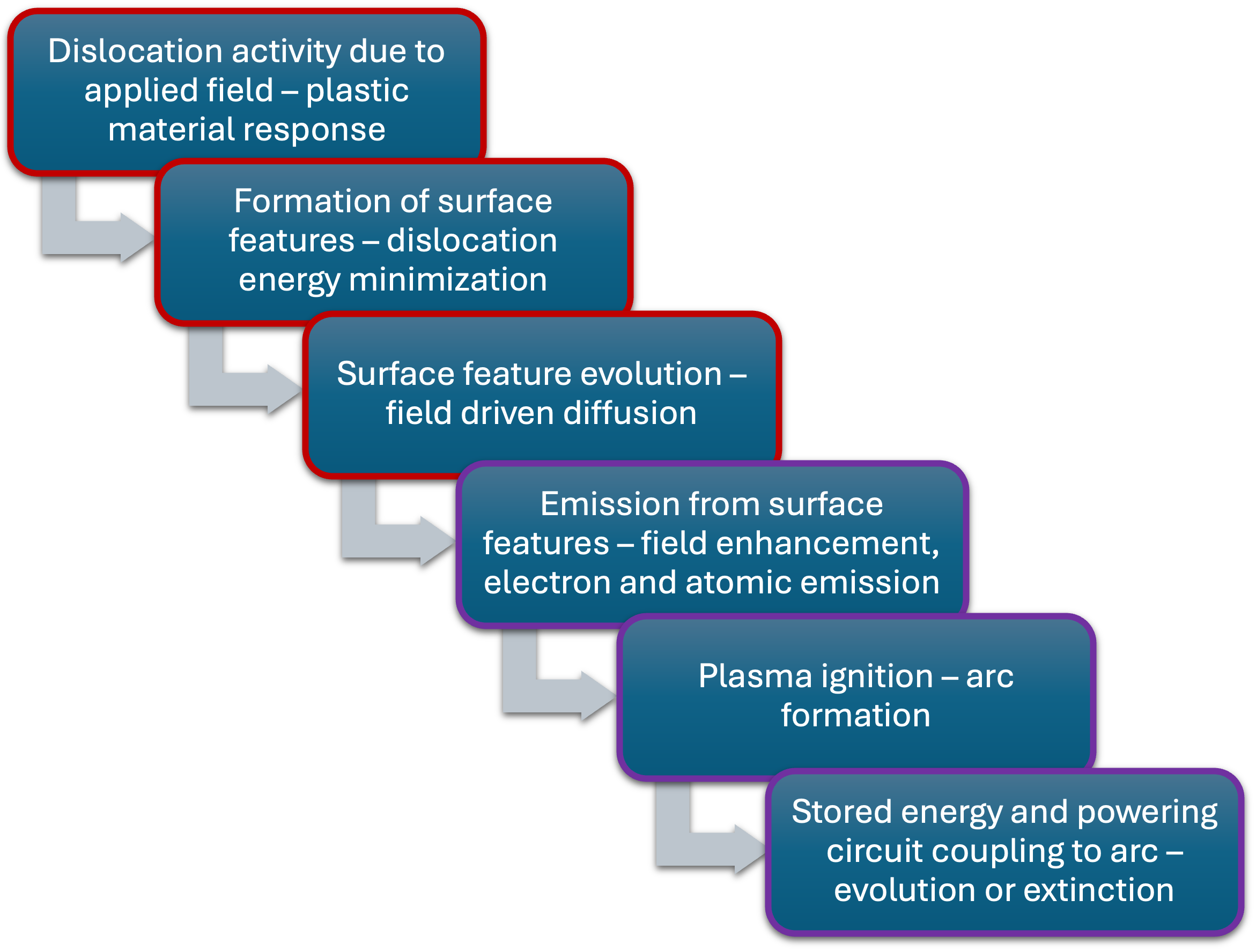}
    \caption{Schematic representation of the arc initiation stages that will be described in this review. The upper three boxes, outlined in red, show the stages during which the electrode, or radio frequency structure, material responds to applied fields and surface features form. These drive the key experimentally observed dependencies, such as breakdown rate on field and temperature. The bottom three boxes, outlined in blue, show the stages that are consequences of the surface features, during which the initial vacuum is populated with electrons and atoms, ionization occurs and an avalanche process begins. In these stages the initially insulating vacuum is filled with a conducting plasma. These are the stages which determine the dependency of achievable field on the electrode and radio frequency structure geometry and powering system. }
    \label{fig:overall scheme}
\end{figure}

A schematic representation of the main stages of the initiation of an arc is shown in Fig.~\ref{fig:overall scheme}.  In general terms, we are considering a system with a starting configuration of a metal electrode or radio frequency structure in vacuum to which electromagnetic fields are applied. The types of structures that are considered in this report are made from copper and typically have some tens of nanometer level surface finish. The fields act on the entire structure surface setting in motion processes, dislocation dynamics. in the material and on its surface, surface diffusion, that occur everywhere that fields are high. The consequence of these processes is the formation of small, 10s of nm scale, features which enhance local fields. The local field enhancement induces further processes, the emission of electrons and neutral copper atoms. Occasionally, one of these features falls into an avalanche condition, which results in the formation of a conductive plasma that effectively short-circuits the system. This is what is observed as a breakdown. These breakdown initiation stages are relevant for a highly conditioned structure, that is one that has been exposed extensively to high fields. The mechanisms are intrinsic so are determined, and limited, by the characteristics of the material. In lower-field devices, breakdown initiation may be dominated by contaminants, that is, extrinsic mechanisms. The different stages with distinct physical processes have been identified and are represented by the different boxes in Fig.~\ref{fig:overall scheme}. We continue by elaborating these stages briefly in the rest of this section. The different stages are discussed in depth in sections \ref{s:dis} to \ref{s:circ}.  

Surface electric fields apply a tensile stress to a metal (conducting) surface, that extends into the bulk metal of the electrodes. This tensile stress leads to dislocation activity within the bulk. This activity, also referred to as plastic response, subsequently leads to the formation of the surface structures where breakdowns are initiated. Plastic response occurs at levels well below the stress limit of the material, in the form of dynamics of point-like defects. Specifically, it drives the creation, movement, and annihilation of dislocations. Due to energy minimization, dislocations project to the electrode surface, creating the surface features that will eventually evolve to breakdown sites. Dislocation dynamics are the origin of the statistical nature of the experimentally observed dependencies of breakdown rate (the number of breakdowns divided by the number of applied field pulses over a long operation period) on field level, temperature and material. They also explain an important part of the conditioning process, by which the field holding capability, i.e. the maximum field a system can hold without breakdown, can be improved by repeated or extended application of high voltage. 

On the surface of electrodes, in particular on the cathode, several mechanisms occur when a high electric field is applied. In the discussion of surface dynamics, it is useful to distinguish between intrinsic and extrinsic mechanisms. Intrinsic mechanisms are related to the dynamics of material from which the electrodes are made, for example copper in most cases in this report. Extrinsic mechanisms are related to materials other than that of the bulk of the electrodes, for example oxides, carbon or dust-like contaminants. In general terms, extrinsic mechanisms dominate the early stages of operation of a high field system while intrinsic mechanisms dominate later stage behavior and determine the ultimate operating conditions, as the high field limit of the surface is approached late in the conditioning process.  

The most important intrinsic dynamic surface processes driven by high surface electric fields include the biased diffusion of surface atoms, localized stress due to local field enhancement and heating due to field emission. Biased diffusion is the migration of surface atoms in the direction of increasing surface field. Biased migration occurs because the surface electric field polarizes surface atoms and consequently induces an attractive force towards an increasing field. This means that surface atoms preferentially diffuse toward areas of locally higher electric field. Such a higher local field occurs when there are surface features so the biased diffusion gives a feedback mechanism to sharpen an electrode surface topography. The local field enhancement gives locally enhanced electron field emission, which results in localized heating. The geometrical field enhancement and subsequent heating due to field emission result in increased local stresses, which contribute further but locally to the plastic mechanisms.

The next aspects of the breakdown process are the mechanisms through which the vacuum is filled with neutral and charged particles, which give rise to the insulating-to-conducting transition that causes applied fields to collapse, the most directly observable aspect of breakdown. Here we also add electron field emission, a low-level current that occurs at sufficiently high fields, even when a breakdown does not occur. When a surface feature exceeds a certain threshold, either through growth driven by underlying surface dynamics projecting to the surface or surface dynamics, localized field-emission driven heating drives atomic evaporation. Field emitted, and then accelerated electrons, as well as the applied field itself ionize atoms, creating a back-bombardment on the cathode surface, liberating further atoms. This then gives rise to the avalanche-like process that forms the current-carrying plasma above the cathode surface.  

The insulating-to-conducting transition occurs very quickly, on a nanosecond timescale, and energy is consumed by the plasma formation processes described in the previous paragraph. If enough local power is available a breakdown can form. If not, then a breakdown will extinguish before it fully evolves. This has the consequence that the strength of the coupling of energy between the powering system and the locally forming breakdown results in a strong dependence of achievable surface electric field on the rf structure, or electrode, geometry and the powering circuit. 

This report covers these different aspects of vacuum breakdown formation and initial evolution in the following way. In Sec. \ref{s:exp} the main experimental systems that were used for the data that underlie this report are described. The key experimental results are presented in Sec. \ref{s:behave}. We then proceed with theoretical analysis; plastic response and dislocation dynamics in Sec. \ref{s:dis}, surface molecular dynamics in Sec. \ref{s:surf}, electron emission and atomic evaporation in Sec. \ref{s:emit}, plasma ignition and evolution in Sec. \ref{s:plasma}, post breakdown surface damage in Sec. \ref{s:crater} and the dependence on high-field surface geometry in Sec. \ref{s:circ}. We conclude and provide our view on prospects and future applications in other fields in the summary Sec. \ref{s:conclusions}.

Finally, table \ref{tab:glossary} in the Appendix gives a comprehensive list of the specialized terms, abbreviations, and symbols used throughout the paper for easy reference.

\section{\label{s:exp}Experimental systems and observations}
\subsection{\label{ss:exp-intro}  Overview and context}

In this section, we introduce the application, linear colliders, that has motivated this in-depth study of breakdown, and then describe the main experimental systems that have provided the fundamental results are the input for the development of the theoretical models described in this review. The requirements of linear colliders motivate the study of breakdown under very high surface fields. 

There are a wide range of systems in which breakdown occurs, as summarized in the introduction, but generally, there is very little quantitative data available from them. This lack is due to various reasons that include operation of devices limited to a narrow range of specified parameters, lack of instrumentation, lack of systematic measurements, concern about damaging equipment and protection of intellectual property (which is often the case in commercial devices). The detailed study and construction of a theoretical understanding of a phenomenon as complicated as vacuum breakdown, however, requires high-quality measurements of relevant physical quantities. Dependencies and repeatability must be determined as well as the range of applicability. Ideally, equivalent data should be taken by different groups in different setups. 

Fortunately, in recent years, such a body of measurements has become available thanks to extensive experimental programs carried out as part of the development of high-energy linear colliders, as well as smaller-scale projects that have common high-field holding requirements. The linear collider programs have provided the motivation, but the results and understanding are important for the broader high-field community.

Linear colliders are a class of possible future high-energy physics facilities that aim for electron-positron collisions with center of mass energies in 250 GeV, extending into the TeV range. The electron positron linear collider project with the highest targeted collision energy is CLIC, which foresees an initial stage with 380 GeV center of mass collisions, and extensions to energies up to 3 TeV. In addition to CLIC, other linear collider projects include the NLC/JLC which was formally ended in 2004, C3 and the ILC, International Linear Collider. The latter is based on superconducting cavities, so it has very different technological challenges and is not directly relevant for this report. References to the  NLC/JLC and CLIC projects can be found here NLC/JLC \cite{larsen_next_2001} and CLIC \cite{aicheler_multi-tev_2012}.

The very high collision energy needed for a linear collider motivates a push towards a very high accelerating gradient to minimize the length, and consequently the cost of a facility. Taking CLIC as an example, the design accelerating gradient is 72 MV/m for the initial 380 GeV stage and 100 MV/m for the higher energy stages up to 3 TeV. This results in over 5 km of active accelerating linac length at 380 GeV, with a corresponding total facility length of 11 km, and over 30 km active length at 3 TeV. 

Vacuum breakdown is a major limitation to achieving such accelerating gradients. This can be seen by considering the surface electric fields that occur inside the accelerating structures. The geometry of linear collider structures is such that the peak surface electric fields are approximately 2 to 2.5 times higher than the accelerating gradient. Thus for accelerating gradients the range of 70 to 100 MV/m peak surface electric fields are in excess of 200 MV/m. This is well into the range of  surface electric field where vacuum breakdown can occur. In addition, accelerating structures operating at such high gradients can also be subject to progressive deterioration due to pulsed surface heating, \cite{laurent_experimental_2011}. Developing accelerating structures that can operate reliably at such high surface electric field, and over the multi-decade lifetime of high energy facilities, has been a major challenge. 

Given the importance of a high accelerating gradient for CLIC, the project has invested in prototype accelerating structure development program coupled with a multi-disciplinary study into the fundamental processes of breakdown. The main objective of the prototype development program has been to validate the feasibility and practicality of operating accelerating gradients at the design values of 70 to 100 MV. The main objectives of the study into fundamental processes has been to improve performance through understanding, including increasing achievable gradient through radio frequency design, through material and process choices, optimization of the overall facility (in power consumption and cost), reduction in conditioning time and optimization of operational issues such as recovery from breakdown. All these can be better addressed through an improved understanding of the fundamental processes of breakdown. 

To carry out the prototype development program and the fundamental studies, the CLIC project has made a major investment in experimental infrastructure, in particular in test stands where high-field experiments can be carried out in controlled and well instrumented manner. They consist of radio frequency test stands referred to as the XBoxes, driven by klystrons operating at 11.994 GHz and 2.9985 GHz, and pulsed dc test stands, driven by Marx generators. The radio frequency test stands are capable of powering accelerating structures to full specifications but are rather costly, as are prototype accelerating structures. The pulsed dc systems have been designed to allow directly comparable experiments to those carried out in the radio frequency test stands but using an apparatus that is much simpler, cheaper and more easily instrumented and make tests on relatively simple electrodes. There is an overlap, or near overlap in some cases, in important parameters while an even larger range of certain parameters, for example repetition rate. Because of the overlap, protocols can be reproduced and results can be directly, and quantitatively compared. 

In addition to the high-gradient studies carried out by the CLIC study, there have been a number of complementary efforts by other linear accelerator development programs. As the linear collider projects showed prototypes operating at high-gradients, other types of accelerator applications have adopted similar technology and have in addition become involved in studies of the fundamentals of high gradients. 

The studies have identified important dependencies, effects and measurables and produced an extensive body of well documented experimental data from measurements made on test electrodes and radio frequency structures. The quantities include field emission, breakdown rate, conditioning and dependence on preparation, material, temperature, to name just a few. The behaviours and dependencies will be addressed in depth in section \ref{s:behave}.

The two subsections that follow describe the high-field test stands at CERN that were built and operated for the CLIC study. The different test stands across projects are very similar but the focus here is on the CLIC test stands for clarity of presentation. The emphasis on them is not meant to exclude others. 

Subsection \ref{ss:exp-rf} describes the radio frequency test stands in the hardware, the type of objects that have been tested, diagnostic equipment, control as well as conditioning and operating protocols. The subsection \ref{ss:exp-dc} describes the same for the pulsed dc systems.

\subsection{\label{ss:exp-rf} Radio Frequency experiments}

The experimental equipment, operating conditions, and procedures used to determine the high-gradient performance and long-term operating behavior of radio frequency accelerating structures, as well as experiments dedicated to the fundamental understanding of high-field phenomena, are described in this section. The radio frequency infrastructure was designed to allow testing of prototype linear collider components, in particular accelerating structures, at full operating parameters. As testing proceeded, key behaviors, dependencies and experimental measureables including field emission, breakdown, conditioning and breakdown rate were identified. These effects will be described in the next section, \ref{s:behave}. 

The radio frequency experiments and experimental infrastructure are described in some detail in order to provide a solid basis for understanding the data used for developing the concepts described in this report. In addition, the description of the radio frequency system is important as an introduction to section \ref{s:circ}, which describes the critical role the coupling of power has on the evolution and, it turns out, the onset level of breakdown. The result is an important variation of maximum achievable surface fields as a function of structure geometry in the case of a radio frequency structure, and powering circuits in the case of dc systems. 

A short description will now be made of the basic operating principle, geometry and electromagnetic field pattern of an rf accelerating structure. This can be helpful for many readers since radio frequency structures do not have clearly separated anodes and cathodes as there are in the dc high-field systems that have been the focus of most breakdown studies. The properties of radio frequency structures are described in more depth in section \ref{s:circ}. 

\begin{figure}
    \centering
    \includegraphics[width=1\linewidth]{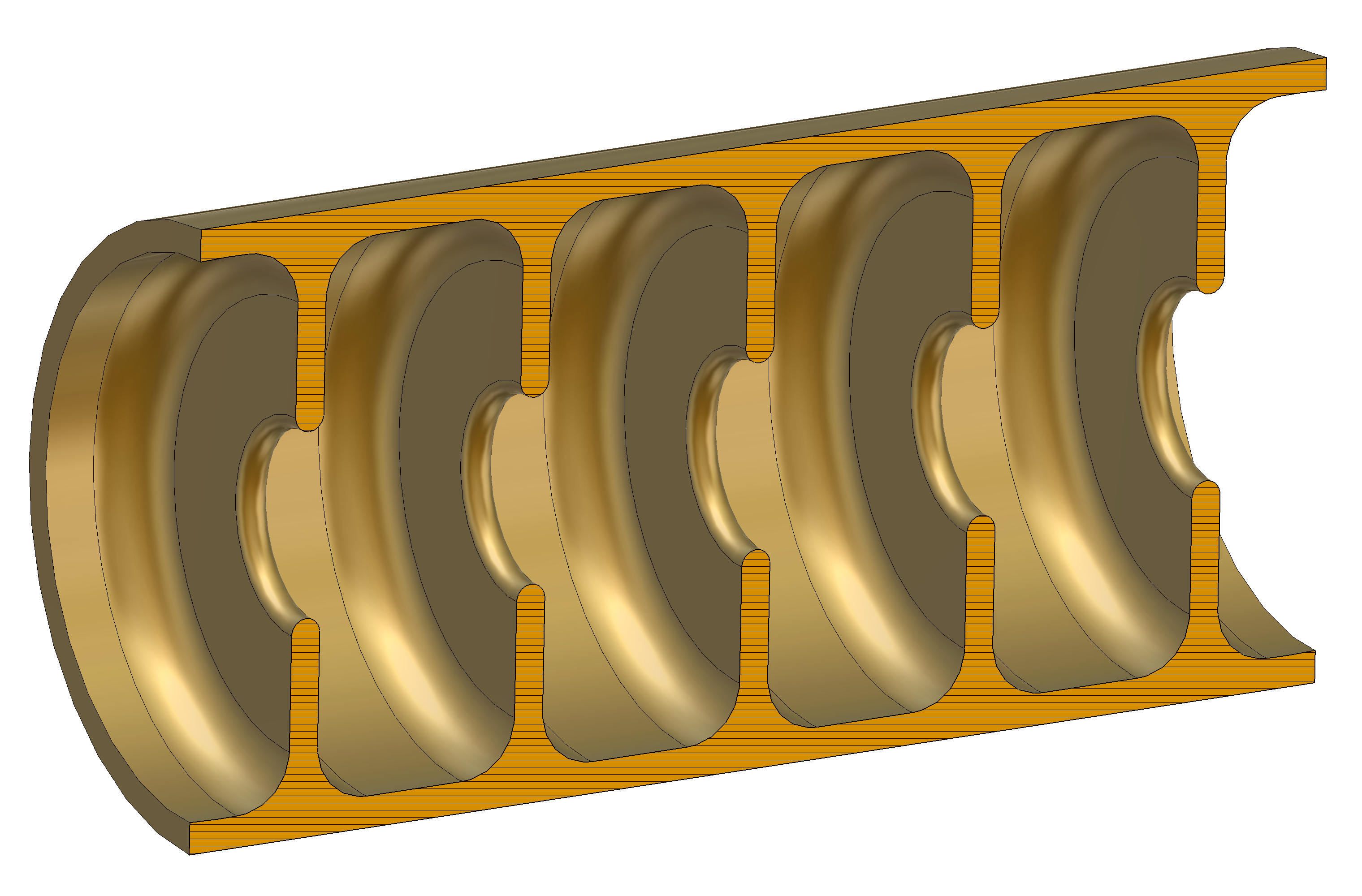}
    \includegraphics[width=1\linewidth]{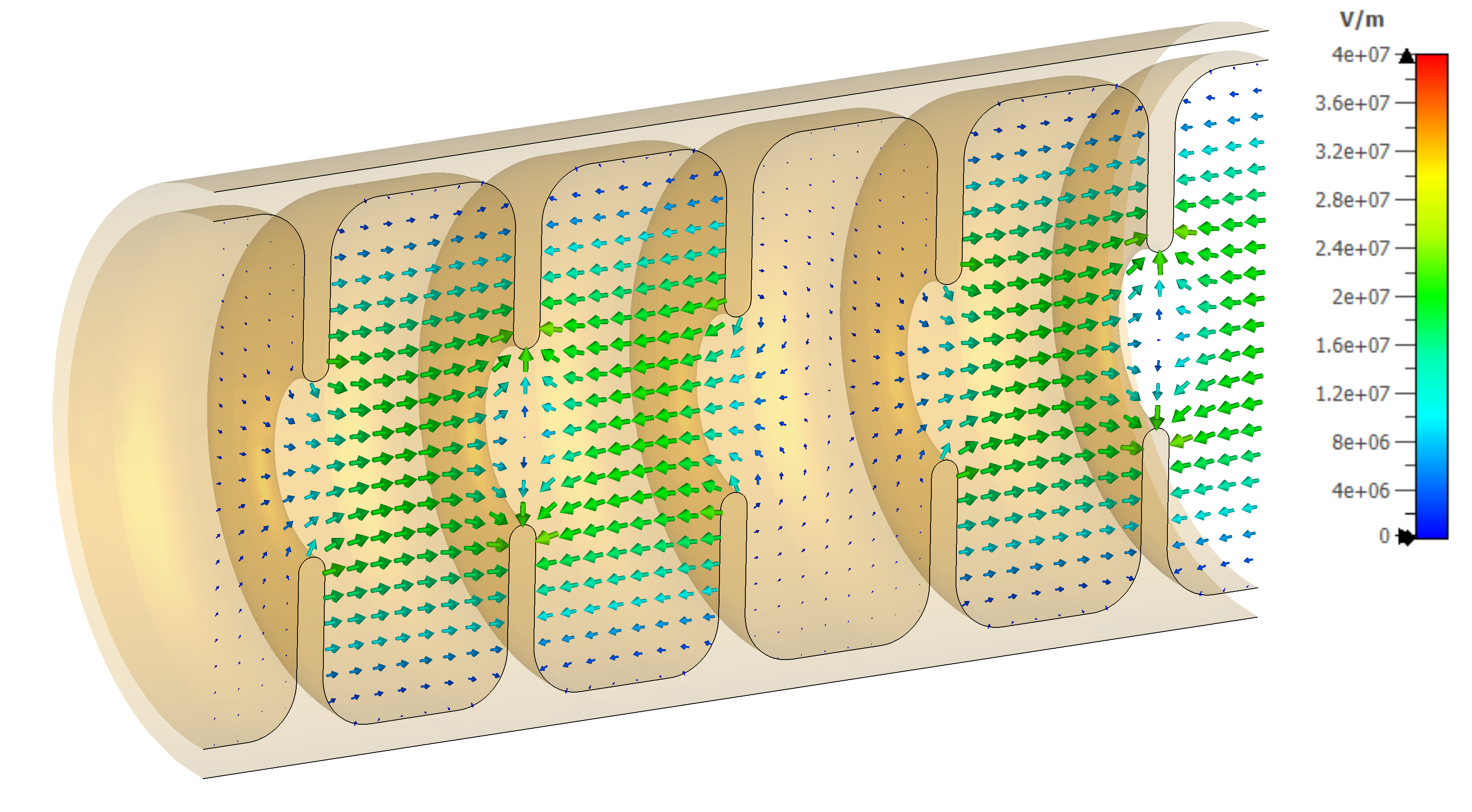}
    \includegraphics[width=1\linewidth]{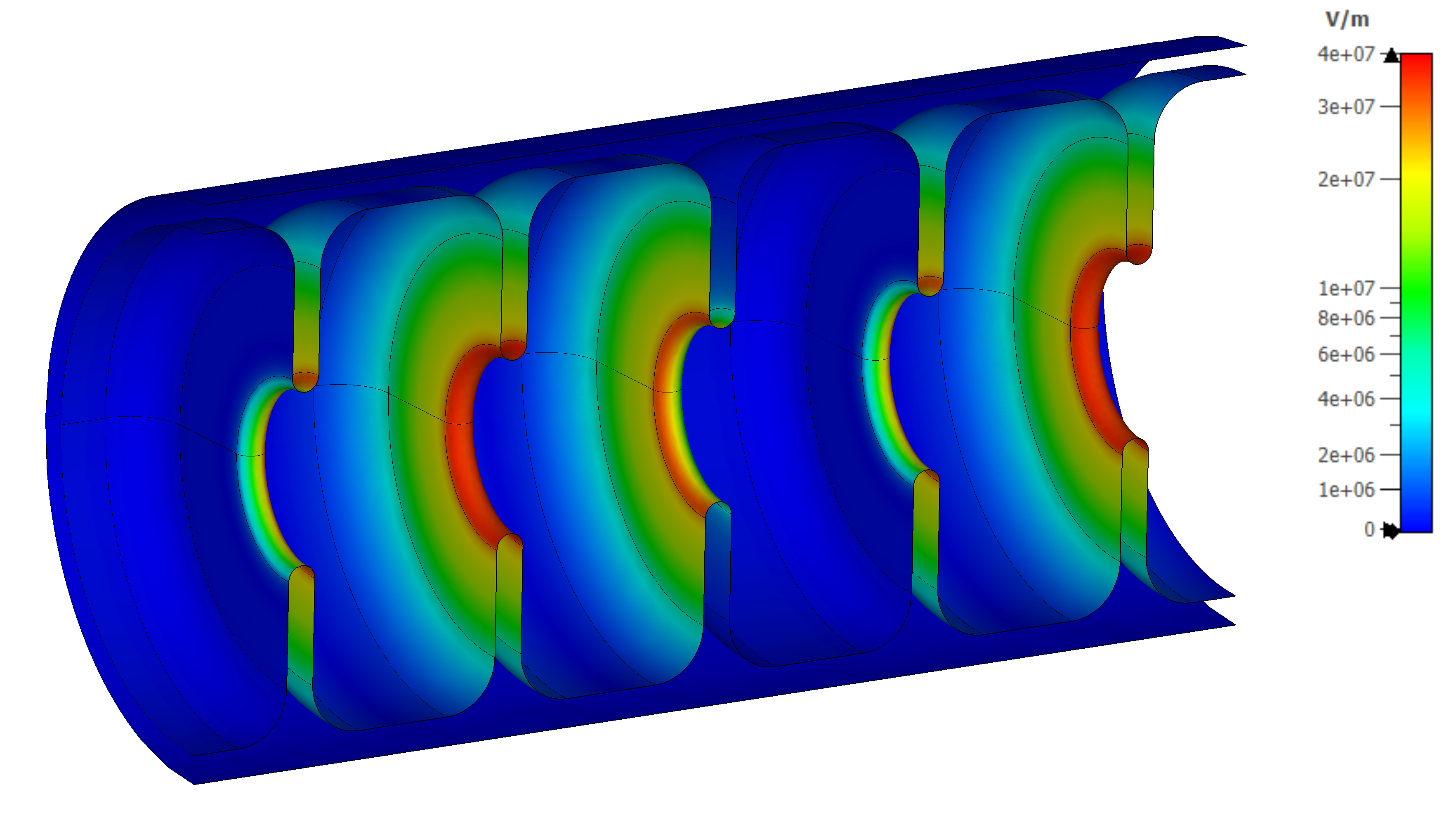}
    \caption{A typical geometry of coupled accelerating cavities is shown in the upper image. The electric fields of a traveling wave $2\pi/3$ mode (wavelength of three cells) in the volume of the cavity are shown in the middle image. The on-axis fields accelerate the beam. The lower plot shows the surface electric field. The highest surface electric field is on the coupling irises, shown in red . This is the area where most breakdowns occur. Figure courtesy of L. Millar}.
    \label{fig:accel geo field}
\end{figure}

The basic configuration of an accelerating radio frequency structure is a set of longitudinally coupled pillbox-like cavities, operating in the TM$_{010}$ mode. The coupling irises also allow for passage of the beam. A schematic examples of the geometry and electric field in accelerating cells are shown in Fig.~\ref{fig:accel geo field}. The electron field, the middle image in Fig.~\ref{fig:accel geo field} provides the accelerating force on a bunched charged particle beam when the beam arrives with the correct phase. When coupled together with the correct phase relationship, a synchronous acceleration is obtained over multiple cavities. Multiple cavities can be coupled either in a so-called standing wave configuration, in which there is a single input power feed, and traveling wave, in which there is both an input and an output power feed. In a standing wave structure, the field builds up over the entire structure with time after the beginning of a power pulse, while in a traveling wave structure, the power pulse travels along the length of the structure.

The phase relationship between successive cells, the phase advance, must be matched to the velocity of the particle beam in order to provide synchronous acceleration over multiple cells. The phase advance in both standing wave and traveling wave structures is determined by the frequency of the cavities, which is primarily a function of the diameter of the cavity, and the coupling strength between the successive cells, determined primarily by the diameter and thickness of the coupling iris. The coupling between the cells also determines the mode separation in standing wave structures and the group velocity in traveling wave structures. 

The geometry of an rf accelerating structure is consequently strongly influenced by the requirements given by synchronous acceleration. Further influence on the geometry is given by the efficiency and power-effectiveness of the acceleration. A fundamental understanding of high-field phenomena can have a significant role in resolving this optimization. We will also see that the field and power coupling that determines the power flow characteristics in the structures also similarly determines the coupling of the power to an incipient breakdown, and through this determines the field-holding capability of a structure. This idea will be elaborated in depth in section \ref{s:circ}. 

The surface electric field in an accelerating structure is concentrated on the inner edge of the coupling iris as shown in the lower image in Fig.~\ref{fig:accel geo field}. Because of this electric field concentration, most vacuum breakdowns are concentrated on the inner edges of the coupling irises. 

\begin{figure}
  \centering
  \includegraphics[width=8.5cm]{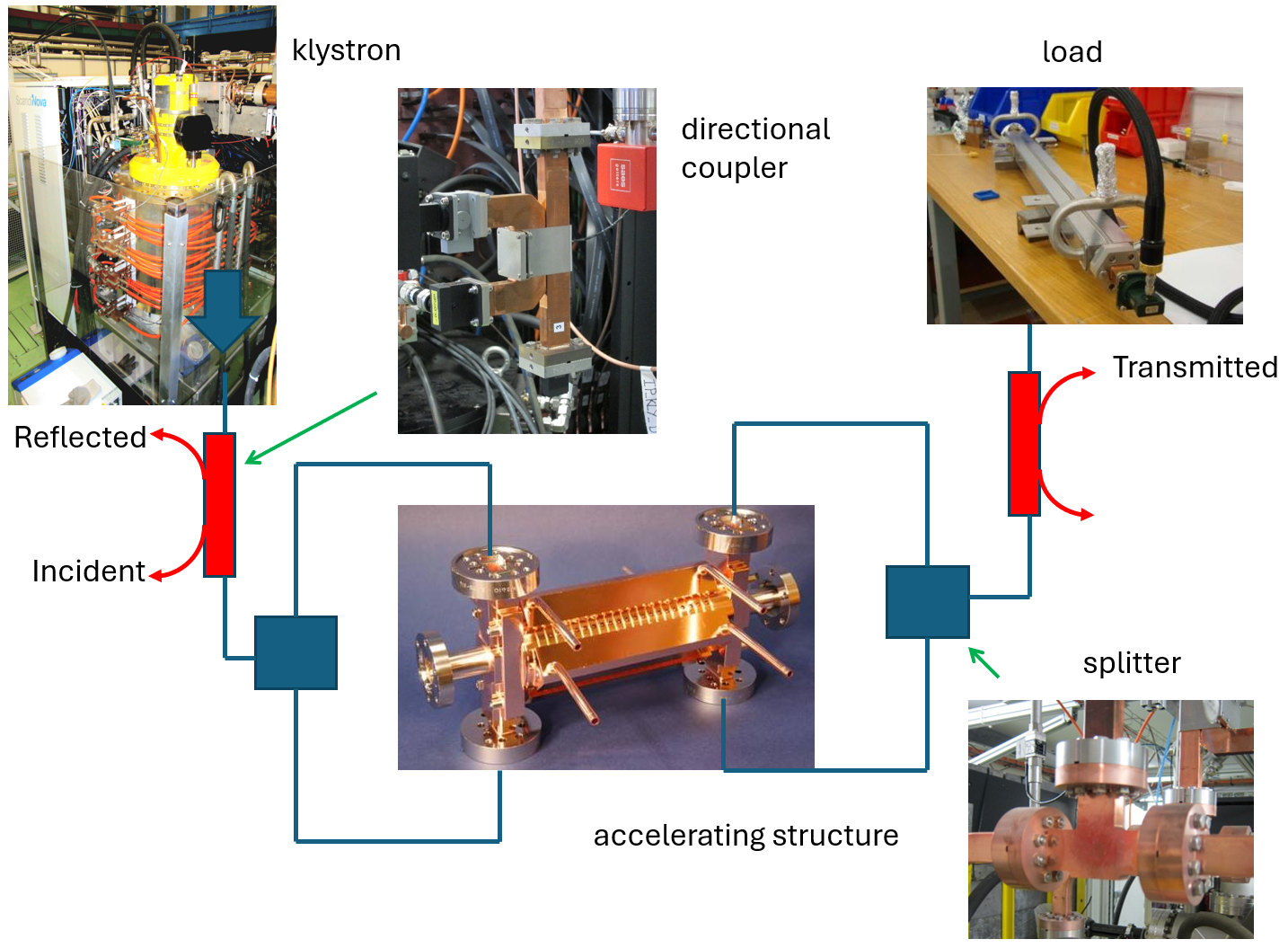}
  \caption{Schematic of the main elements of an rf test stand and also a linac radio frequency unit. Radio frequency power is produced in a klystron, is transported to the accelerating structure and afterwards is absorbed in a load. Directional couplers are used to monitor the rf pulse before and after the accelerating structure.}
  \label{fig_test_stand_schematic} 
\end{figure}

We now turn our attention to the configuration of an rf accelerating system as a whole.  Both linear accelerator systems and the test stands use the same basic configuration and a schematic of such an radio frequency system is shown in Fig.~\ref{fig_test_stand_schematic}.  The basic configuration consists of a MW-range power source called a klystron, a pulse compressor, the structure under test, and a high-power load in the case that the structure is traveling wave, all connected through a waveguide power transmission network. 

The operating frequency of linear accelerators and test stands typically lies in the S, C, and X-bands, along with Ka-band during the first stage of the CLIC project. The electromagnetic fields inside an accelerating structure are established by feeding it with radio frequency power via a transmission line and through an impedance-matching coupler. The transmission line is typically a hollow rectangular waveguide in high-frequency systems. 

In high-gradient experiments and linacs in general, the rf system operates in a pulsed mode. In the case of the CLIC linear collider the radio frequency pulses going into the accelerating structures have a power and pulse length of around 50 MW and 200 ns respectively, for a pulse energy of around 10 J, and are produced at repetition rates in the range of 50 to 100 Hz. This report focuses on vacuum breakdown in pulsed operation, although connections to systems with continuous field exposure are discussed in the section \ref{s:behave}. 

The electric and magnetic field levels inside a radio frequency device depend on the geometry of the device and are proportional to the square root of the input power. These radio frequency properties  can be determined using commercial electromagnetic solvers such as HFSS and CST Microwave studio. The discussion in this section will focus on traveling wave structures since they have made up the bulk of the test devices that underlie analysis described in this review. Power flow considerations in the case of breakdowns in standing wave structures is taken up in the breakdown power coupling section \ref{s:circ}.

Vacuum breakdowns affect the transmission through and reflection from the accelerating structure. In particular, breakdowns cause abrupt, nanosecond-scale changes in power flow with suppression of transmission and large reflected signals. We will return to this point in subsection \ref{ss:behave-breakdown pulse}. Consequently, power flows are one of the main experimental signals. They are sampled through directional couplers or similar devices and are monitored using detector systems that typically consist of power-sensitive diodes, or heterodyne receivers, and signal digitization. The acquisition generally is done pulse by pulse, with a sequence of pulses stored in a buffer.  
 
\begin{figure}
  \centering
  \includegraphics[width=8.cm]{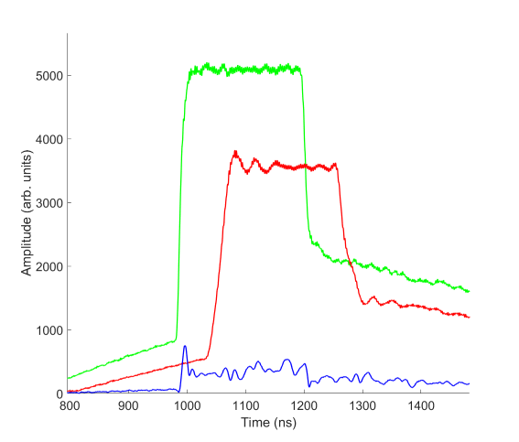}
    \includegraphics[width=8.cm]{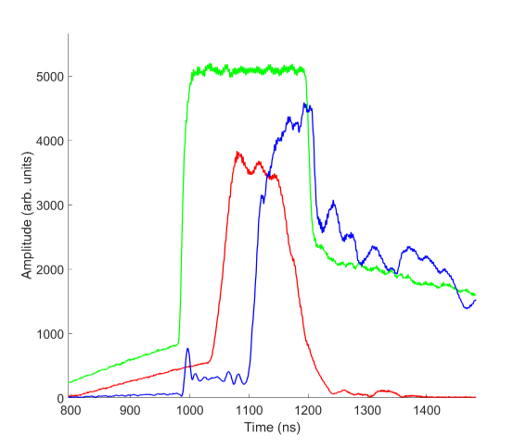}
  \caption
  {
    The power waveforms of a rf pulse without (upper) and with (lower) breakdown in a traveling wave accelerating structure. The incident power is in green, the transmitted in red and the reflected in blue.
  } \label{fig_rf_pulses} 
\end{figure}

Power traces of incident, transmitted and reflected radio frequency pulses of a traveling wave accelerating structure with and without breakdown are shown in Fig.~\ref{fig_rf_pulses}. These are analogous to the voltage and current traces of a dc high-voltage system as we will see in the next subsection \ref{ss:exp-dc}. The drop in transmitted power and concurrent rise in the reflected one are typical of vacuum breakdown in radio frequency structures. In an operating machine, the drop in transmitted power means a collapse in the accelerating field and a fault in the accelerator operation.

One of the most important measures of the operation of an accelerating structure is its breakdown rate. The breakdown rate is defined as the number of pulses with breakdown divided by the total number of pulses in the measurement window. The breakdown rate is strongly dependent on field level, as will be described in depth in section \ref{ss:behave-Statistics}.
Pulsing is typically inhibited after a pulse with a breakdown has been detected. A short time, typically of the order of a second, is given for vacuum levels to return to normal  and pulsing starts again, initially at a lower power level which is gradually returned to nominal. 

When a breakdown occurs a large current is emitted from the breakdown site. As we will see in the breakdown power coupling section \ref{s:circ}, this current accelerated in the radio frequency structure fields is what absorbs incident power and causes reflected power. Most of the breakdown current is accelerated into the radio frequency structure walls, however some of it is accelerated with a longitudinal component and exits the input and output ends of the structure. The measurement of this breakdown current is detected using Faraday cups that are placed outside an accelerating structures under test along the beam axis. On-axis Faraday cups can only be implemented in test stands that do not accelerate an externally injected beam.

The on-axis Faraday cups are also used to measure dark currents. The high-fields in the iris area result in field emission when the phase of the local field makes the surface cathode-like. The high fields in the cavity volume accelerate the emitted currents. Most of the field emitted current collides with the cavity walls however some fraction of the current is swept onto the structure axis, and is then accelerated down the structure length and out the structure. This current can be detected by the same on-axis Faraday cups used to detect breakdown currents.

The radio frequency devices described in this review typically operate in the range \(10^{-7}\) to \(10^{-9}\) mbar. The vacuum level must be continuously monitored since it tends to be high during initial operation, as described in subsection \ref{ss:behave-conditioning} and pressure bursts are caused by breakdowns, as described in Subsec. \ref{ss:behave-breakdown pulse}. 

\subsection{\label{ss:exp-dc} dc experiments}

Pulsed dc experiments have been carried out to complement radio frequency experiments using systems that are simpler, less expensive and easier to instrument. In addition pulsed dc experiments can produce results more quickly because electrodes are simpler to fabricate than a radio frequency structure and in later versions, were carried out at a much higher repetition rate. Overall the parameters of the pulsed dc experiments were chosen to overlap with the radio frequency in important aspects for example electrode material, heat treatment and machining and pulse length but also to extend beyond to provide a broader parameter space for benchmarking the theoretical work. Examples of the latter include pulse length, repetition rate and electrode temperature. 

In a dc experiment voltage is applied across two clearly defined electrodes: cathode and anode. This is an obvious advantage when investigating breakdown compared to rf experiments where the dynamics have a further degree of freedom due to the fast polarity changes. In addition, the electrode geometry can be simpler than an rf cavity so localizing breakdown is simpler. 

\begin{figure}
  \centering
  \includegraphics[width=8.5cm]{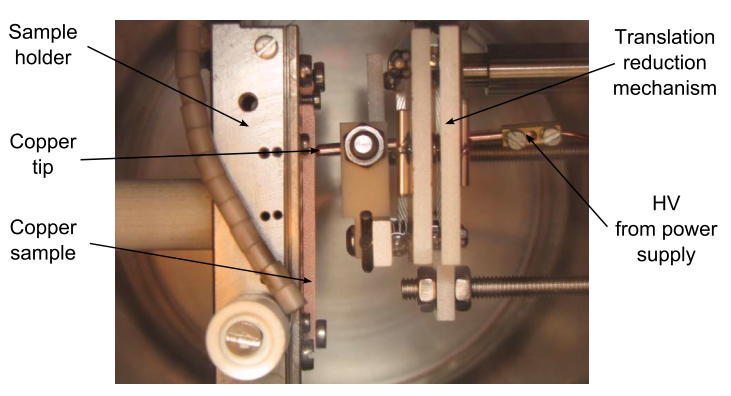}
  \caption
  {
    \label{fig_tip-sample} 
    Close-up image of the anode tip system \cite{kovermann_comparative_2010}.
  }
\end{figure}

The first dc setup of the CLIC study consisted of a tip facing a plane sample, with the former as the anode and the latter as the cathode. In such a configuration, and with a tip of the order of the size of a breakdown plasma spot, the breakdown location is defined by the tip shape and dimensions. The experiments performed at CERN were performed with an anode tip with an apex radius of the order of \SI{100}{\um} separated from the sample by a gap of \SI{20}{\um}, accurately regulated with a micro-manipulator (Fig.~\ref{fig_tip-sample}). The tip and sample were placed in an all-metal ultra-high vacuum (UHV) system, pumped with turbo-molecular pumps down to less than \SI{e-9}{\milli\bar} and instrumented with standard vacuum gauges and a residual gas analyser \cite{kildemo_new_2004}. This type of experiment draws inspiration from the large body of previous literature, in particular for the choice of having the test sample setup as the cathode. Evidence in the literature, as in this review, is that breakdowns start on the cathode and that critical aspects such as the threshold field depends solely on the cathode material. This will be discussed extensively later in this review.  

\begin{figure}
  \centering
  \includegraphics[width=8.5cm]{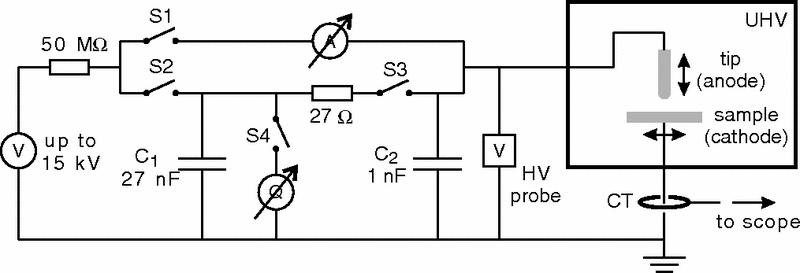}
  \caption
  {
    \label{fig_capacitor_bank} 
    Circuit of the power supply of the dc system. \cite{descoeudres_dc_2009}
  }
\end{figure}

A typical setup of the anode tip system consisted of a capacitor bank charged by a dc power supply, as shown in Fig.~\ref{fig_capacitor_bank} (represented in the lower branch). Once the capacitor bank was charged, it was connected either by mechanical or electronic switches to the electrode allowing for a fast rise of the potential (often in the \SI{}{\ns} range), accompanied by a displacement-current surge due to the intrinsic and stray capacitance of the system and its connecting cables. The high voltage was held for a defined time interval, typically from a few hundreds of \SI{}{\ns} to hundreds of \SI{}{\us}, or even longer intervals up to seconds, depending on the type of experiment, and on the circuit hardware. When a breakdown happens, it manifests itself through a discharge current of several tens or hundreds of \SI{}{\ampere}, which is easily detected by the control electronics and found typically to be in coincidence with pressure spikes in the vacuum system \cite{descoeudres_investigation_2009} and with optical emission from the plasma \cite{kovermann_comparative_2010}. In this case the capacitor bank is entirely discharged through the arc, and the experimental cycle restarted by charging again the capacitor bank typically at a lower voltage. In case an arc did not develop during the programmed pulse duration, the capacitor bank is discharged on a dumping resistance. The cycle is then restarted, either at the same field in case of breakdown rate experiments, or higher field in case of ramp-to-breakdown experiments. The arc current depends mostly on the energy stored on the capacitor bank (which in turn depends also on the applied voltage) and on a possible matching series resistance \cite{timko_energy_2011}.  The voltage across a vacuum arc, and thus its electrical resistance depends on the cathode material and is of the order of \SI{20}{\volt} for copper \cite{anders_cathodic_2008}. Repetition rate of such cycles ranged from several seconds per cycle in the earliest experiments, to several \SI{}{\hertz} when more performing solid-state electronics was adopted.


An important result was the demonstration of the similarity observed by scanning electron microscopy (SEM) \cite{kildemo_breakdown_2004, timko_energy_2011} of the breakdown craters on the cathode of the pulsed dc system and on the surface of rf cavities for a similar stored energy, as will be discussed later in subsection \ref{ss:behave-microscopy}. This confirmed the relevance of these experiments to the study of high-gradient rf accelerating structures.


Another important experiment that was routinely performed is the measurement of field emission current. In this case, the power supply is connected directly to the system via a limiting resistor and the FE current is read with a multimeter (represented in the top circuit branch of Fig.~\ref{fig_capacitor_bank}). The voltage is increased from zero to the value producing a FE current of $10^{-7}$~\SI{}{\ampere}, which is a safe value. With such a low FE current, a breakdown is not triggered and damaging the multimeter is avoided.
From the field emission it is then possible to extract standard theoretical parameters such as the Fowler-Nordheim field enhancement factor $\beta$. This measurement is performed at various stages of the experiments, and in some cases even before and after each high-voltage pulse in order to identify the evolution of $\beta$ at each pulse and breakdown \cite{descoeudres_investigation_2009}. Field enhancement factor $\beta$ for various metals has been measured, and ranges typically between 30 and 50 for different metals and at different conditioning stages \cite{kildemo_breakdown_2004, descoeudres_dc_2009}. Field emission data is presented in subsection \ref{ss:behave-field emission}.



Total and partial pressure measurements in the vacuum system allowed identification and quantification of the amount of the different gases released during a breakdown \cite{descoeudres_investigation_2009, levinsen_quantitative_2013}. These quantities are of great importance when translated into a real accelerating structure, since the gas released may have an impact on the beam stability in a real accelerator if not adequately pumped away from the vacuum system before further particle bunches traverse the machine \cite{aicheler_multi-tev_2012}.

\begin{figure}
  \centering
  \includegraphics[width=1\linewidth]{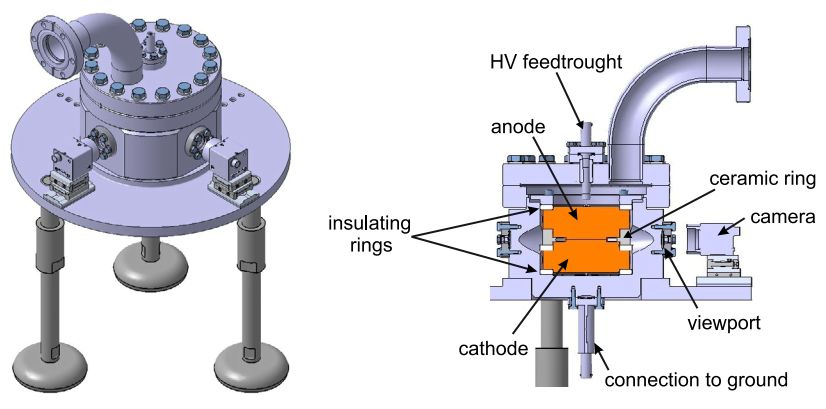}
  \caption
  {
    \label{fig_LES} 
    Image of the LES system \cite{profatilova_breakdown_2020}.
  }
\end{figure}

Moving from mechanical switches to fast solid state electronics in the subsequent generations of the test setup allowed pulse repetition rates increase from sub-Hz to several Hz and later to several kHz. At the same time, the micron-accuracy manufacturing technologies devised for the rf structures had become mature enough to allow fabrication of large components at moderate cost. As a consequence, a test system where two large flat electrodes were constructed. The electrodes are kept separated by a gap of a few tens of microns by an insulating ceramic spacer. The electrodes are exposed to high-repetition dc high-field pulses for exploring the breakdown rate behaviour of large surfaces, adding the statistical and stochastic nature of breakdown development to the scope of the study. This Large Electrode System (LES), also known as the pulsed dc system, has been described in detail in \cite{profatilova_breakdown_2020}, and is illustrated in Fig.~\ref{fig_LES}. 

\begin{figure}
  \centering
  \includegraphics[width=1\linewidth]{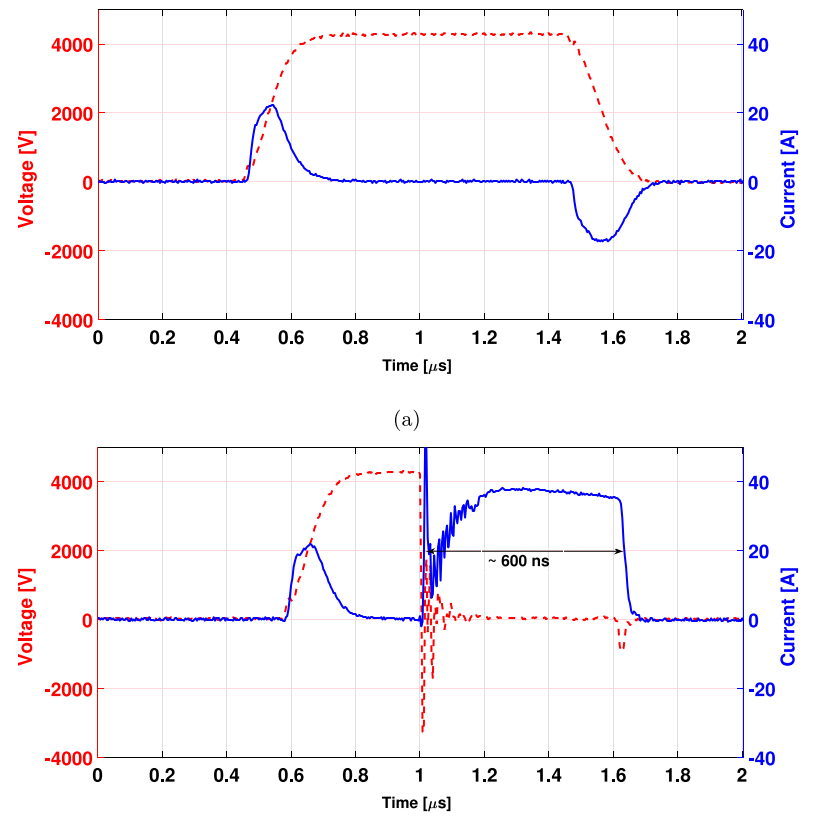}
  \caption
  {
    Voltage and current traces of a pulse in the LES system. The top panel shows the voltage (red) and current (blue) traces in normal operation; the current pulses are due to displacement currents charging and discharging  the electrode. The bottom panel shows voltage and current traces in the case of a breakdown~\cite{profatilova_breakdown_2020}.
  }     \label{fig_LES_pulse} 

\end{figure}

\begin{figure}
  \centering
  \includegraphics[width=0.85\linewidth]{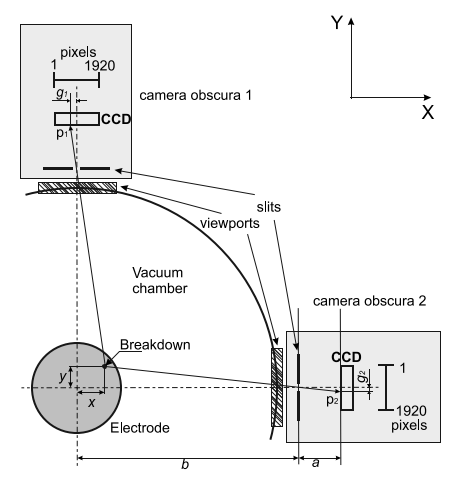}
  \includegraphics[width=1\linewidth]{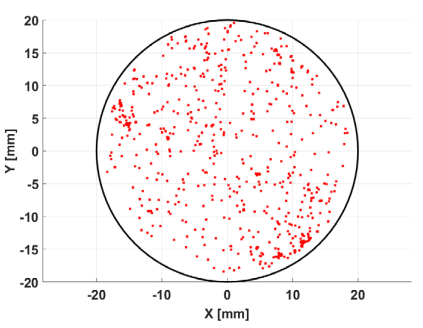}
  \caption
  {
    \label{fig_LES_cameras} 
    Layout of the cameras in LES system used for breakdown positioning through triangulation, upper image. An example of the distribution of reconstructed breakdown locations, lower image. Localization resolution is~\SI{15}{\um}, smaller than the average crater size~\cite{profatilova_breakdown_2020}.
  }
\end{figure}

\begin{figure}
  \centering
  \includegraphics[width=1\linewidth]{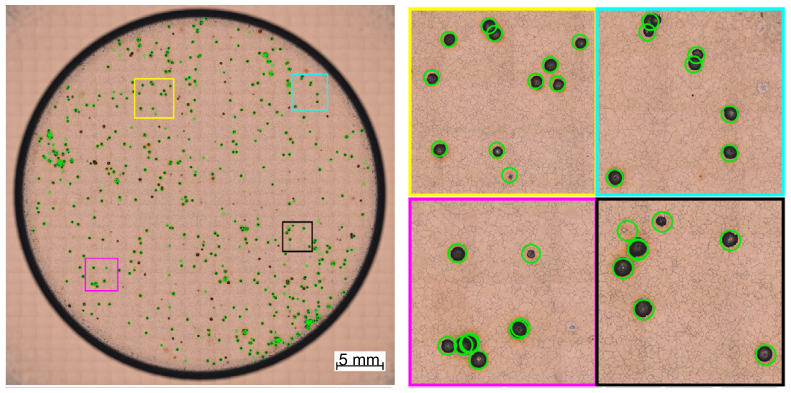}
  \caption
  {
    \label{fig_LES_BDs} 
    In this figure the optical microscope image of the breakdowns (black spots) has been superposed with the breakdown locations measured with the cameras, indicated by the green circles (size of the circles is about~\SI{150}{um})~\cite{profatilova_breakdown_2020}.
  }
\end{figure}

The pulsed dc system consists of a UHV chamber where two large electrodes (of various diameters depending on the experiments, up to \SI{60}{\mm} of active surface) are hosted, separated by a precision machined ceramic spacer maintaining a gap between the two active surface (typically \SI{60}{\um}, although several spacers are available allowing gaps ranging from \SI{20}{\um} to \SI{100}{\um}). The most recent version of this test system is supplied by a Marx pulse generator, specially developed for the purpose \cite{redondo_solid-state_2016}. Depending on the experiment, either electrode can be connected either to positive or negative polarity, and pulses of arbitrary duration from a few 100s of ns to several 100s of \SI{}{\us} can be generated with a rise-time of the order of \SI{100}{\ns}, as illustrated in Fig.~\ref{fig_LES_pulse}, where typical current and voltage traces are shown in the two cases when a breakdown is observed or not. The availability of very fast electronics with pulse repetition rate up to \SI{10}{kHz} allowed working in very low breakdown rate regime ($10^{-6}$ or lower) and reproducing the conditioning process as performed in typical rf tests. In particular, a conditioning algorithm similar to what developed for rf X-band tests \cite{wuensch_experience_2014} could be developed and adapted to pulsed dc tests. Moreover, the system has been designed with viewports on the sides, which allowed the installation of CCD cameras to detect the light emitted by breakdowns, and to reconstruct their position through triangulation, as shown in Fig.~\ref{fig_LES_cameras}. The location of breakdowns can then be easily correlated with their counterparts acquired through microscopic imaging of the electrodes surfaces, as illustrated in Fig.~\ref{fig_LES_BDs}, and with the electrical signals acquired by the pulse generator to analyze for example their size and their spatial and temporal distributions. These diagnostics are complemented by the usual vacuum pressure recordings discussed for the anode tip original test system.

\begin{figure}
  \centering
  \includegraphics[width=8.5cm]{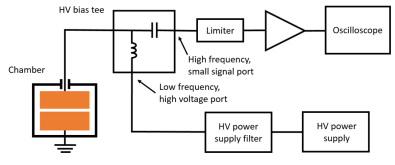}
  \caption
  {
    \label{fig_LES_fluctuations} 
    Circuit for measuring fluctuations with the LES system \cite{engelberg_dark_2020}.
  }
\end{figure}

Field emission measurements are performed with this setup with a different circuit, by measuring the voltage drop over a large resistance in series with the gap. 
Measurements are in this case performed in pulsed mode, with pulses of the order of \SI{100}{\micro\second} to allow for sufficient stabilization time of the electronics. However, a further development to analyze current fluctuations has been introduced \cite{engelberg_dark_2020}, to allow large bandwidth high-frequency fluctuations to be detected while performing a dc field emission scan. This is performed through a purpose-built high-voltage capacitive coupling setup, illustrated in Fig.~\ref{fig_LES_fluctuations}, which guarantees a sub-\% sensitivity when measuring sub-$\mu \textrm{s}$ spikes on top of a dc signal, as will be discussed later.


\section{\label{s:behave}Behaviour of high-field systems }
\subsection{\label{ss:behave introduction} Introduction}

In this section we present a selection of the most important measurement results from the high-power radio frequency and pulsed high-voltage dc systems. The results are grouped in subsections that represent the main categories of experimental signatures of breakdown and different aspects of breakdown initiation and evolution. This grouping also reflects the different aspects of the theoretical analysis of breakdown as are described in the different sections in this report. Both rf and dc results of a particular type are presented together in order to highlight similarities but also differences in behaviour between the two types of systems. 

\subsection{\label{ss:behave-breakdown pulse} Signals Produced During Breakdown}

In this subsection we present data that illustrate features that are characteristic of breakdowns taken individually. These features include how applied fields collapse, the contemporaneous burst-like signals of electron, radiation and light emission, as well as vacuum spikes. These features are important for understanding how breakdowns evolve, the bottom three boxes in Fig.~\ref{fig:overall scheme} and the subject of sections \ref{s:emit}, \ref{s:plasma} and \ref{s:circ}. The data shown are by necessity selected examples since most of the breakdown features are random and noisy. The examples are chosen to illustrate features clearly but with care so as not to bias the discussion. Some categories of randomness in the features has been identified to be important and these will be subsequently covered in the statistics subsection \ref{ss:behave-Statistics}.  

As we have seen in the description of radio frequency systems, section \ref{s:exp}, a traveling wave accelerating structure, has input, transmitted and reflected power flows, as shown in Fig.~\ref{fig_test_stand_schematic}. From a macroscopic perspective, a breakdown within a traveling wave device acts like a localized short circuit. The short circuit is formed by the large electron currents created as the breakdown site develops its plasma spot and resulting plasma sheath driven current as described in section \ref{s:plasma}. The liberated electrons are accelerated by the incoming radio frequency fields and thus acts like a short circuit, resulting in a near full reflection of incoming power and near full suppression of transmitted power as is seen in Fig.\ref{fig_rf_pulses}. The corresponding feature of breakdown formation in the pulsed dc system is the sudden collapse of the  voltage applied across the electrodes with a corresponding rise in the current flowing through the electrode gap. Here also the breakdown appears as a short circuit, but in this case occurring across the electrode gap. The corresponding $V$ and $I$ traces for the pulsed dc system are shown in Fig.~\ref{fig_rf_pulses} in section \ref{s:exp}.

From the length of the drop-time in transmission and rise-time in reflection, as shown in Figs.~\ref{fig_rf_pulses} and \ref{fig_LES_pulse}, one can determine that the development time of a breakdown short circuit is very short, less than 10 ns from all available data. The time is given as an upper limit since the measurement is bandwidth limited by the structure, detection electronics and the signal acquisition. 

\begin{figure}
    \centering
    \includegraphics[width=1\linewidth]{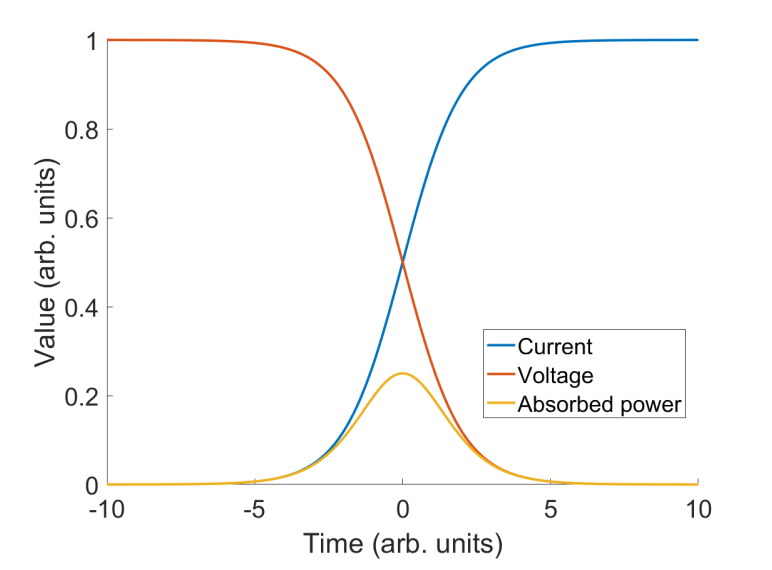}
    \caption{Idealized plot of the voltage, current and power absorbed during a breakdown transient. The power absorbed by a breakdown is maximum during the transient, when the applied voltage has partially collapsed and the breakdown current is still rising. Taken from \cite{paszkiewicz_studies_2020}.}
    \label{fig:power during breakdown}
\end{figure}

The breakdown formation transient is critical since this is when the most power is absorbed by the breakdown. Since power absorbed is $IV$, where $I$ is the instantaneous breakdowns current and $V$ is the local voltage, the maximum power absorbed is when the current has risen but the applied field has not yet collapsed. An idealized example of this transient power coupling is shown in Fig. \ref{fig:power during breakdown}. A corresponding power spike is observed in radio frequency power, through the calculation of so-called missing energy. This missing energy on a pulse is the difference between total incoming energy minus that transmitted and reflected. It peaks during the time when transmission is collapsing and reflection is building. A high gradient limit based on an analysis of what occurs during the breakdown transient will be developed in section \ref{s:circ}.

The electron current emitted during breakdown, the one causing the field collapse described above, can also be observed directly in rf accelerating structure experiments by Faraday cups installed along the beam axis. Most of the breakdown current is accelerated and then lost on the cavity walls, but a fraction is captured by the accelerating rf and is transported along the central beam axis, the beam aperture, out of the structure. It can then be directly measured using on-axis Faraday cups and connected to an oscilloscope. Example of such Faraday cup signals on normal pulses and pulses without breakdown are shown in \cite{obermair_explainable_2022}. 

\begin{figure}
    \centering
    \includegraphics[width=1\linewidth]{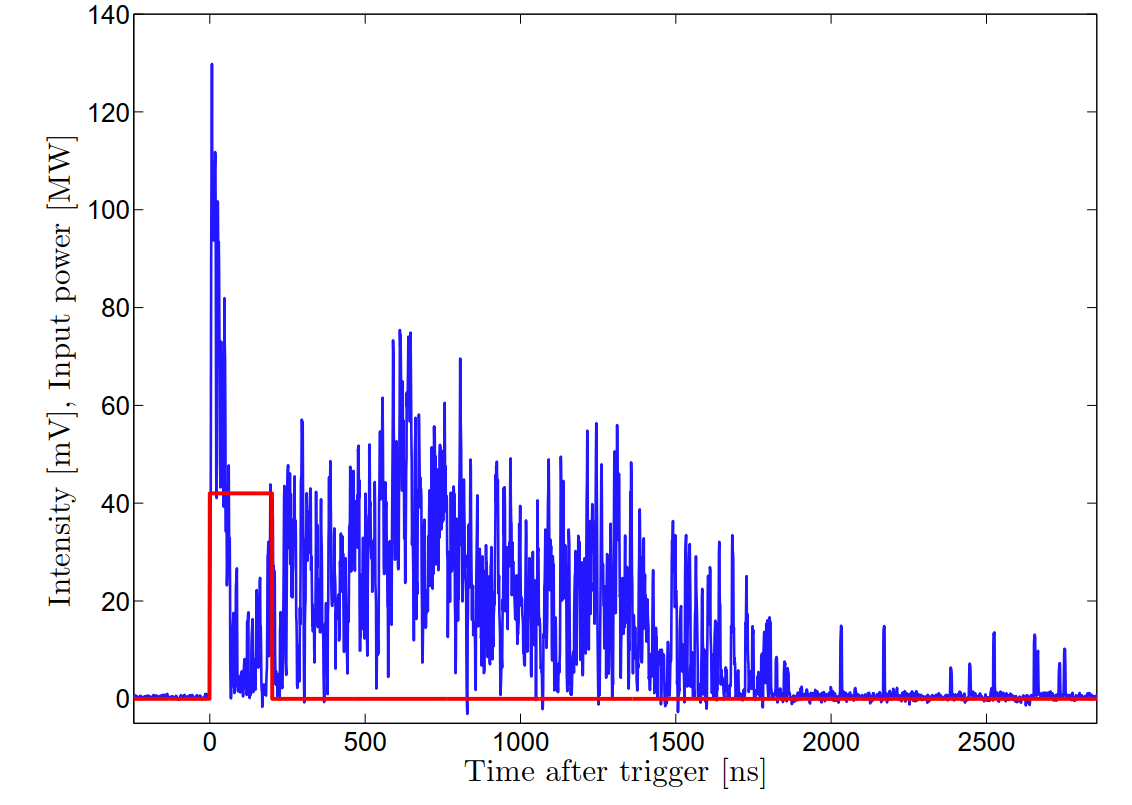}
    \caption{The time dependence of light emitted by a breakdown in a radio frequency accelerating structure. The data shows a fast initial spike followed by a long emission. The initial spike has a broad continuous optical spectrum, while the long emission is composed of a forest of discrete spectral lines, mostly of copper transitions. The long emission lasts much longer than the time of the applied radio frequency power which is shown in red. Taken from \cite{kovermann_comparative_2010}. }
    \label{fig:time light}
\end{figure}

Another important experimental signature of breakdowns is the emission of light. The light has many features and carries temporal, spatial and spectral information. An example of the time behavior of light is shown in Fig. \ref{fig:time light}. The data show a light spike associated with the breakdown transient and a long emission associated with the breakdown plasma. The light is emitted from a small area, so it can be imaged and used to determine the position of a breakdown as shown in the pulsed dc system, Fig. \ref{fig_LES_cameras}. Examples of positioning in an rf structure are presented in \cite{jacewicz_spectrometers_2016}. Extensive data on the optical spectrum emitted during a breakdown in both rf and dc systems are compiled in \cite{kovermann_comparative_2010}. Optical emission also occurs during field emission conditions and will be discussed further in the field emission subsection \ref{ss:behave-field emission}. 

\subsection{\label{ss:behave-Statistics} Breakdown Statistics}

We now consider the statistical characteristics of breakdowns which emerge when considering an ensemble of breakdowns. One of the most directly observable and important statistical properties of a pulsed high-field device is that it operates with a breakdown rate. Stable operation of a large-scale facility like the CLIC linear collider creates the requirement that breakdown rate of individual components, in particular the accelerating structures, be very low $O(10^{-7})$. The origin of breakdown rate will emerge from the dislocation model described in Sec. \ref{s:dis} Plastic Material Response. 

\begin{figure}
    \centering
    \includegraphics[width=1\linewidth]{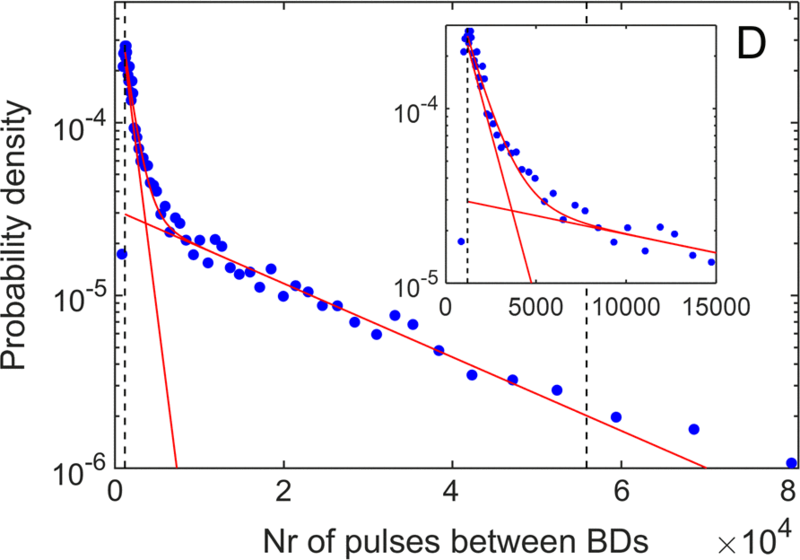}
    \includegraphics[width=1\linewidth]{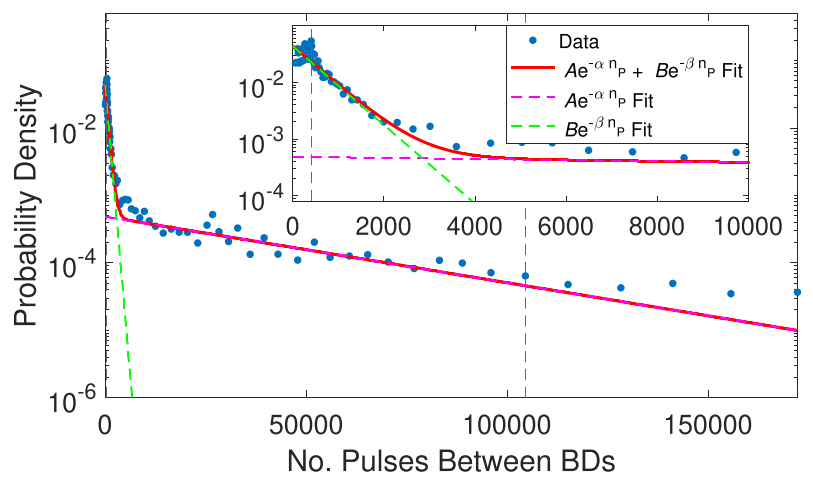}
    \caption{The distribution of intervals of the number of pulses between successive breakdowns in data taken from the pulsed dc system, upper, and a traveling wave radio frequency accelerating structure, lower. The double exponentials are fits to the data. The longer interval of the distributions shows primary breakdowns, while the shorter intervals show follow-up breakdowns. From \cite{wuensch_statistics_2017,millar_high-power_2023}.}
    \label{fig:BD statistics double fit}
\end{figure}

When a high field device operates for an extended period, a distribution of intervals between successive breakdowns emerges that can be visualized by a histogram of the number of pulses between breakdowns. Typical examples of such histograms, one dc and one radio frequency, are shown in Fig.~\ref{fig:BD statistics double fit}. The histograms are not simple Poisson distributions, which would be the case if breakdowns occurred randomly, but rather have two parts, each Poisson-like. Breakdowns occurring after a long interval are caused by the underlying, long-term breakdown rate mechanism and are referred to as primary breakdowns. Those occurring after a short interval, typically a few thousand pulses, are caused by the after-affects of a previous breakdown and are referred to as follow-up breakdowns. The source of a follow-up breakdown can be the droplets or sharper features seen in breakdown craters as shown described in Sec. \ref{ss:behave-microscopy}. After the most unstable features either initiate a follow-up breakdown or are eroded by field exposure, the system returns to the long-term breakdown rate. The long-term breakdown rate is given by dislocation dynamics, as will be described in Sec. \ref{s:dis} Plastic Material Response. More data and analysis of this phenomenon can be found in \cite{rajamaki_vacuum_2016,wuensch_statistics_2017,saressalo_effect_2020, millar_high-power_2023}. 

There is also a distribution of the distance between successive breakdown locations, and we shall see that it is related to the distribution of the number of pulses. The localization of breakdowns in the pulsed dc system is made using triangulation of emitted breakdown light as we have seen subsection \ref{ss:exp-dc}. The longitudinal position of breakdown positions in a traveling wave radio frequency structure can also be determined. The technique measures the breakdown position from the relative timing of the rising edge of the reflected wave and falling edge of the transmitted wave with respect to the beginning of the incident pulse along with the group velocity in the structure. This measurement is feasible due to the steeply rising/falling edges of the reflected/transmitted wave.

\begin{figure}
    \centering
    \includegraphics[width=1\linewidth]{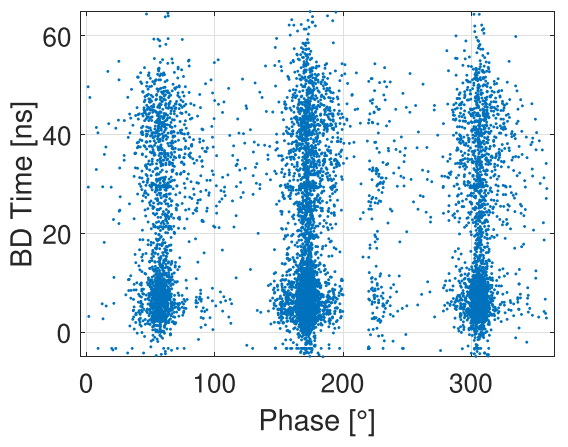}
    \caption{Breakdown position determination in a traveling wave accelerating structure using both timing, vertical axis, and reflected phase, horizontal axis. The breakdowns can be seen to cluster into three distinct phases, which here reflect the 2$\pi$ cell-to-cell phase advance of the structure. The roughly uniform density of spread of points in the timing indicates that breakdowns are uniformly spread over the length of the structure. Taken from \cite{millar_high-power_2023}.}
    \label{fig:TW BD position}
\end{figure}

Complementary position information is contained in the relative phase between the incident and the reflected wave from the breakdown. A traveling wave accelerating structure has a characteristic phase advance. One can see this phase advance in plots of the reflected waves, which are grouped into distinct phases that correspond to two times the cell-to-cell phase advance in the structure. This indicates that the breakdowns and associated short circuits are localized to the same location in each cell. This is consistent with post-test imaging, as we will see below, that show that breakdown craters are concentrated on the coupling iris in this type of structure. Combining timing and reflected phase gives an enhanced accuracy in determining the longitudinal position in a traveling wave accelerating structure, as shown in Fig.~\ref{fig:TW BD position}. Determination of breakdown position in a traveling wave accelerating structure technique is described in more depth in \cite{rajamaki_vacuum_2016}. 

\begin{figure}
    \centering
    \includegraphics[width=1\linewidth]{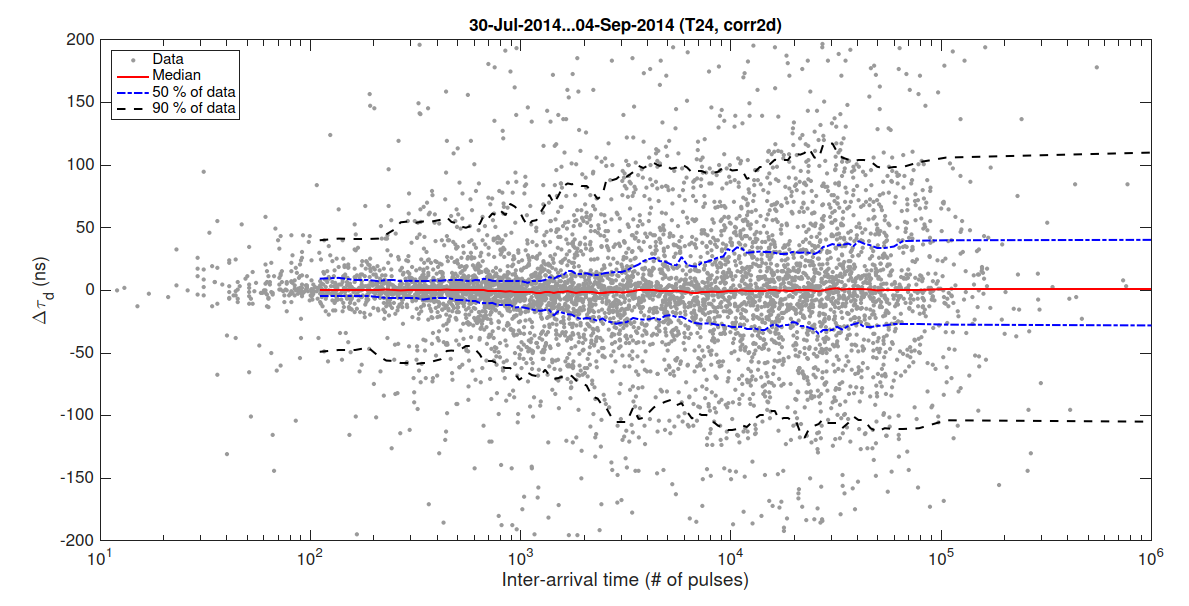}
    \caption{The spatial and temporal correlation of successive breakdowns in a traveling wave accelerating structure. Breakdowns that are closer in time are also closer in space. Taken from \cite{rajamaki_vacuum_2016}.}
    \label{fig:spatial temporal TW}
\end{figure}

Temporal and spatial distributions can now be combined into a single plot, in order to motivate into the physical explanation for the two part Poissonian distribution seen in Fig. \ref{fig:BD statistics double fit}. Such a combined plot produced from data taken in high power testing of a traveling wave accelerating structure is shown in Fig.~\ref{fig:spatial temporal TW}. The plot shows that successive breakdowns separated by relatively fewer pulses tend to occur closer together, and conversely those separated by a large number of pulses tend to be further apart. The most plausible explanation for the short timescale effect is that breakdowns produce surface features, for example crater-like bumps and splashes of metal as we will see in the post-test microscopy discussion below, that result in an enhanced propensity to breakdown. The enhanced tendency to break down decreases with time, either because features are eliminated during the subsequent breakdowns or because the features are smoothed without breakdown over a certain number of pulses. Once the aftereffects of a breakdown are neutralized, the high-field surfaces return to their underlying breakdown rate which is due to the mechanisms described in section \ref{s:dis}. 


The likelihood of having a follow-up breakdown is influenced by how fields are adjusted following a breakdown. In many systems the field level is reduced on the pulse that follows immediately after a breakdown. The field is then ramped back up to the niminal operating value in a pre-defined sequence on subsequent pulses. Such a breakdown-recovery ramp typically takes a hundred to a thousand pulses in the pulsed systems described in this review.  A comparative study of different post breakdown ramps, was carried out in \cite{saressalo_linear_2021, saressalo_effect_2020} with the objective to find a ramp which minimized the number of follow-up breakdowns.

\begin{figure}
    \centering
    \includegraphics[width=1\linewidth]{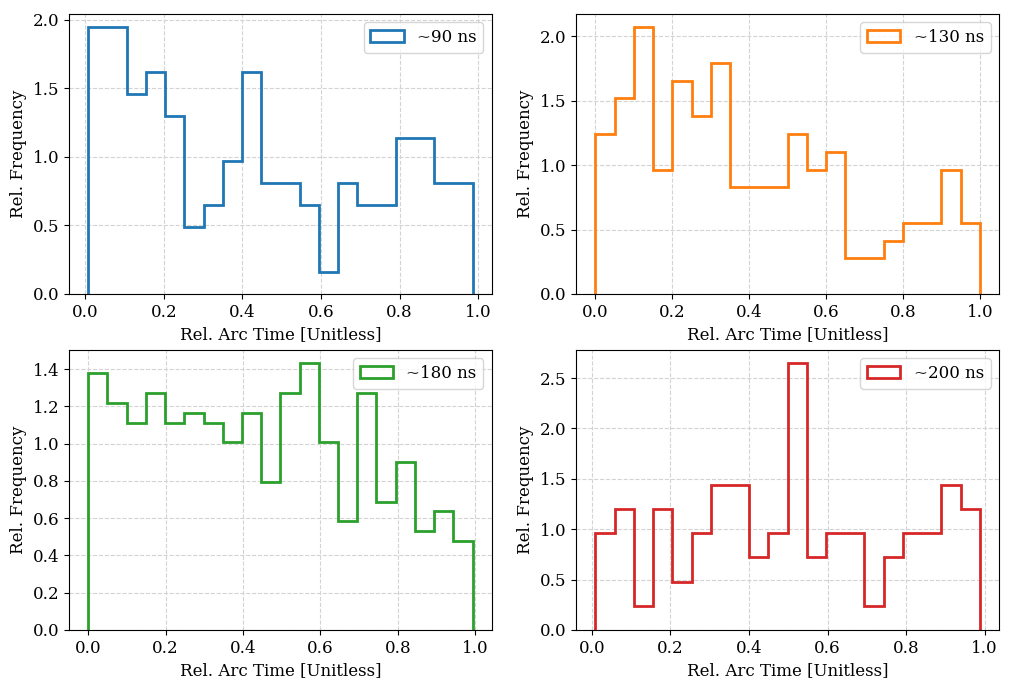}
    \caption{These plots show the distribution of when inside a pulse breakdowns occur. The data are from an X-band accelerating structure and were taken at different pulse lengths.}
    \label{fig:BD in pulse dist}
\end{figure}

Another important statistical behavior that is observed during pulsed mode operation, is the time-to-breakdown within the pulses. An example of such a distribution is shown in Fig. \ref{fig:BD in pulse dist}. The striking feature of this distribution is that it is relatively flat, for all the different pulse lengths and is not, for example, rising towards then end of the pulse. This would be the case if breakdown initiation was the result of a heating process occurring within individual pulses. The observed overall $\tau^5$ dependence is clearly not produced by a higher density of breakdowns at the end of the pulse. This means that pulse length dependence is produced by preceding pulses, which which we can refer to as a pulse-to-pulse memory effect. Further time-to-breakdown data can be found in \cite{martinez-reviriego_high-power_2024, descoeudres_investigation_2009}. The time-to-breakdown within pulses is discussed further in Sec. \ref{ss:dis pulse length}. 

\subsection{\label{ss:behave-conditioning} The conditioning process}

Conditioning, in the context of high-field systems, is the process by which a newly fabricated device must be initially operated at reduced parameters but is then gradually brought up to full operating values through controlled operation. The necessity for conditioning is that as-manufactured devices typically need to be initially operated at reduced parameters because they exhibit enhanced vacuum degassing and breakdown. Remarkably however, the field-holding level of a device can be improved by operating it with applied field in a controlled way, because the act itself of applying the field improves the device, allowing it to run at progressively higher fields. This effect, called conditioning, and its consequences are the subject of this subsection. 

Conditioning is often a time-consuming process, so costly, and a risky one since irreversible damage can occur if for example, fields are increased too rapidly, an excessive number of breakdowns occur and the device surface is damaged. Almost all high-field devices require applied-field conditioning despite efforts to fabricate in ways that are favorable for direct high-field operation, for example by high-quality machining, heat treatment, chemical cleaning, limiting exposure to air and dust, etc. Conditioning is necessary for all of the high-field radio frequency and dc devices described in this report. Conditioning also applies to the field recovery process that is often necessary following anomalous operating conditions, for example after breaking of vacuum and exposure to the atmosphere during a repair or operational failure.

Understanding what is actually changing in a device as it progresses from an unconditioned to a conditioned state has benefits that are both theoretical, important input for the understanding the different mechanisms that initiate breakdown, and practical, for guiding the development of fabrication methods that reduce conditioning time.

\begin{figure}
  \centering
  \includegraphics[width=8.5cm]{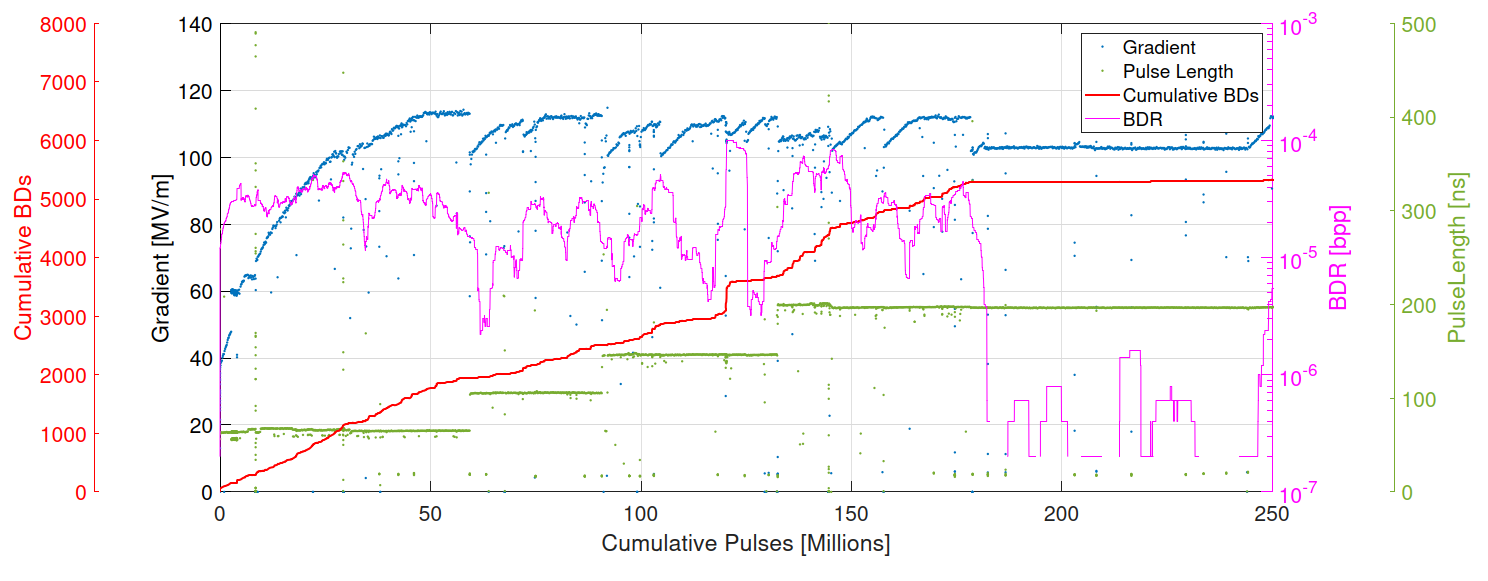}
  \caption{An example of a conditioning curve of a radio frequency structure. The blue points show the operating gradient as a function of pulses applied to the system, the red points show the accumulated number of breakdowns experienced by the system, the pink curve shows the breakdown rate over a sliding $10^6$ pulse window and the green points show the pulse length. From \cite{millar_operation_2021}.}
  \label{fig_conditioning curve}
\end{figure}

A device is typically conditioned by exposing it first to lower fields than nominal and often also to shorter pulse lengths. Then the device is operated while gradually increasing field and pulse length in a controlled way. Field and pulse length conditioning was used in all of the devices covered in this report. An example of the conditioning history of a high-gradient radio frequency accelerating structure is shown in Fig. \ref{fig_conditioning curve}. The plot shows how the structure starts from initial operating conditions of reduced field level and pulse length and progressively, over in this case hundreds of millions of pulses, reaches nominal operating conditions. 

The overall shape of the conditioning curve, that of an initially steep rise followed by an asymptotic approach to the maximum stable field holding field is a typical one. However the details of the curve can be affected by the series of specific decisions including about how quickly to raise the field, how to recover stable operation following a breakdown and when to raise pulse length. We will refer this collective set of decisions as the conditioning strategy. All of the rf structures, including that shown in Fig. \ref{fig_conditioning curve}, and dc electrodes referred to in this report have been conditioned using conditioning strategies that have been implemented as computer controlled algorithms. This algorithmic control is essential for comparing results from different structures and electrodes in a quantitative way and also for producing reproducible device performances. In many domains of high-field systems, conditioning is controlled by a human operator, so cannot be precisely defined, which makes comparison of results difficult. 

The basic conditioning algorithm used for the radio frequency structures and dc electrodes referred to in this review is described in \cite{wuensch_experience_2014, profatilova_breakdown_2020}. Even with the common basic algorithm, various parameters, such as the target breakdown rate, are set-able. In addition, the optimizing the conditioning strategy has been investigated and an improved strategy will be described below.

The initial stages of conditioning, when fields are applied but before the first breakdown is observed, is characterized by an increased vacuum level. This is caused by electron stimulated desorption (ESD) outgassing of the structure surface by the applied fields, typically caused by field-emitted electron bombardment of the inner surface of the cavity, emitting also UV and X-Rays photons resulting indirectly in additional photon stimulated desorption. The applied field induced vacuum rise decreases as conditioning progresses as well as the field-off base vacuum level. This is because the high-field operation has an effect similar to a vacuum bake-out for the surfaces exposed to the direct and indirect effects of high fields. 

The vacuum behavior raises the interesting question of the role of residual gas in breakdown initiation itself. No significant trigger effect from these gases, such as a larger release prior to a breakdown, has been observed \cite{descoeudres_dc_2009}. In fact, the amount of ambient gases surrounding the electrodes seems to have only a limited effect on the breakdown triggering. Experiments performed by injecting different gases (Ar, H$_2$, CO and air) demonstrated that pressure in excess of \SI{e-5}{\milli\bar} are needed in order to lower the breakdown field from its typical saturated value \cite{ramsvik_influence_2007}. This pressure value is orders of magnitude larger than the background vacuum level of the test system as mentioned earlier and also of the vacuum level of a typical radio frequency accelerating structure. Consequently ambient gasses do not appear to have a direct role in  breakdown triggering. It is as yet unclear whether the observed breakdown field decrease is due to enhanced ionisation of the ambient gas by field emitted electrons, which seems unlikely due to the mean free path being orders of magnitude larger than the gap distance, or to enhanced surface adsorption of gases, and their subsequent release upon applying electric field playing a role as an early trigger for breakdowns.

We will now refer to Fig. \ref{fig_conditioning curve} to describe the subsequent stages of conditioning. After the initial vacuum-limited operation, the vacuum level will no longer rise to critical values with fields applied, so the field can be raised to a level where the breakdowns start to occur. In Fig. \ref{fig_conditioning curve} the first breakdowns of this type occur at an accelerating gradient in the range of 10 to 30 MV/m, which corresponds to a peak surface electric field of about 25 to 75 MV/m for this type of structure. In this stage of conditioning the breakdowns produce large vacuum bursts, i.e. abrupt increases in the pressure. It is believed that these initial breakdowns are initiated primarily by extrinsic factors, such as residual dust, which contain a significant amount of non-metal material that is vaporized by the breakdown. Like the initial steady state vacuum pressure decreases during operation, the magnitude of the pressure bursts caused by the breakdowns decreases with continued operation. This transition is probably because the earliest breakdowns are caused by extrinsic factors, while the later, and vast majority in a high-field system, breakdowns are caused by the intrinsic mechanisms. Because the intrinsic mechanisms only involve copper, and copper will immediately stick to the first surface it encounters and will not propagate to the vacuum gauge. 

Once conditioning has reached the stage when breakdowns dominate, the breakdowns limit the rate  at which the field in the device can be raised. The control algorithms used on the devices described in this report ramps up the field level automatically at a speed which is limited by feeding back to maintain a maximum breakdown rate. This was of the range of $10^{-5}$ to $10^{-4}$ for the data shown in Fig. \ref{fig_conditioning curve}. The feedback on the maximum allowable breakdown rate is reflected in the level of the pink line in Fig. \ref{fig_conditioning curve}.  

\begin{figure}
  \centering
  \includegraphics[width=1\linewidth]{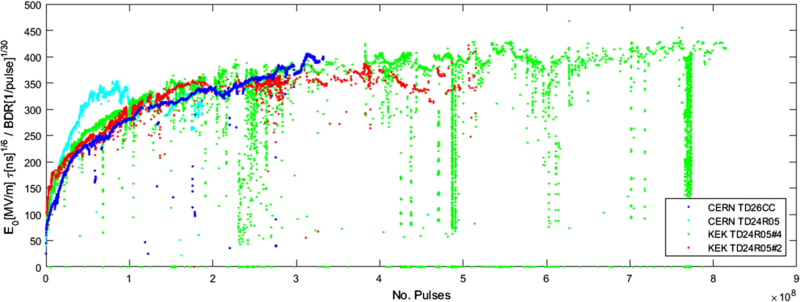}
  \includegraphics[width=1\linewidth]{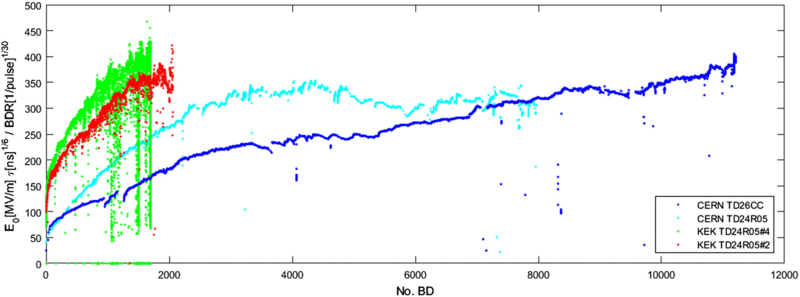}
  \caption{Comparison of the scaled gradient vs number of accumulated pulses for several structures. Despite the different conditioning approaches, the curves for the scaled gradient are similar (upper). Comparison of the scaled gradient vs number of accumulated breakdowns for several structures. When plotted with respect to the total accumulated number of breakdowns, the curves of the scaled gradient diverge significantly. From \cite{degiovanni_comparison_2016}}
  \label{fig_pulse and BD comparison} 
\end{figure}

A comparison of conditioning data from different structures of the same type leads to the observation that progress in conditioning proceeds with the number of pulses, not the number of breakdowns. This observation is illustrated in the plot in Fig. \ref{fig_pulse and BD comparison} which is taken from \cite{degiovanni_comparison_2016}. The similarity of the conditioning curves of different structures of the same type when plotted against the number of pulses - and their dissimilarity when plotted against the number of breakdowns - suggests that conditioning is primarily driven by the number of pulses. This argues against the supposition that conditioning progresses through the removal by breakdown of extrinsic elements on the device surface. It argues rather that the applied surface field improves the surface, while simultaneously creating breakdown nucleation points. 

The breakdown dominated phase of conditioning is initially carried out typically at a reduced pulse length. This is done partly to minimize damage to the device by reducing the pulse energy. It is also done because it appears to decrease the conditioning time. This is because the conditioning effect of a pulse on the surface is greater at a higher field level, we will return to this point below, and the breakdown rate is lower at shorter pulse lengths as discussed in Sec.~\ref{ss:behave-dependencies}. Thus at a lower pulse length the field level can be ramped up at a higher rate. Pulse lengths are usually increased step-wise after the maximum field has been approached. When a pulse length increase is implemented, it is necessary to reduce the field in the device to maintain the target breakdown rate. We will discuss the dependency of breakdown rate on gradient in subsection \ref{ss:behave-dependencies}. The field is then ramped up at the new pulse length, as before. This increasing pulse stage of conditioning can be seen in the drop-then-ramp structure in Fig. \ref{fig_conditioning curve} from about 50 to 150 million pulses. 

The ultimate gradient of a particular type of structure, is determined by both its design and by the techniques that were used in its fabrication. The final stage of conditioning must be done very carefully because pushing the field too close to the ultimate value can result in irreversible damage. This damage is observed as it forces the gradient to reduced significantly, as much as 20\%, in order to reestablish operation at the typical target breakdown rates of $10^{-5}$ to $10^{-4}$. In some cases certain materials will operate for some time at a high gradient, but eventually crash as discussed in \cite{peacock_experimental_2023}. The maximum practical operating field, including an appropriate safety margin, can typically be established only after iterative testing on a given device type.  

Conditioning is also needed after a vacuum high-field system has been exposed to air, for example during maintenance or an accidental venting of a system. In such circumstances the system must again be initially be operated at reduced field and be raised with the same techniques as described above. In the case of exposure to atmosphere however, the system can be ramped much more quickly than during the initial conditioning. The explanation is that the exposure to air results in extrinsic contaminants, for example dust and water, sticking to the surface. These can be driven off by the applied fields comparatively easily. After this relatively short period of 'reconditioning' the device will return to the longer-term conditioning trend. 
 
In addition to exposure to air, a smaller but similar effect is observed when the structure is idle for an extended period. Although the vacuum may be very good in a high-field system, there is still a finite presence of gas. While the high field surfaces become extremely clean due to the action of the applied fields, the rest of the vacuum system surface has the adsorbed gas associated with the base vacuum pressure of the system. During an idle period, gas is transferred from the non high-field regions to the high-field ones. Extrinsic-associated vacuum and breakdown activity occur once fields are applied again. This is the same effect which is proposed to explain the dependency of breakdown rate on repetition rate and pause time that will be discussed subsection \ref{ss:behave-dependencies}.

A few important insights can be made from these observations from conditioning. One is that there is not a unique type of breakdown trigger, but rather there are both those associated with extrinsic surface contaminants and those associated with intrinsic features coming from the fundamental properties of the device material surface or bulk. Long-term conditioning is determined by intrinsic material properties and takes many hundreds of millions of pulses and is preserved even when a device is exposed to air. Because the dislocation model predicts the observed field and temperature dependencies of breakdown rate, one may suppose that an evolving dislocation structure underlies this long-term conditioning. However differences in dislocation structure between conditioned and unconditioned areas has not yet been directly observed. 

\begin{figure}
  \centering
  \includegraphics[width=1\linewidth]{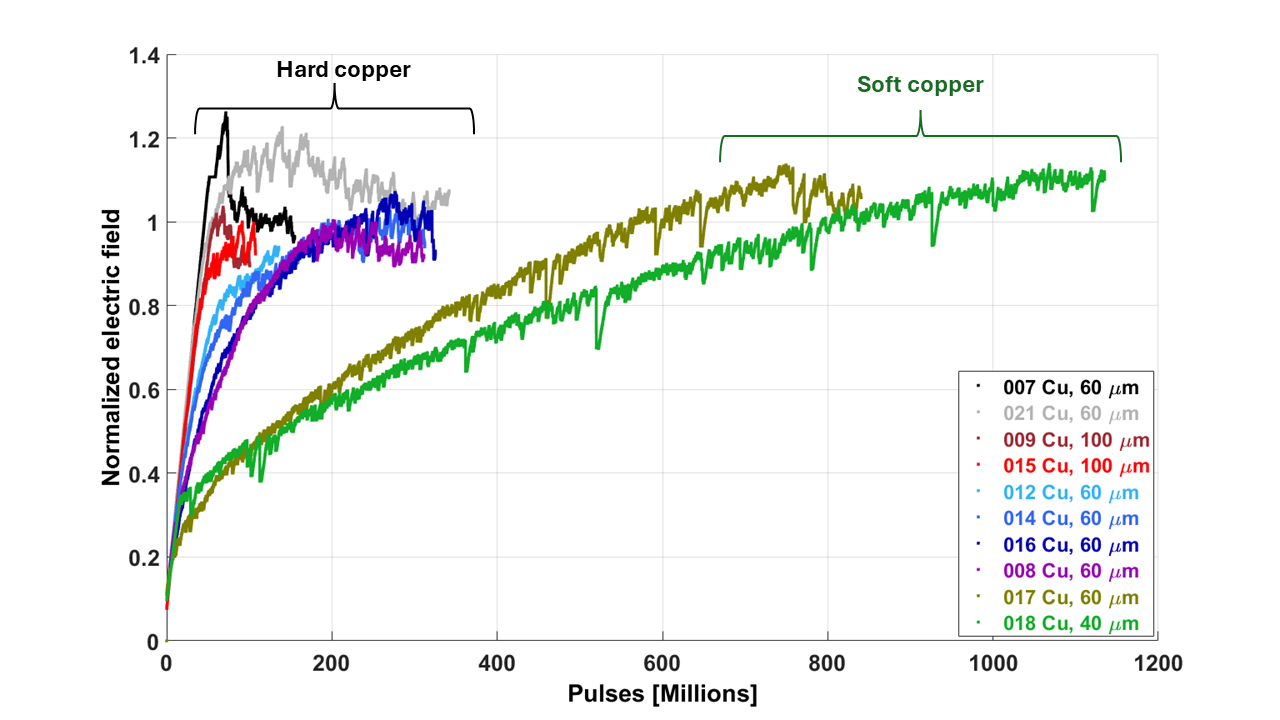}
  \caption{Conditioning curves for soft (heat treated) (green and brown curves) and hard (as machined) copper. Data from test of pulsed dc electrodes \cite{profatilova_behaviour_2019}. The applied field normalisation is further discussed in Sections \ref{ss:behave-dependencies} and \ref{ss:powerCoupling}. }
  \label{fig_SoftHardCu}
\end{figure}

Different materials, and materials in different metallurgical states condition in different ways and to different levels. For example, hard copper conditions faster than soft copper as shown in Fig.~\ref{fig_SoftHardCu}. This supports that long-term conditioning is driven by intrinsic properties of the system. We will return to dependence on material in subsection \ref{ss:behave-material}. 

\begin{figure}
    \centering
    \includegraphics[width=1\linewidth]{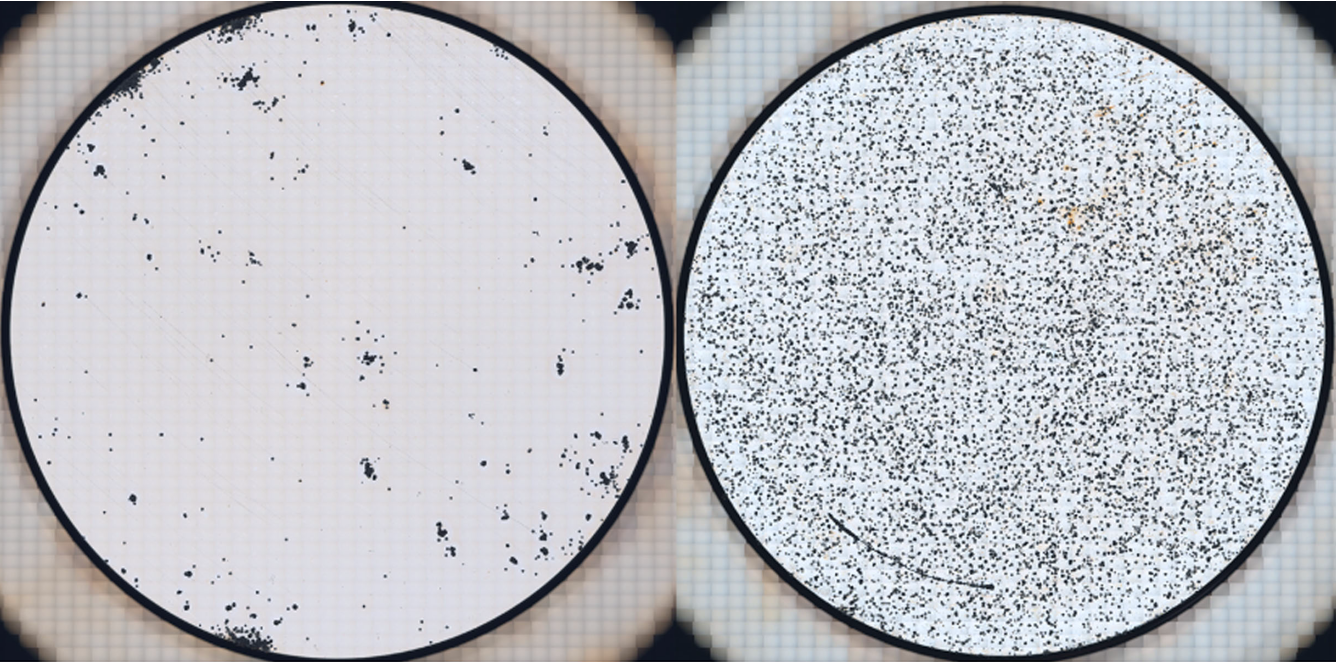}
    \caption{Comparison of breakdown locations after conditioning of hard (as machined), left, and soft (heat treated to 1040~$^\circ$C), right. The breakdowns have accumulated on the edge of the hard electrode and are uniformly distributed on the soft electrode. From \cite{korsback_vacuum_2020}.}
    \label{fig:hard soft cond comp}
\end{figure}

A crucial question for optimizing a conditioning strategy is if the conditioning speed is dependent on the level of applied field. As we have seen conditioning is correlated to the number of pulses as shown in Fig. \ref{fig_pulse and BD comparison} but the overall rate may be higher with applied field. Evidence that conditioning proceeds more quickly with higher applied field for dc systems is reported in \cite{korsback_vacuum_2020}, with key evidence reproduced here in Fig. \ref{fig:hard soft cond comp}. The anodes and cathodes in this test had the same diameter, which resulted in a field enhancement at the outer edge of about 7.5\%. For the hard electrode, on the left in Fig. \ref{fig:hard soft cond comp}, breakdowns populate the outer edge because the hard electrode is in a more conditioned state, as can be seen in the faster conditioning shown in Fig. \ref{fig_SoftHardCu}. On the other hand, breakdowns populate the soft electrode (on the right in Fig. \ref{fig:hard soft cond comp}) quite uniformly and are not concentrated on the outer edge despite the field enhancement. The explanation for this is that the higher field on the outer edges results in a higher conditioning speed compared to the center of the electrode, which offsets the strong dependency of breakdown rate on applied field. 

Taking this observation on the interplay between conditioning speed and breakdown rate through applied field further, and in order to better determine the mechanisms involved in conditioning a Monte Carlo simulation tool has been developed \cite{millar_monte_2022}, also inspired by previous work \cite{levinsen_statistical_2009} . The dual role of the applied field in both driving conditioning and the breakdown rate is a key element of understanding the breakdown process.

\subsection{\label{ss:behave-dependencies} Dependencies of breakdown rate}

Breakdown rate depends very strongly on field level, pulse length and temperature. These dependencies are very important for device operation, and are essential input for the dislocation model described in Sec. \ref{s:dis}.

\begin{figure}
    \centering
    \includegraphics[width=1\linewidth]{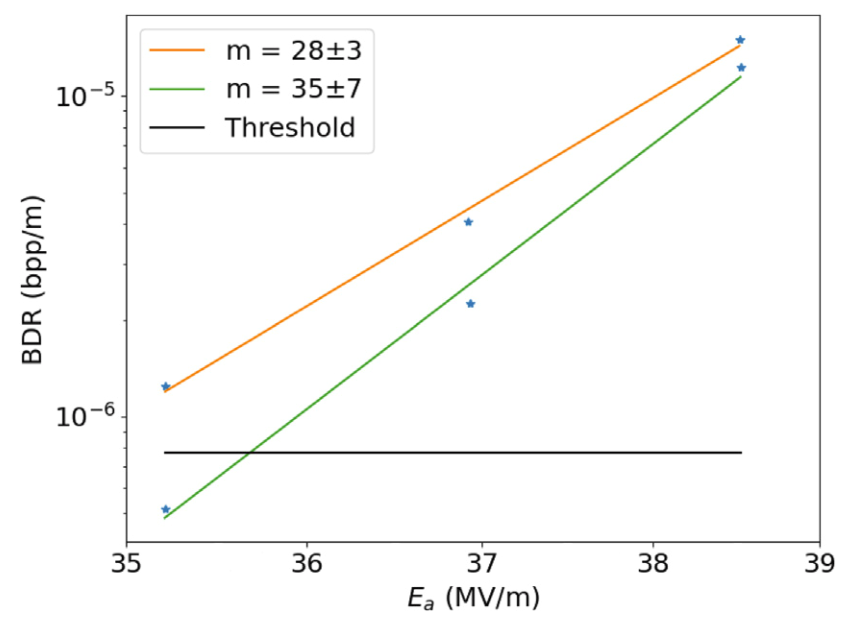}
    \caption{Breakdown rate as a function of accelerating gradient in a 3 GHz radio frequency structure. The peak surface electric field is 4.4 times higher than the accelerating gradient. The green and orange lines are polynomial fits to the experimental data points, with values indicated by the variable m in the legend. From \cite{martinez-reviriego_high-power_2024}.}
    \label{fig:BDR RF}
\end{figure}

\begin{figure}
    \centering
    \includegraphics[width=1\linewidth]{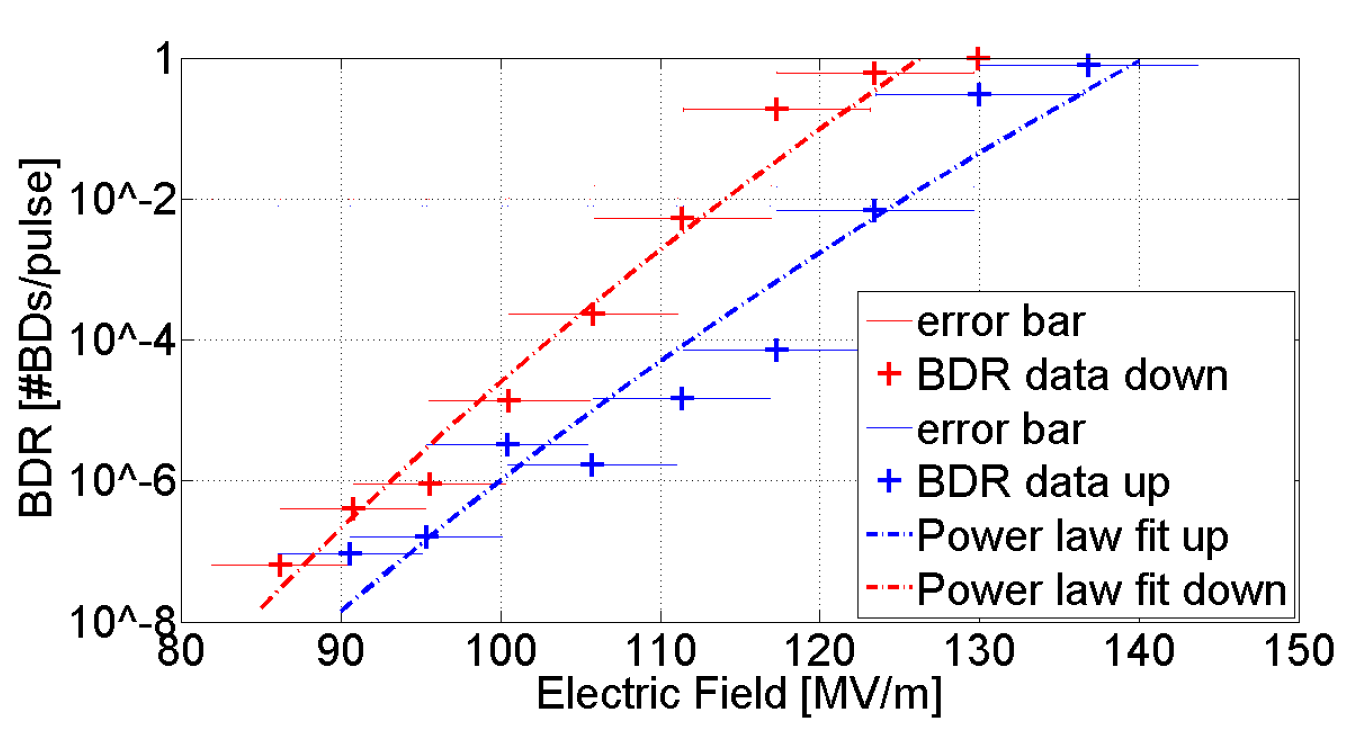}
    \caption{Breakdown rate as a function of applied field in the pulsed dc system. Up and down data refer to the direction of the voltage scanning. From \cite{shipman_experimental_2014}.}
    \label{fig:BDR_DC}
\end{figure}

The breakdown rate of a high-field system exhibits a very strong dependence on applied field level. They are an important input for the dislocation models described in Sec. \ref{s:dis}.In both pulsed rf and dc systems, this dependence is typically,
\begin{equation}
  R_{BD}\propto E^{30} 
  \label{BDRE30}
\end{equation}
This has been reported for both radio frequency \cite{grudiev_new_2009} and  pulsed dc systems \cite{shipman_experimental_2014}. Example of breakdown rate vs. gradient measurements are shown in Figs.~\ref{fig:BDR RF} and \ref{fig:BDR_DC}. Breakdown rate data for radio frequency experiments is typically presented in units of $R_{BD}/m$, with the values computed by dividing the device breakdown rate by the active length of the structure. This is done so that the breakdown rate of a single accelerating structure can be extrapolated to a linac containing many structures. Breakdown rate data from the pulsed dc system is typically presented for the device, without normalizing for example for the surface area of the electrode. 

This dependence is important both for practical reasons, since breakdown rate is an important specification for reliable operation of a facility, and for theoretical reasons since the field dependence has led to significant insights into the breakdown initiation mechanism, in particular to the dislocation-based model described in Sec. \ref{s:dis}. In addition, the breakdown rate dependence is often necessary to compare different experiments, as data is often taken at different breakdown rate conditions.

\begin{figure}
    \centering
    \includegraphics[width=1\linewidth]{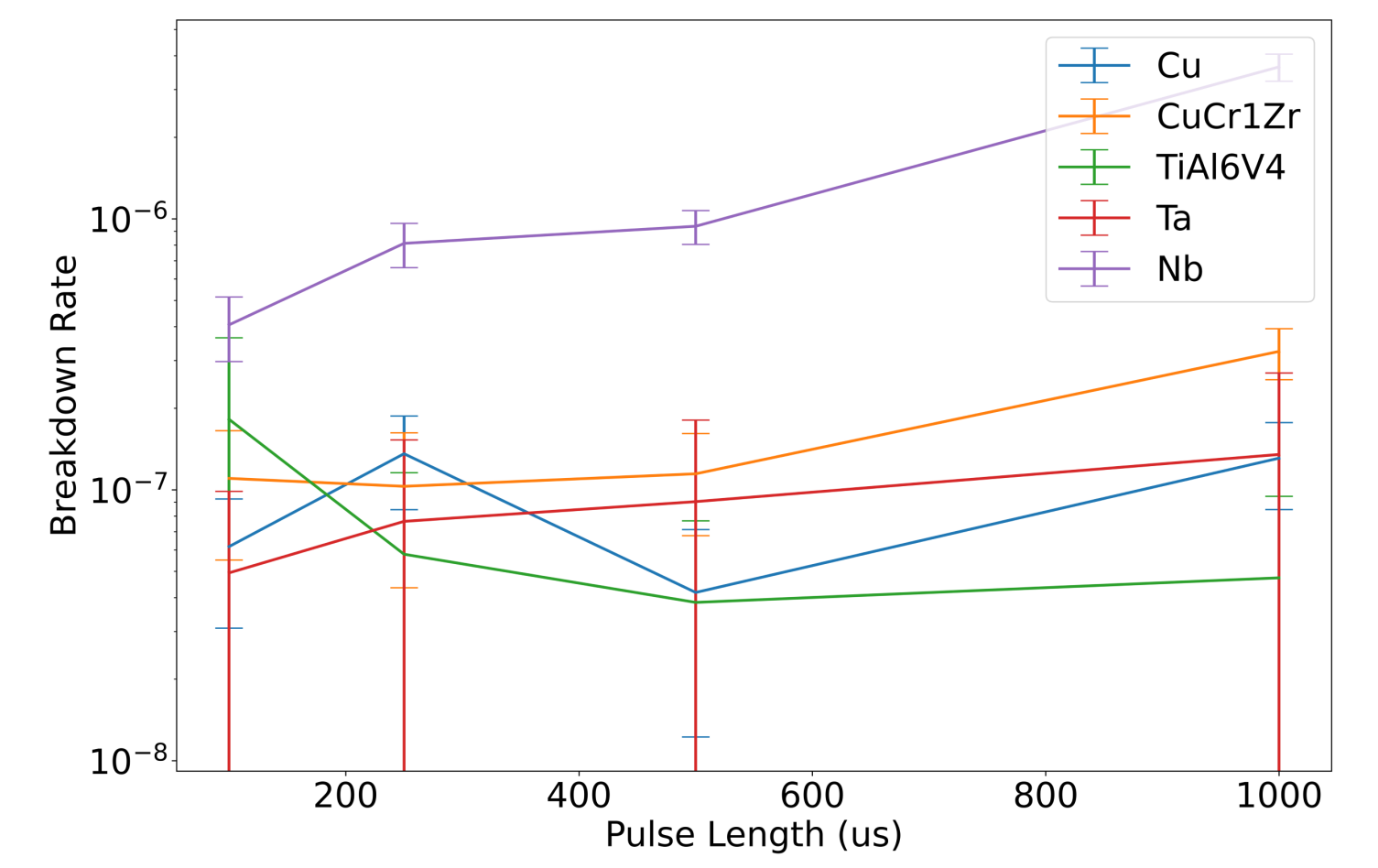}
    \caption{Breakdown rate as a function of pulse length for long pulses and different materials. The $\tau^5$ dependence seen at shorter pulse lengths has saturated. From \cite{peacock_experimental_2023}.}
    \label{fig:BDR long pulse}
\end{figure}

The breakdown rate also displays a dependence on pulse length. In the traveling wave rf accelerating structures tested in the linear collider programs, measured breakdown rates typically have shown a dependency of,
\begin{equation}
  R_{BD}\propto \tau^{5} 
  \label{BDRt5}
\end{equation}
where $\tau$ is the pulse length, as for example reported in \cite{grudiev_new_2009}. An example where this dependency can be observed is by comparing the shapes of the conditioning curves shown in Figs.~\ref{fig_conditioning curve} and \ref{fig_pulse and BD comparison}. In both cases conditioning has been carried out at successively longer pulse lengths. The change of pulse length can be clearly seen in Fig. \ref{fig_conditioning curve}. In Fig. \ref{fig_pulse and BD comparison} the electric field is normalized by $\tau^5$, the conditioning curve becomes smooth showing that the pulse length scaling is correct. It is important to note that these  pulse lengths for the accelerating  structure tests ranged from about 50 to 300 ns, with the minimum time give by the filling time of the structures (tests were rarely carried out below the filling time) and the maximum time is given by the power limit of the klystron-based power sources driving the structures. Breakdown rate as a function of pulse length measurements have also been carried out using the pulsed dc system, but here pulse lengths were in the range of around 500 ns up to 1 ms. In this system and pulse length range  the pulse length dependence was indiscernible as shown in Fig. \ref{fig:BDR long pulse}. It appears that the pulse length dependence saturates at around 500 ns. The mechanism of pulse length dependence is discussed further in subsection \ref{ss:dis pulse length}. Finally, breakdowns can even occur for pulses that short compared to the plasma formation time, 10~ns and less. Very short pulse breakdowns are not covered further in this report but data is available in \cite{braun_frequency_2003, dal_forno_rf_2016,tan_demonstration_2022}.

An important insight into the character of the breakdown mechanism has been found when combining the observation that the breakdown timing distribution is flat, as shown in Fig. \ref{fig:BD in pulse dist} and discussed in subsection \ref{ss:behave-Statistics} and even while a strong pulse length dependence is observed in rf tests. Since the breakdown rate dependence is not produced by a progressively greater density of breakdowns at the end of the pulse, the conclusion must be that the system displays a memory of the pulse length of previous pulses. This is analogous to a fatigue process where a material will tend to fail after a certain number of stress and de-stress cycles, but fails at a random moment within a cycle. This memory effect is an important motivation for the dislocation-based breakdown initiation model described in Sec. \ref{s:dis}.

Another observed dependence of breakdown rate is on temperature. For this dependence, the breakdown rate at a given field level goes down significantly when a high-field system is cooled to cryogenic temperatures. Alternatively one can say that system supports a higher field for a given breakdown rate when cooled. This type of measurement has required construction of specialized versions of the rf and pulsed dc systems. 

\begin{figure}
  \centering
  \includegraphics[width=8.5cm]{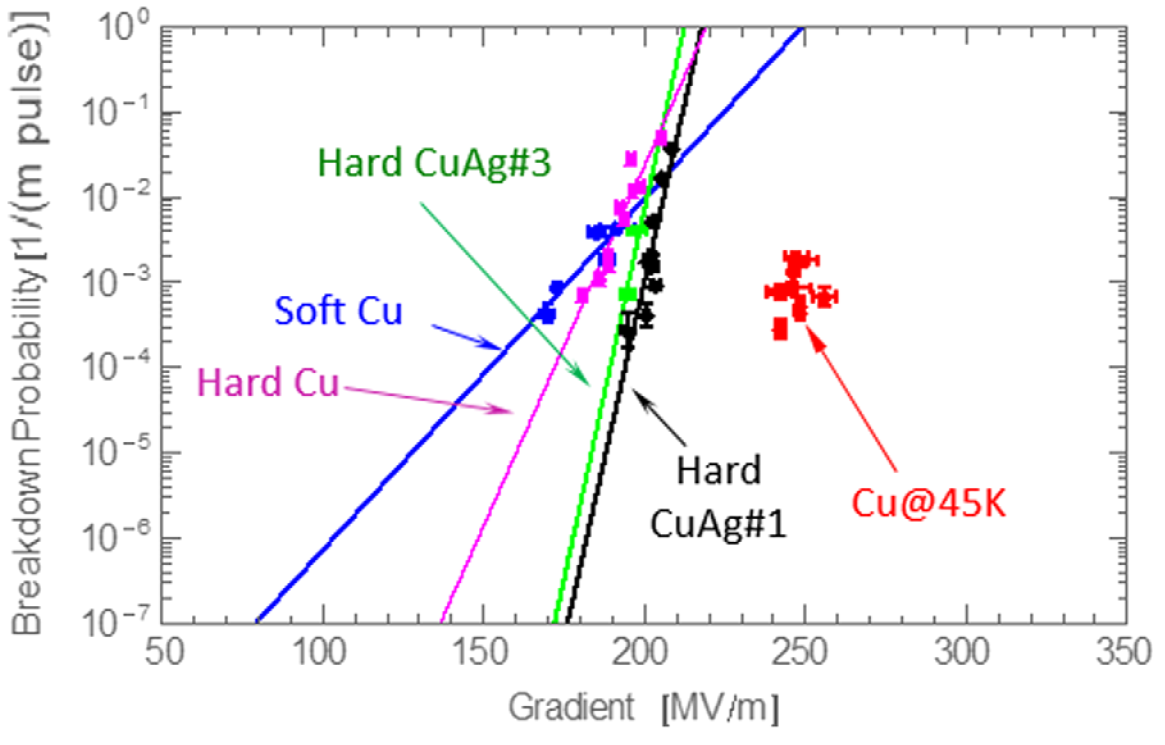}
  \caption{Improvement of field holding capability of a cryo-cooled copper rf structures. From \cite{cahill_high_2018}.}
  \label{fig_cryo rf}
\end{figure}

An early high-field measurement of the effect of cooling on field was carried out in the CTF2 facility using a beam-driven cavity cooled to liquid nitrogen temperature \cite{braun_frequency_2003}. This experiment did not show any temperature dependence, however only the maximum field holding level was measured, that is the field level where the breakdown rate approaches 1. A subsequent rf experiment was carried out using a cryogenically cooled cavity as described in \cite{cahill_rf_2018}, in which the effect of temperature on field and breakdown rate was measured at low, $10^{-4}$ range, breakdown rates. The experiment used an rf test stand similar to the type described in Sec. \ref{s:exp} but where the cavity was operated at cryogenic temperatures using a cryo-cooler. When cooled to 45~K, the cavity showed a very clear, approximately 20$\%$, increase of field level for a breakdown rate of approximately $10^{-4}$. A plot of the effect of temperature on gradient is shown in Fig. \ref{fig_cryo rf}. The publication also reports that field emission at the increased field levels enabled by the cooling was strong enough to result in a measurable dynamic decrease of cavity $Q$. 

\begin{figure}
  \centering
  \includegraphics[width=8.5cm]{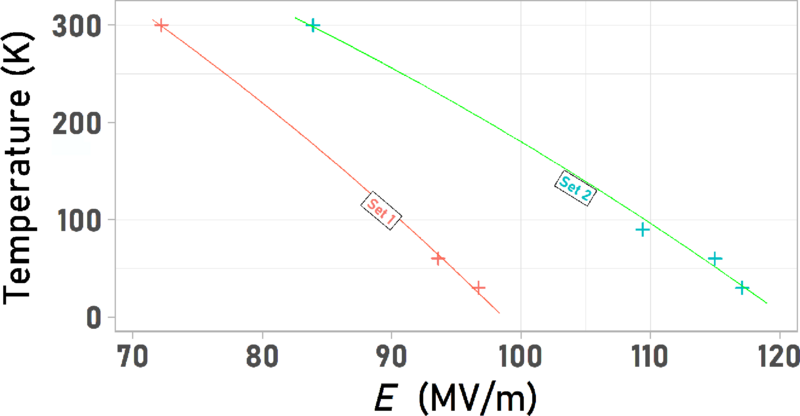}
  \caption{Temperature dependence of field holding, for a normalized breakdown rate, as a function of electrode temperature in a pulsed dc system. The crosses are measured data points and the lines are fits based on the dislocation model described in \cite{engelberg_stochastic_2018}. From \cite{jacewicz_temperature-dependent_2020}.}
  \label{fig cryo dc}
\end{figure}

A similar increase of field for a given breakdown rate at lowered temperatures has also been observed in pulsed dc experiments \cite{jacewicz_temperature-dependent_2020}. In this case, the experimental setup consisted of a pulsed dc system similar to that described in Sec.\ref{s:exp} but where the electrodes were cooled to cryogenic temperatures. Here also the field increased by approximately 20$\%$ for a normalized breakdown rate when cooled to a temperature of around 60~K and conditioned, with further improvement as the system was cooled further to around 30~K. Data showing the increase of field for a normalized breakdown rate in the pulsed dc system is shown in Fig. \ref{fig cryo dc}. The dc systems also showed enhanced field emission at the elevated field levels, and we will return to this in subsection \ref{ss:behave-field emission}.

\begin{figure}
    \centering
    \includegraphics[width=1\linewidth]{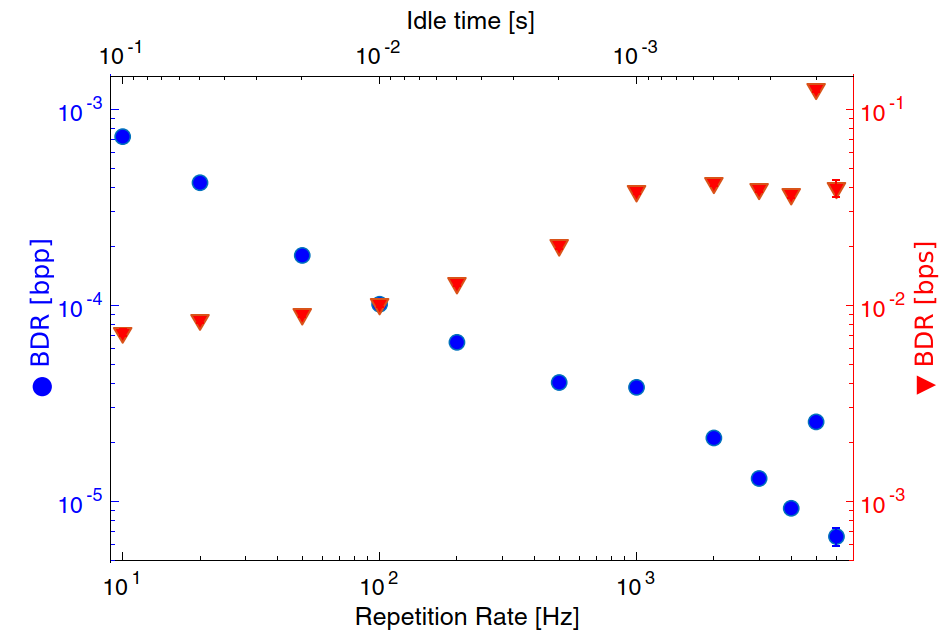}
    \caption{Breakdown rate as a function of repetition rate for data taken with the pulsed dc system. The breakdown rate goes down as the repetition rate goes up. This is thought to be caused by less time for adsorption of residual gas in the system between voltage pulses when the repetion rate is high. From \cite{saressalo_effect_2020}.}
    \label{fig:BDR vs repetition rate}
\end{figure}

Another dependence of breakdown rate that has been observed is on repetition rate and more generally, the time interval between pulses. In a pulsed dc experiment described in \cite{saressalo_effect_2020}, the breakdown rate was measured as the repetition rate was varied from 10 Hz to 6 kHz and is shown in Fig. \ref{fig:BDR vs repetition rate}. The breakdown rate decreases as the repetition rate increases. A related measurement is reported in \cite{saressalo_effect_2020} for which pulsing was paused for intervals between tens of seconds to the order of a day. An enhanced breakdown rate was observed following the pause, with a greater enhancement for longer pauses. The explanation given in \cite{saressalo_effect_2020} for this effect is that gas is adsorbed during the field-off interval between pulses and during pauses. The longer the time between pulses, either through a lower repetition rate or through inhibited pulsing, the more residual gas, such as water vapor, is re-adsorbed. This adsorbed gas then results in an increased probability of breakdown.  

\begin{figure}
    \centering
    \includegraphics[width=1\linewidth]{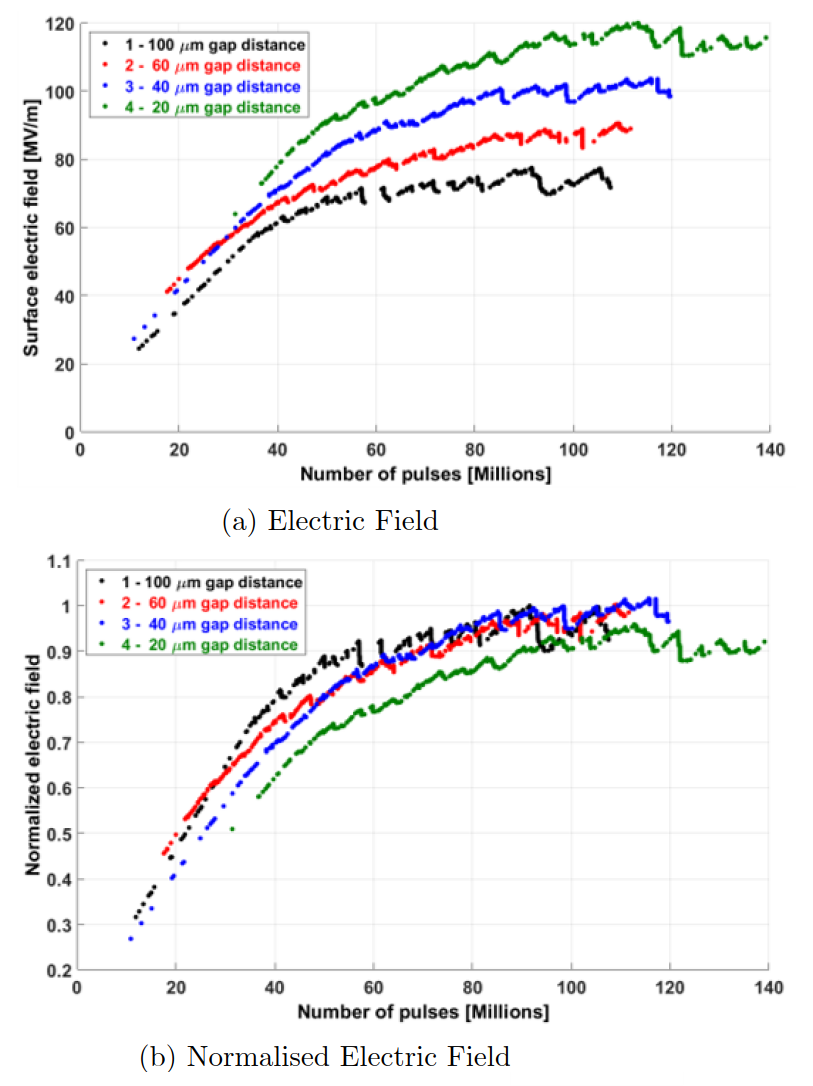}
    \caption{Surface field as a function of the number of pulses in the pulsed dc system during the conditioning phase as described in section \ref{ss:behave-conditioning}. The different colors are data taken with different gaps between the electrodes. The upper plot shows the surface electric field and the lower plot shows the normalized electric field $E^*$ defined by $E^*=V/d^{0.7}$ where $V$ is the applied voltage and $d$ is the gap distance. Taken from \cite{peacock_experimental_2023}.}
    \label{fig:gap dependence DC}
\end{figure}

Finally, we discuss the dependence of the breakdown rate that has been observed in dc systems on the distance $d$ between electrodes. The surface electric field $E$ is $E=V/d$, where $V$ is the applied voltage. It has been observed that most experimental results obtained at various electrode separations the maximum voltage scales less than linearly with the gap, as $d^{0.7}$. Thus the maximum surface electric field decreases for increasing inter-electrode gap. This is illustrated in Fig.~\ref{fig:gap dependence DC}. This phenomenon has already been identified in the literature, both in dc \cite{maitland_new_1961, alpert_initiation_1964} and at high frequency \cite{little_electrical_1965} as being an implicit effect of the applied voltage. The theory which also explains the $d^{0.7}$ gap dependence will be developed in Section \ref{s:circ}.

\subsection{\label{ss:behave-material} Dependence on material}

There are preferred metals for different high-field applications, for example, copper in radio frequency devices, stainless steel in large high-voltage systems as well as focusing elements in scanning electron microscopes, and alloys of copper in circuit breakers and X-ray tubes. These have usually been chosen through long experience and trial-and-error testing. The demand for a very high accelerating gradient in linear colliders motivated a renewed look at the field holding capabilities of different materials to see if a better performing metal than copper could be found. In addition, comparisons of the results of systematic tests of different materials have yielded fundamental insights into the breakdown mechanism. The material tests carried out by the linear collider and related collaborations summarized here include those carried out using rf accelerating structures and using both the anode tip and pulsed dc systems. 

Tests of the field holding capability of copper, molybdenum and tungsten have been carried out using Ka and X-band accelerating structures using both clamped irises and quadrant type structures. The results show that in certain tests, molybdenum and tungsten reached higher accelerating gradients than copper. However, the improvement was not seen in all cases, which is likely due to the difficulties of machining the refractory materials to the shapes and surface finish required for rf structures as well as the difficulty in joining parts. In addition, the sometimes observed field holding improvement were not sufficient overall to overcome the poorer electrical and thermal conductivities. These results are described in \cite{wuensch_high-power_2004,wuensch_high-gradient_2006,dobert_high_2004}.

Compared to radio frequency structures, dc systems are better adapted for comparative material tests since electrodes have simpler geometries, and thus are easier to fabricate and do not have current carrying joints. In addition, low electrical conductivity materials can be tested in the dc system.

\begin{figure}
  \centering
  \includegraphics[width=1\linewidth]{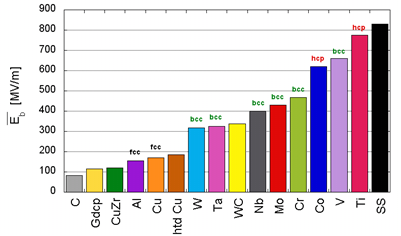}
  \caption{Histogram of maximum field level reached by different materials in tests with the anode tip dc system. From \cite{descoeudres_cobalt_2009}.  }
    \label{fig_cobalt}
\end{figure}

The anode tip system has been used to test the maximum field holding capacity of a large number (16) of materials, resulting from averaging several ramp-to-breakdown experiments. The materials show a large spread in maximum field holding strength, as shown in Fig. \ref{fig_cobalt}. The tests yielded the insight that pure metals are ranked by their crystal structure with FCC, BCC and HCP holding progressively higher surface fields. These crystal structures also have progressively higher energy barriers to dislocation movement, providing evidence that dislocations play a major role in breakdown initiation. Cobalt, a relatively soft and low melting point material but with an HCP crystal structure, was only tested once the crystal structure dependence was observed. It held a high field as predicted. The results are reported in \cite{descoeudres_cobalt_2009}. This was an important step in the development of the dislocation model as described in Sec. \ref{s:dis} Plastic Material Response. 

Samples of different type in terms of material, surface finishing, heat treatment, etc. have been tested with the anode tip system, \cite{kildemo_breakdown_2004,descoeudres_dc_2009}. The ramp-to-breakdown test protocol was to apply a progressively increasing voltage until a breakdown occurred, giving the threshold field. Voltage was then reduced to the initial value, and the cycle restarted until next breakdown. A steady state, with some level of fluctuation, was reached after few or several breakdowns, depending on the material. The breakdown field varies from pulse to pulse but an average maximum  field holding capacity could be computed. This type of experiment has been performed for many sample-tip pairs of the same material and reported in several publications \cite{kildemo_breakdown_2004, descoeudres_dc_2009, descoeudres_cobalt_2009}. One particular experiment involved cathode-anode pairs made of two different materials, namely Ti and W. In this case, the maximum breakdown field corresponded to that of the cathode material when tested in a same-material cathode-anode pair \cite{descoeudres_dc_2009}. This supports observations that breakdown starts on the cathode and that the maximum field is determined by the cathode material.

\begin{figure}
  \centering
  \includegraphics[width=1\linewidth]{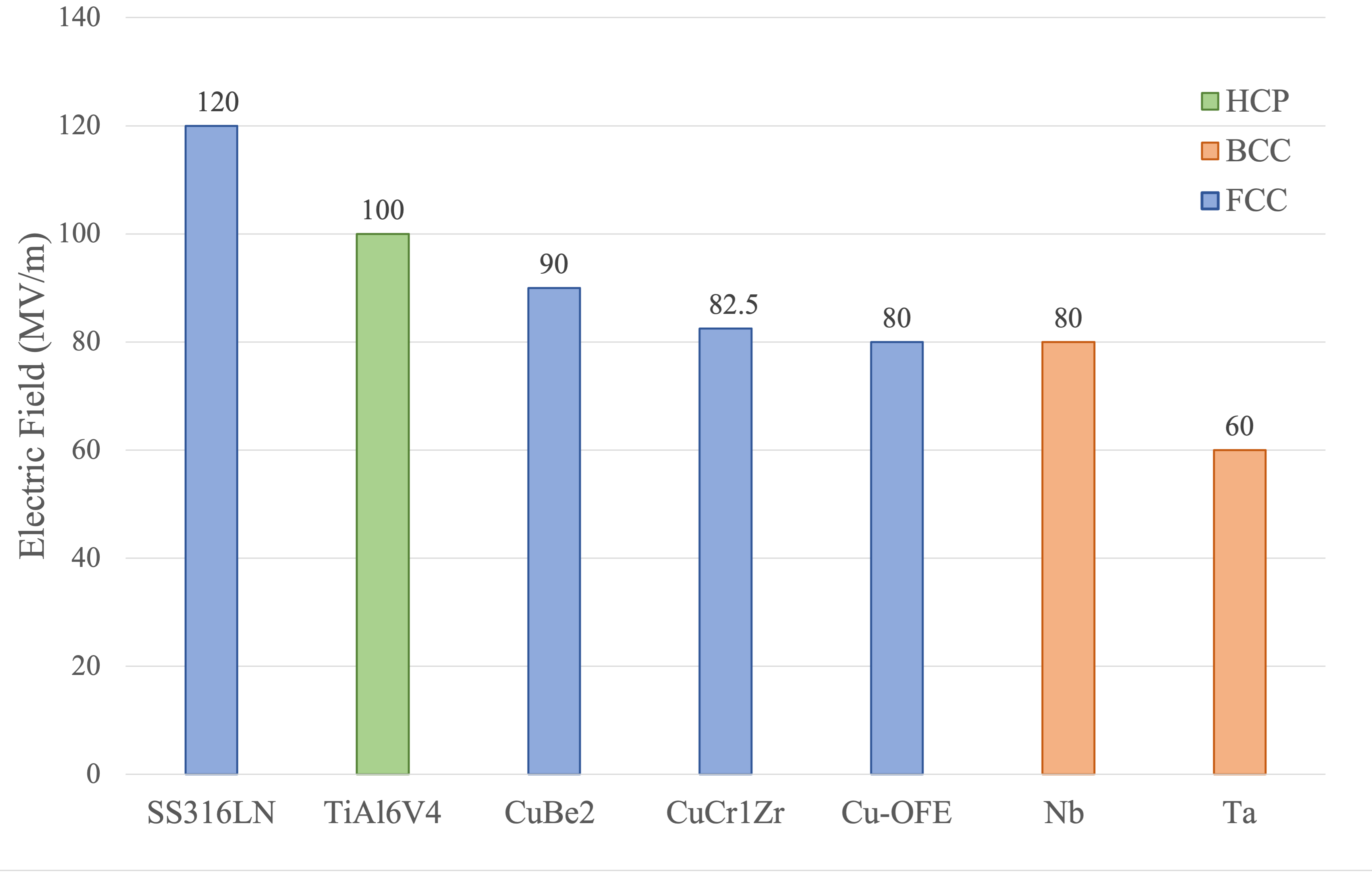}
  \caption{Maximum stable field reached at a normalized breakdown rate for different materials, adapted from \cite{serafim_h-_2024}.}
  \label{fig_LES-barplot} 
\end{figure}

The pulsed dc system has been used to test the field holding properties of a range of materials. Compared to the anode tip tests, the pulsed dc test focused more on alloys of copper rather than pure metals. This choice was motivated by the high electrical conductivity requirement for potential use in a radio frequency structure. In addition, some materials with a high solubility of hydrogen were tested for use in radio frequency quadrupoles, a linear accelerator device used to accelerate and bunch low energy particles such as protons and ions, which are sometimes subject to H$^-$ beam loss \cite{serafim_investigation_2024}. The results are shown in Fig.~\ref{fig_LES-barplot}. These results are for low breakdown rate operation performed with large electrodes, while those shown in Fig.~\ref{fig_cobalt} are for ramp-to-breakdown operation performed with the anode tip system, as described in Sec.~\ref{ss:exp-dc}. In ramp-to-breakdown operation the voltage is raised until a breakdown occurs, giving an effective breakdown rate of 1. The smaller range of surface field shown in Fig.~\ref{fig_LES-barplot} may be due to the choice of materials and possibly to the difference between low breakdown rate and ramp-to-breakdown testing.


There have also been systematic tests of single-cell rf cavities made from different materials. These have the same design as those used for the cryogenic tests described above and reported in \cite{cahill_rf_2018}. These have included cavities built from different copper alloys and different hardness states, and some results appear in Fig.~\ref{fig_cryo rf}. The general conclusion is that harder materials hold higher gradients. In the case of the comparison of heat-treated and not heat-treated copper, this result can be compared to the conditioning results shown in Fig.~\ref{fig_SoftHardCu} where hard copper conditions much more quickly than soft copper, but both arrive at the same limiting value. A further publication is \cite{dolgashev_design_2023}. 

\subsection{\label{ss:behave-field emission} Field emission}

Electron field emission is another phenomenon that occurs when high surface electric fields are applied to conducting surfaces. In this section, we will present experimental measurements of field emission in both rf and dc conditions.  There are both theoretical and practical motivations for measuring and understanding field emission. An example of a practical consideration is found in the field of high-gradient accelerators, where field emission can interfere with operation. The problem occurs when field-emitted electrons, often referred to as dark currents, are captured by rf fields and accelerated over long distances inside the accelerator. These accelerated electrons can create an unwanted background signal for beam instrumentation and can also lead to X-ray radiation and material activation when they hit accelerator components. 

The measurements of field-emitted currents presented here were made to contribute to understanding the breakdown initiation process. One aspect has been to provide experimental data to complement the simulation studies of the direct role of field emission in the breakdown initiation as described in Sec. \ref{s:emit}. These simulations study breakdown initiation through localized heating of field emission enhancing features on cathodic surfaces. Another aspect has been to try to provide direct experimental evidence of the fluctuations that are predicted in the analysis of dislocation dynamics described in Sec. \ref{s:dis}. In addition, an attempt has been made to study field emission current signals using machine learning techniques to see if a breakdown precursor could be identified. Although inconclusive, such a study is described in \cite{obermair_explainable_2022}.

The radio frequency field emission measurements presented here were taken in pulsed mode, just as the breakdown measurements described throughout this review. The dc field emission measurements were taken with continuously applied fields, whereas the breakdown measurements were taken in pulsed mode. These two modes of the pulsed dc system are described in subseciton \ref{ss:exp-dc}. One typical type of measurement is the $IV$, current vs. voltage curve.

\begin{figure}
    \centering
    \includegraphics[width=1\linewidth]{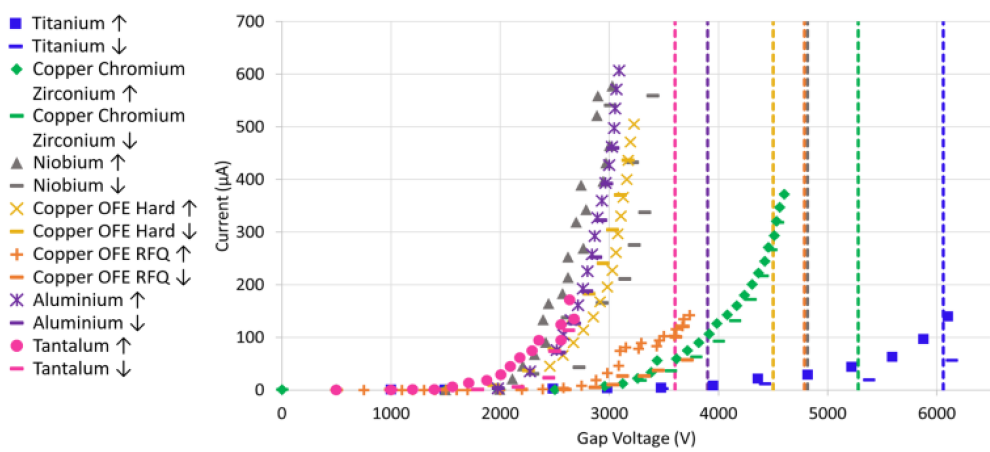}
    \includegraphics[width=1\linewidth]{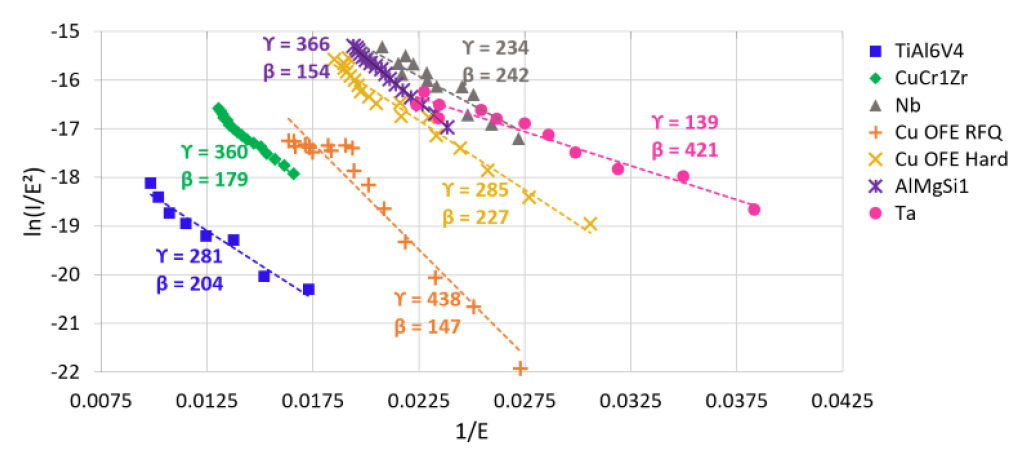}   
    \caption{The upper plot shows current vs. voltage traces of field emission from different materials in the pulsed dc system. The vertical dotted lines represent the maximum field achieved at a breakdown rate of $10^{-5}$ during conditioning. The lower plot shows the Murphy-Goode representation of the same data, as explained in Sec. ~\ref{sec:WorkFunction}. $\beta$ values in the range of 100 to 500 are observed.  From \cite{peacock_experimental_2023}}
    \label{fig:FE different materials}
\end{figure}

The wide variety of conditions in which breakdowns have been studied have given opportunities for investigation of field emission as well, and new perspectives on the relationship of the two processes. An example of such a measurement is a comparison of the field emission behavior in the same system for different materials, as shown in the upper plot of Fig.~\ref{fig:FE different materials}. A striking feature of this plot is that, at equivalent field, materials that reached higher fields showed much lower current compared to those that achieved lower fields. Field emission and field holding measurements of different materials were made to better understand how the two processes are related. 

$IV$ data can also be plotted in Fowler-Nordheim and Murphy-Goode forms in order to extract the so-called field enhancement factor $\beta$. 
A field enhancement factor is a way of matching measured field emission current to the applied field dependence predicted by the Fowler-Nordheim  \cite{fowler_electron_1997}. 
This correction factor generally falls in the range from 50 to a few hundred. To the knowledge of the authors, such large correction factors are required for \textit{all} field emission measurements made on macroscopic systems. The only exception where a correction factor is not required is for nanometric-scale tips, for example \cite{yanagisawa_laser-induced_2016}. The correction factor is of the order of 30 to 60 for the high-gradient accelerating structures described in this review, and of the order of 100 to 200 for the electrodes used in the pulsed dc system. The $IV$ data shown in the upper part of Fig.~\ref{fig:FE different materials} is re-plotted in Fowler-Nordheim form, along with extracted field enhancement values in the lower part of Fig.~\ref{fig:FE different materials}. The field enhancement factor is discussed further from a theoretical point of view in Sec. \ref{s:emit}.

\begin{figure}
    \centering
    \includegraphics[width=1\linewidth]{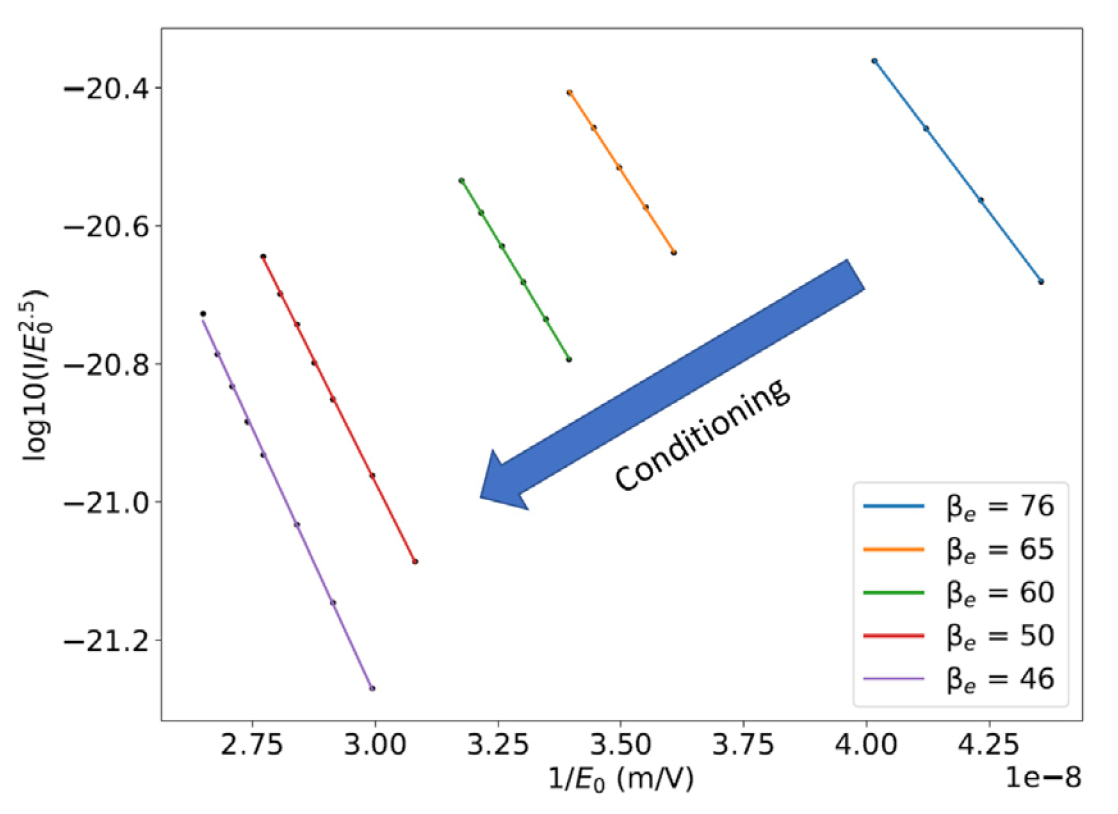}
    \caption{$IV$ data visualized in a Fowler-Nordheim plot showing the evolution of the field enhancement during the conditioning of a 3 GHz accelerating structure. From \cite{martinez-reviriego_high-power_2024}}
    \label{fig:FE during cond}
\end{figure}

\begin{figure}
    \centering
    \includegraphics[width=1\linewidth]{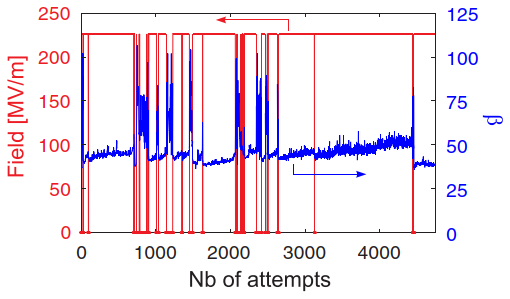}
    \caption{Evolution of the field enhancement coefficient $\beta$ (blue) measured after each applied high voltage pulse, at a constant macroscopic field of \SI{225}{\mega\volt / \meter} (red), in the anode tip system. The $\beta$ increases progressively, and breakdown happens when the local field $\beta E$ reaches \SI{10.8}{\mega\volt / \meter}. From \cite{descoeudres_investigation_2009}.}
    \label{fig:beta_x_E}
\end{figure}

It should be noted that in some cases, high $\beta$ values recorded at the beginning of an experiment can usually be attributed to surface contamination \cite{padamsee_rf_2008}. These may be removed with some surface treatments \cite{lagotzky_enhanced_2014}, or are more effectively destroyed upon first breakdowns, resulting in surface conditioning. However, the decrease in the field enhancement factor observed when a device is conditioned is not only linked to extrinsic properties like contamination, but has some more deep links with intrinsic material properties. An example of this was seen during the high-power test of a 3 GHz backward traveling wave accelerating structure as described in \cite{martinez-reviriego_high-power_2024}. Here the field enhancement factor decreased from 76 to 46 over the conditioning process as shown in Fig.~\ref{fig:FE during cond}. In this test $\beta E$ remained relatively constant over the conditioning process. Another example where $\beta E$ remained constant during operation was in a test with the anode tip system as described in \cite{descoeudres_investigation_2009} and shown in Fig.~\ref{fig:beta_x_E}. In this experiment the system was operated in a breakdown threshold mode, where on each pulse the voltage was kept at the same level, until breakdown occurred. The plot shows that the field enhancement factor measured between pulses was an accurate predictor of the field at which the next breakdown would occur through the relationship \SI{10.8}{MV/m}. This confirms earlier measurements \cite{alpert_initiation_1964, brodie_prediction_1966, kranjec_test_1967}. Other examples showing the evolution of $\beta E$ during conditioning can be found in \cite{banon-caballero_dark_2019, vnuchenko_high-gradient_2020}. These particular experiments indicate that there is an intrinsic material property that determines the local breakdown field threshold $\beta E$. Other experiments have not shown the same relationship.

\begin{figure}
    \centering
    \includegraphics[width=1\linewidth]{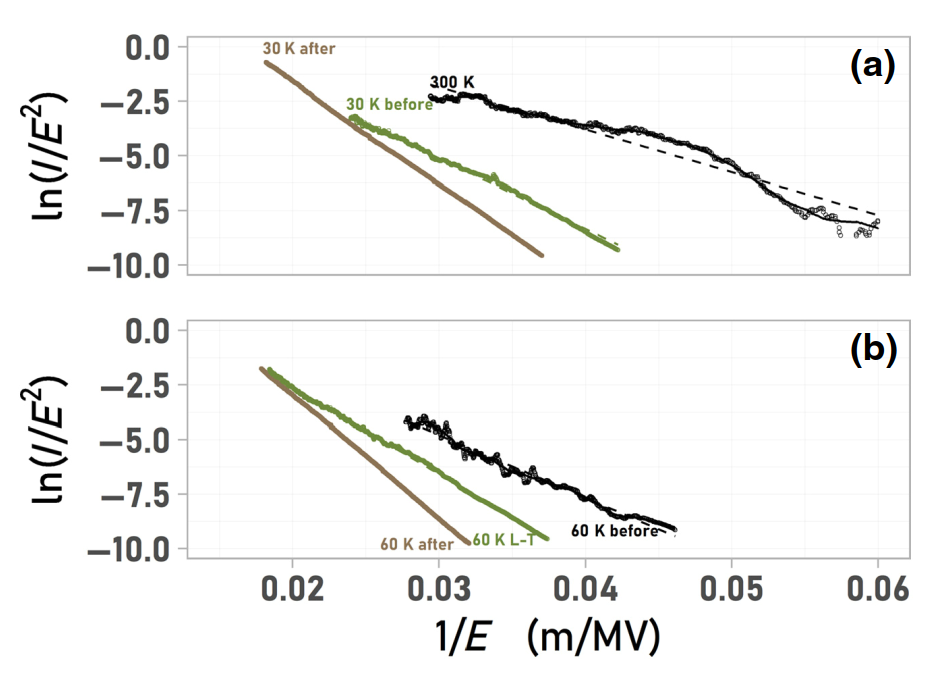}
    \caption{Field emission from cryogenic temperature electrodes. In the upper plot, field emission at 300 K (black), cooled to 30 K (green), and after conditioning at 30 K (brown). In the lower plot cooled to 60 K (black), after conditioning at 60 K, and after 9 day interval at 60 K. From \cite{jacewicz_temperature-dependent_2020}}.
    \label{fig:cryo FE}
\end{figure}

Lower field emission for an equivalent field level is also observed with cryogenically cooled electrodes as described in \cite{jacewicz_temperature-dependent_2020}. We have seen in Fig.~\ref{fig cryo dc} of Sec. \ref{ss:behave-dependencies} a significant increase in field for a given breakdown rate when electrodes are cooled to cryogenic temperatures. Field emission data from the same electrodes show that, \textit{for an equivalent applied field}, a significant drop in field emission current when cooled. On the other hand, the field emission current becomes much higher than at room temperature at the heightened field levels only accessible when the electrodes are at cryogenic temperatures. This is because the field emission current level rises exponentially, which dominates over the reduced field emission at lower fields. This effect was also observed in the rf measurements described in \cite{cahill_rf_2018} in which a dynamic loss of cavity quality factor due to field emitted current beam loading was observed. In addition, the field emission becomes much more stable at cyrogenic temperatures than at room temperature, resulting in close to ideal straight lines in the Fowler-Nordheim plots shown in Fig.~\ref{fig:cryo FE} over a wide range of field emission current. These cryogenic measurements indicates that field emission sites become extremely stable at low temperatures.    

Fluctuations in field emission current have been investigated as a possible signal caused by the dislocation dynamics described in Sec. \ref{s:dis}. Measurements and analysis described in \cite{engelberg_stochastic_2018} indicate that sub-critical current spikes follow the expected field dependence.  Fluctuations in field emission measurements have also been used as a way of extracting the field enhancement factor $\beta$, as described in \cite{lachmann_statistical_2021}. 

\begin{figure}
    \centering
    \includegraphics[width=1\linewidth]{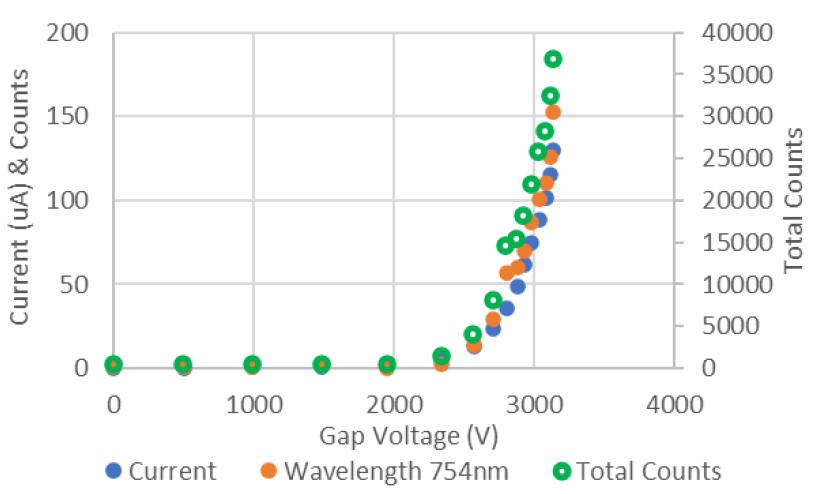}
    \caption{Correlation between field emission current and light intensity taken in the pulsed dc system. The blue points are current in $\mu \textrm{A}$, the green points show total light intensity and the orange points light intensity in a narrow band around a wavelength of 754~nm. From \cite{peacock_experimental_2023}}
    \label{fig:FE current light}
\end{figure}

Another important aspect of field emission is light is emission. Measurements of the light associated with field emission are described in \cite{kovermann_comparative_2010, peacock_experimental_2023}. The intensity of emitted light is closely correlated with emitted current, as shown in Fig.~\ref{fig:FE current light}. The origin of the light is not yet certain, but it is currently believed to be from optical transition radiation emitted when field emitted currents strike the anode. Such a mechanism would explain the observed direct proportionality of light and current compared to, for example, black body radiation from the emitter site which would give a more complex relationship. Optical emission opens new ways of monitoring field emission since it avoids potential damage to electronics which can occur due to the multi-order-of-magnitude current surge that occurs if a breakdown occurs during a field emission measurement.  Mitigating electronic measures, such as using series resistances can interfere with the measurement itself. In addition, the emitted light has a very large bandwidth especially compared to electronic current measurements. This may improve the fluctuation measurements described above. 

\begin{figure}
    \centering
    \includegraphics[width=1\linewidth]{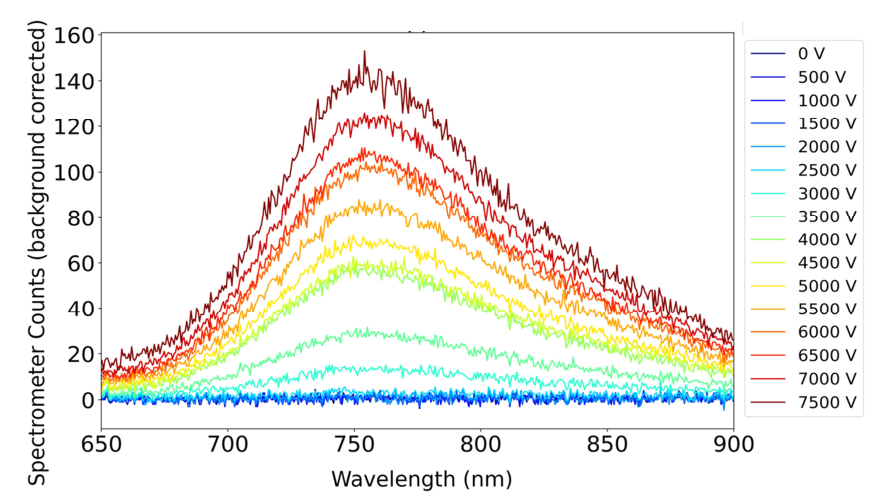}
    \caption{Optical spectra of light emitted during electron field emission measurements made on copper electrodes as a function of applied voltage. From \cite{peacock_experimental_2023}}
    \label{fig:FE light}
\end{figure}

In addition to intensity, optical spectra have been made of the light emitted during field emission measurements \cite{kovermann_comparative_2010, peacock_experimental_2023}. An example of such a measurement done using copper electrodes is shown in Fig.~\ref{fig:FE light}. Different materials show different spectra. Also, repeated measurements show fluctuations in the shape of the spectrum. This could be because the spectrum, or at least part of it, is caused by a mechanism in the emitter itself. Further studies are ongoing.

\subsection{\label{ss:behave-microscopy} Post-test microscopy}

All signals described in the previous sections are measured while the systems are exposed to high fields . 
However, a common feature of all these signals is that they offer indirect observations 
regarding their sources. 
Various attempts to directly observe variations in features and properties of metal samples have been performed, with the objective of trying to identify features that give a field enhancement factor and pertaining to the processes leading up to breakdown \cite{anders_review_2014}. Preliminary attempts at measuring field-emitted electron currents inside a scanning electron microscope (SEM) and observing the emitting sites prior to breakdown did not show any significant surface feature leading to the breakdown, at the resolution level permitted by the SEM \cite{muranaka_instrumental_2011, muranaka_-situ_2011}.

Following the high field exposure, many radio frequency devices and dc samples have been processed to allow direct observation of surface and subsurface features. The most prominent features observed in the samples that underwent breakdowns are craters. The latter are typically accompanied by material splashes outside the main crater formation. The size of the craters depends on various factors such as available energy for the breakdown process \cite{timko_energy_2011} and dimensional constraints; for example, in LES experiments, the small distance between electrodes leads to plasma confinement, which, together with electrical energy constraints, results in a typical size of breakdown crater. In contrast, breakdowns in large-scale rf structures usually are not confined, and variation in electrical conditions during breakdown leads to a wider distribution of crater sizes.

\begin{figure}
    \centering
    \includegraphics[width=0.52\linewidth]{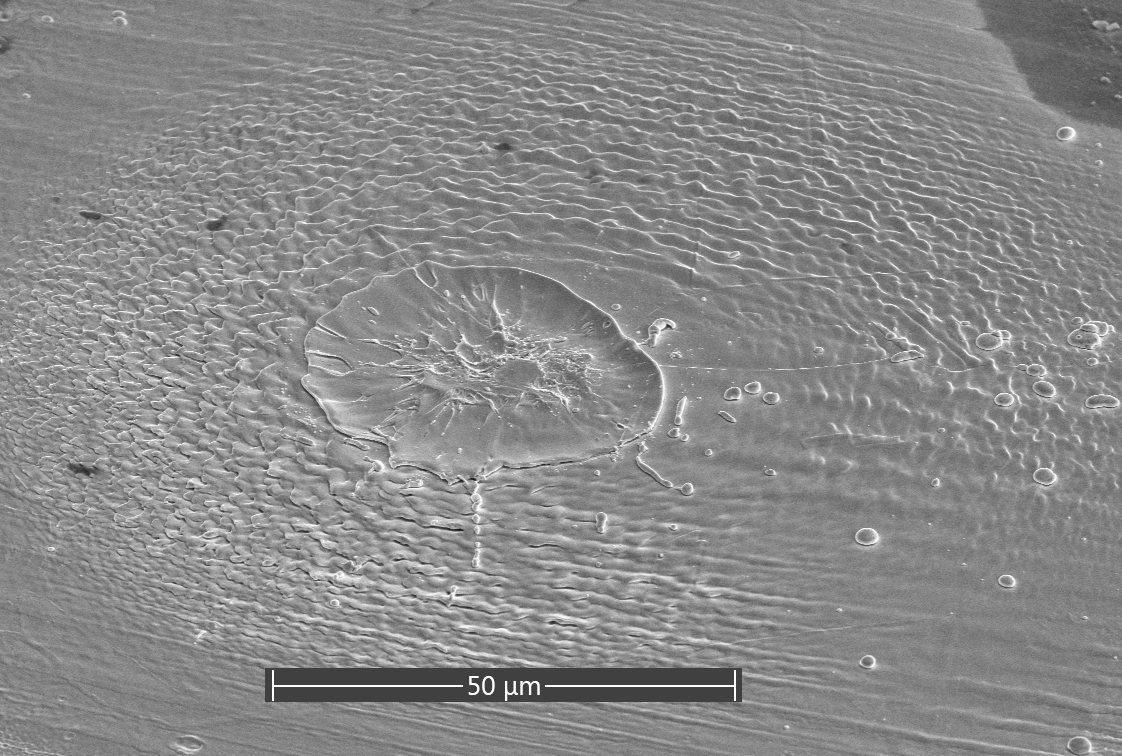}
    \hfill
    \includegraphics[width=0.45\linewidth]{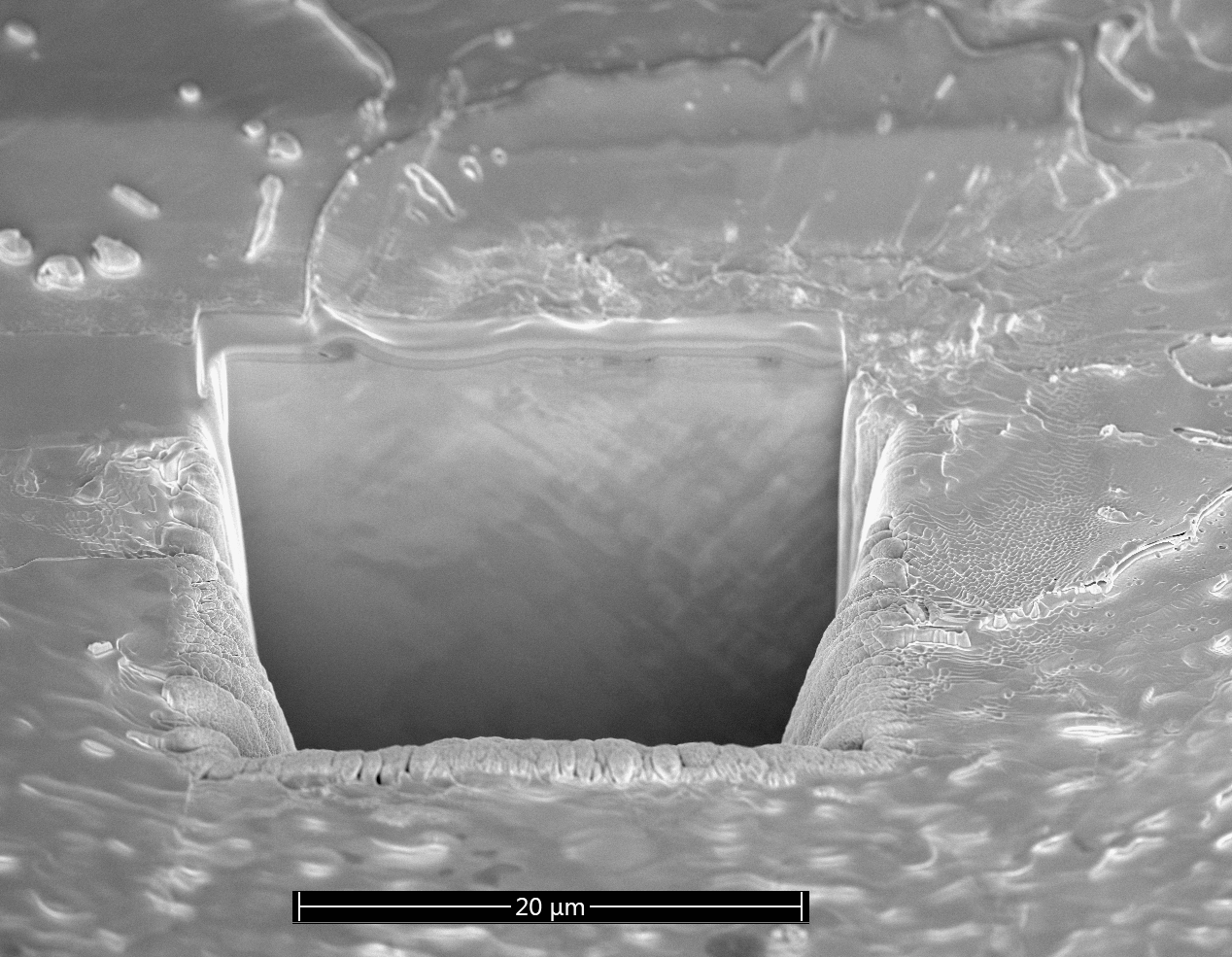}
    \caption{SEM image of a crater formed by a breakdown on an pulsed dc system Cu cathode. On the left is a global view, and on the right is a close-up of the same crater after part of the surface was removed by FIB (Focused Ion Beam) to expose the sub-surface plastic damage; shading is indicative of crystal orientation. \cite{ashkenazy_identifying_2021}}
    \label{fig:FIB_LES_1}
\end{figure}

\begin{figure}
    \centering
    \includegraphics[width=0.59\linewidth]{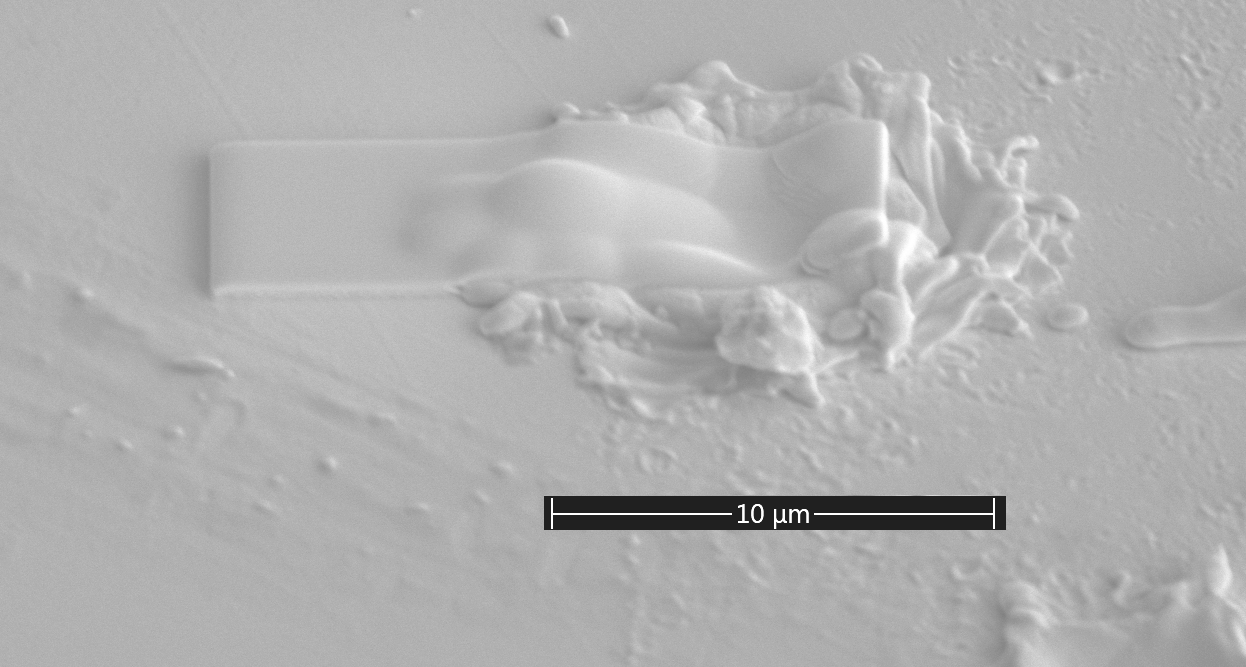}
    \hfill
    \includegraphics[width=0.39\linewidth]{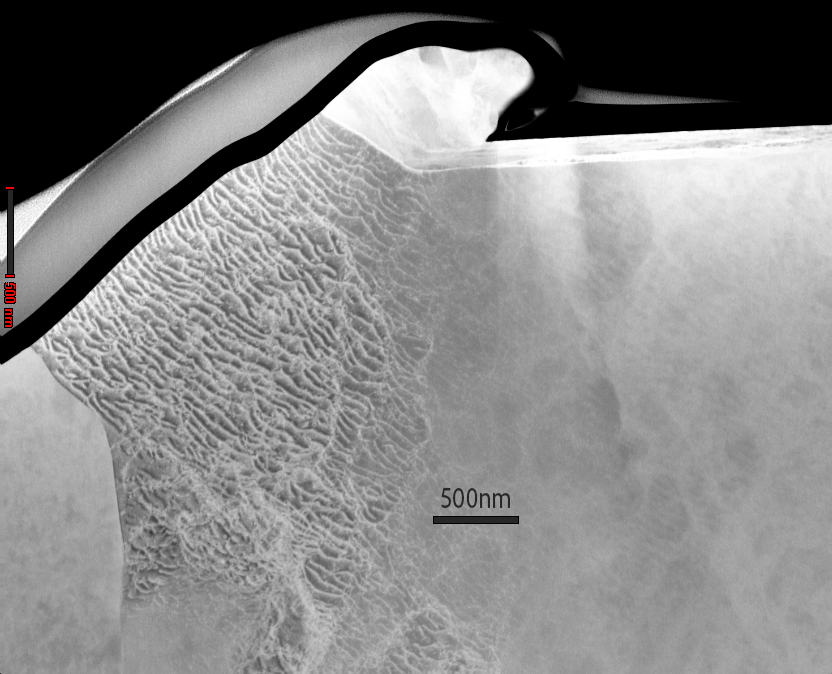}
    \caption{Crater formed by breakdown on a radio frequency device surface. The left picture is a global SEM view, and the right is a close-up STEM on a cross-sectional lamella. the lamella is created using FIB cut below the Pt film, which is deposited to prevent damage during production and is visible as a dark layer in the STEM picture.\cite{ashkenazy_identifying_2021}}
    \label{fig:cut_rf_1}
\end{figure}

Such dependencies can be reproduced in simulations of late-stage plasma evolution 
and theoretical results in that respect reproduced experimental observations of damage properties as described in Sec.  \ref{s:plasma}. 
An example of the surface damage is provided in Fig.~\ref{fig:FIB_LES_1} and Fig.~\ref{fig:cut_rf_1}  where crater and subsurface large-scale plastic evolution are demonstrated. It is clear from the cut below the surface that the energetic process that led to the formation of the surface crater also caused plastic deformation, seen as grain refinement, below the surface at dimensions similar to the radius of the crater. This damage was observed in a dedicated dc experiment with large energies available for plasma discharge. 
The observed damage in radio frequency systems is instead limited by the amount of power delivered to the breakdown plasma. 
These craters are the result of the 
energy deposition during plasma discharge 
and, as such, do not hold direct evidence of processes leading to a breakdown onset. 
Moreover, they may serve to erase any such feature responsible that can be associated with pre-breakdown signals. 


Typically, the surfaces of radio-frequency structures and direct-current electrodes are scrutinized \emph{post-mortem} using 
microscopic techniques after they were exposed to a number of breakdowns 
to address questions related to 
the nature of conditioning (both in terms of surface and material states) as well as to insights into the plasma composition. 
The most prominent features identified on surfaces that underwent breakdowns are craters formed by deposited material from the hot plasma and material displaced due to high temperature and pressure exerted by plasma on the surface.


\section{\label{s:dis} Plastic material response}

The current understanding of the early stages of a breakdown event at high electric fields is that it starts with the ignition of a localized plasma. 
Plasma onset evolves from the triggering of a nucleation event accompanied by a localized increase in field emission and is assumed to be driven by local plastic deformation \cite{engelberg_stochastic_2018}. 
This section discusses the link between the electric field-driven plastic response and the processes controlling the initial event that leads to a localized plasma burst. 
Models based on this link correlate temporal distributions of breakdown and field-induced current emission with sample conditions. These include dependencies on field intensity, sample temperature, and sample structure.
The following chapters treat the ensuing dynamics on and within the surface, which results in plasma formation, and the kinetics of this plasma.

 Even though vacuum breakdown and arcing are dominated by plasma dynamics, the latter, in turn, is a consequence of other processes developing after plasma onset. 
Therefore, we may have various processes at the origin of a breakdown with similar late-stage plasma evolution characteristics. 
Here, we divide breakdowns into three different groups according to their origin: extrinsic, intrinsic, and secondary events. 

Extrinsic events are those caused by external sources such as contaminants and singular preexisting surface irregularities that may cause localized sudden bursts of currents under the effect of an applied field \cite{hassanein_effects_2006, latham_high_1981, latham_high_1995}. 
However, such extrinsic sources are known to be removed efficiently during surface conditioning and are, therefore, ruled out as controlling the long-term asymptotic dynamics, as the breakdown rates tend to stabilize at a specific rate. 

Furthermore, attempts to correlate observations of such features using microscopy before field exposure with the locations of breakdown craters following exposure were unsuccessful. It is especially noteworthy 
that the location of breakdown craters on large-grained OFE copper does not correlate with the most pronounced topographical features, such as 
grain boundaries and triple junctions.

Another differentiation between observed events is the distinction between fast versus slow dynamics; see section \ref{ss:behave-Statistics}.
These two distinct time distributions, observed in field emission currents and breakdowns, were suggested to correspond to two different processes. The slow dynamics represent the underlying intrinsic breakdown processes, while the fast dynamics result from the secondary follow-up events, i.e., events that are likely triggered by remnants of the preceding breakdown event 
\cite{korsback_vacuum_2020, saressalo_effect_2020}. 
According to this hypothesis, plastic evolution within the surface mainly affects initial intrinsic events, so-called primary breakdowns. Thus, observations and models related to these do not apply to breakdowns caused by extrinsic mechanisms, i.e., by contaminants or external damage, even if due to a follow-up to a previous event.
 
\subsection{ \label{ss:material correlation} Link of breakdown rates to material response}

Previous chapters established that the properties of metals, specifically those used as cathodes in dc systems, control the ability of the metal to withstand high electric fields
(see Fig.~\ref{fig_cobalt})
\cite{descoeudres_dc_2008, descoeudres_cobalt_2009}. It was demonstrated that certain metals start developing conducting plasmas above their surfaces in fields much higher than those of the others. Although it was possible to rank different materials in order of the ascending breakdown fields (the average minimum electric fields resulting in an immediate breakdown for a given material), the mechanism by which this effect takes hold remained unexplained. 
Surprisingly, thermodynamic properties such as melting temperature and elastic constants failed to demonstrate the observed correlation \cite{calatroni_breakdown_2010}.

Based on the strong variation of maximal attainable fields with electrode material described in section \ref{ss:behave-material}, potential mechanisms were explored by studying the correlation between specific properties and the observed dependence.
The correlation of the probability of breakdown with the lattice structure was identified in Ref. \cite{descoeudres_cobalt_2009}.
Since dislocation motion is well-known to correlate with crystal structure, this insight was used to motivate a model based on the defect formation and migration properties \cite{nordlund_defect_2012}.

In this model, the variation of the mean time between breakdowns ($\tau = 1/R_{BD}$ where $R_{BD}$ is the breakdown rate) with the applied electric field was associated with the rate of defect formation in the material exposed to the field.  
Assuming a constant activation volume for defect formation and that the tensile stress $\sigma$ is derived directly from the applied electric field $E$ using Maxwell stress: $\sigma\propto\epsilon_0 \cdot E^2$, it follows that the breakdown rate $R_{BD} \propto exp (-\frac{E^2}{k_BT})$ or, in other words, a $ln(\tau) \propto -E^2$, giving the alternative functional form to the $E^{30}$ dependence, proposed earlier to explain the steep dependence of the breakdown rate on the applied field.
The defect formation-based model showed a good fit to a compilation of experimental data \cite{nordlund_defect_2012}. The defect formation probabilities were calculated to be consistent with the nucleation of extended defects, such as dislocation loops, although the ensuing dynamics were not specified. 

The initial proposal in Ref. \cite{nordlund_defect_2012} for a mechanism related to defect nucleation, led to a further search for mechanisms based on sub-surface plastic evolution in materials exposed to high electric fields and leading to the full plasma onset. 
One realization of such a process can be the motion of preexisting and multiplying extended defects - namely, dislocations. 
This proposal corresponds well with the structural correlation observed in Fig.~\ref{fig_cobalt} \cite{descoeudres_dc_2008, descoeudres_cobalt_2009} as dislocation mobility is strongly affected by the metal crystalline structure through its effect on properties such as the number of easy glide planes, the low-energy dislocation structure, dislocation mobility, and the mechanism for release of sessile systems. 

\subsection{ \label{ss:plastic_proc} Plastic processes due to dislocations - from quasi-static to dynamic evolution}

Dislocation dynamics is known to control plastic deformation of metals \cite{anderson_theory_2017}.
These dynamics are stochastic and the deformations depend on the finite probability of releasing various dislocation structures. Thus plastic deformation can occur below the deterministic yield point of the metal \cite{ngan_transition_2010} 
with a diminishing probability as driving forces are decreased. 
The stochastic nature of the process is significantly pronounced when the low stresses are applied as a cyclic load, which offers the repetitive sampling of the finite probability of dislocation release. 
 The collective dynamics of dislocations, which are highly affected by interactions between
mobile and sessile ones, can lead to strain localization along specific slip systems \cite{lukas_dislocations_1968}. These are known as persistent slip bands and can create significant protruding structures on the surface at the intersection between the slip band and the surface \cite{man_extrusions_2009}. 
Such micron-scale features were repeatedly observed in fatigued copper structures using SEM \cite{aicheler_evolution_2011}. 

\begin{figure}
  \includegraphics[width=0.49\linewidth]{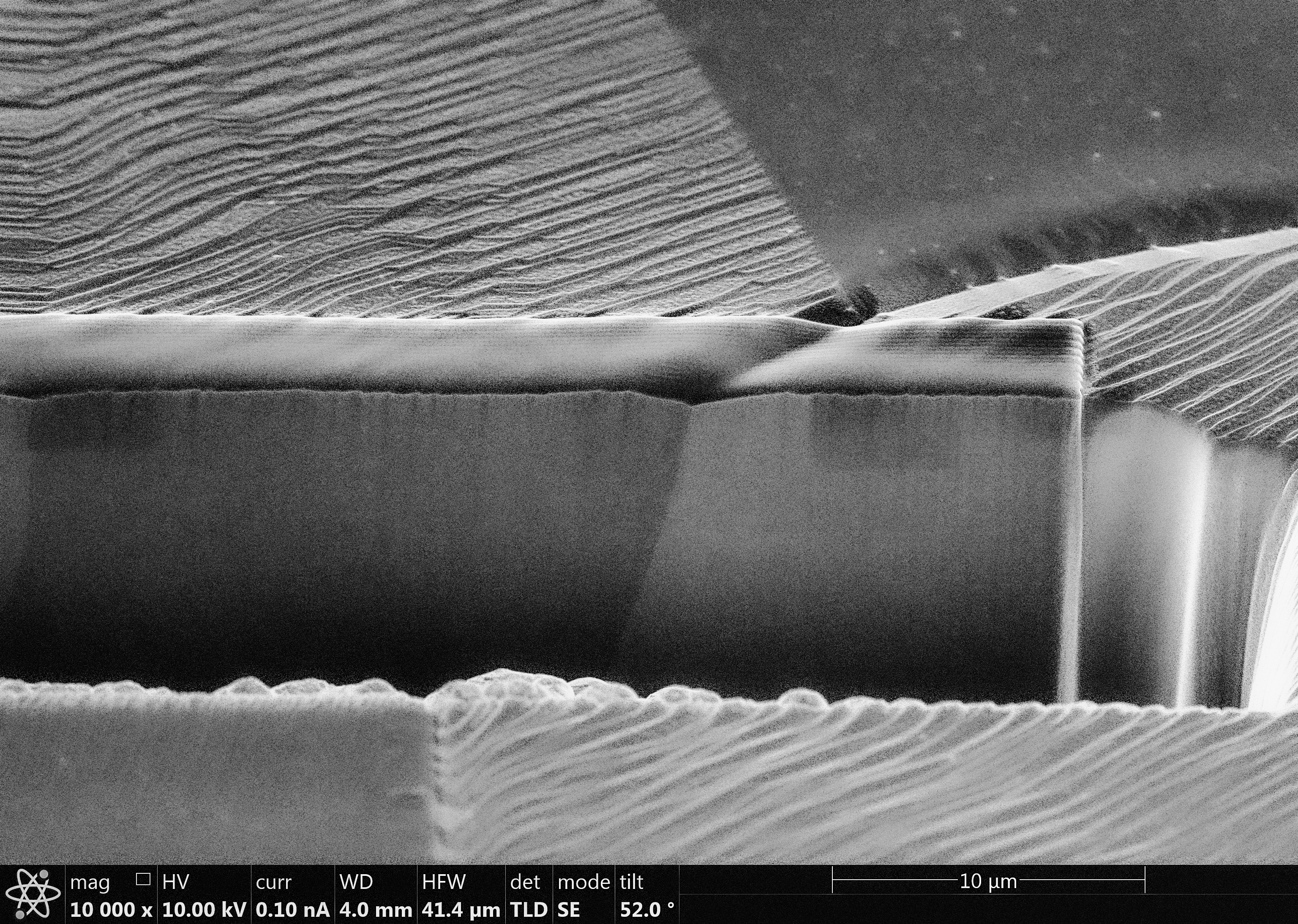}
  \hfill
  \includegraphics[width=0.49\linewidth]{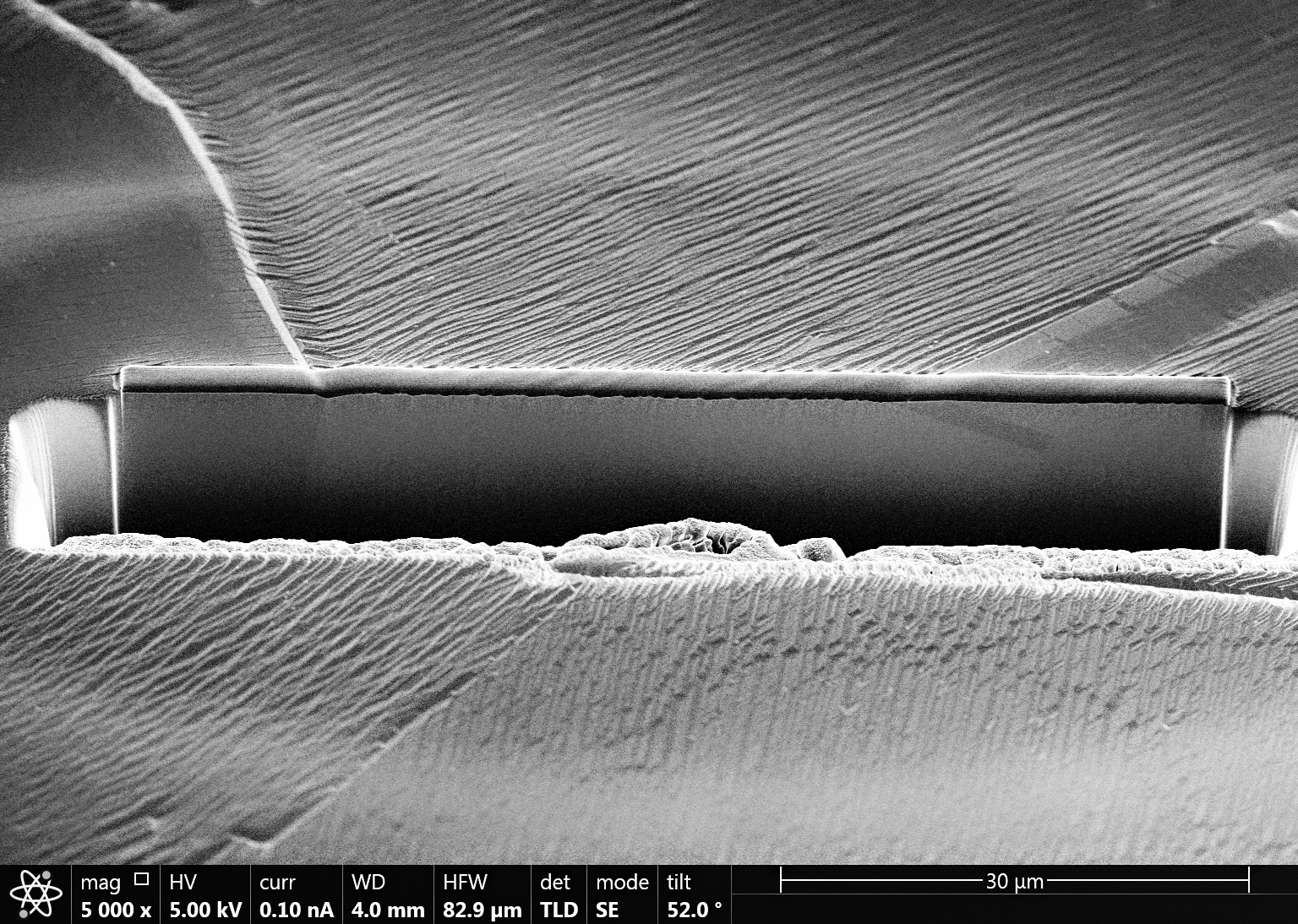}
  \caption
  {
    \label{fig_conditioned_surface}
    SEM of surface oxygen-free high-conductivity Cu electrode. The left panel shows a surface conditioned by repetitive exposure to a high field in a region between breakdown craters created during conditioning. The right panel shows how a similar scale view of how the surface looks for the same sample, but in the area that was not exposed to a high field. 
   }
\end{figure}

The structures identified in these cases involve stable extended periodic structures that can be observed after formation using SEM \cite{basinski_copper_1989,polak_profiles_2017}. 
However, detailed studies of surfaces exposed to high electric fields do not show similar features. Even conditioned surfaces (i.e., exposed to a long repetitive set of high-field pulses as described in Sec.~\Ref{ss:behave-conditioning}) retained their as-produced original global surface structure as exemplified in Fig.~\ref{fig_conditioned_surface}.
This is markedly different from the surfaces where persistent slip bands evolved, \textcolor{red}{as shown in Fig. 11} in \cite{polak_profiles_2017}. 
In addition, if a breakdown indeed nucleated due to the gradual deterioration of the surface or accumulation of damage leading to the growth of surface features, it should have been possible to identify the marking of these features outside the specific region of the breakdown itself. The gradual evolution over a wide area should produce independent sub-failure features also in similarly field-affected regions that are outside the particular breakdown nucleation site. 
As such features are not observed (see subsection \ref{ss:behave-microscopy}), a causal relationship between plastic surface damage and strong variation in surface response to the electric field is expected to be linked to a sudden localized event.
That such features are not observed by post-exposure microscopy is consistent with
measured field emission currents from the surface, which are indicative of a stable surface structure. 
While field exposure does affect surface resilience to breakdown (the aforementioned conditioning effect) 
gradual accumulation of damage leading to breakdown, even if unobserved using post-exposure microscopy, would lead to a strong memory effect in breakdown rates even in conditioned surfaces, as the accumulated damage reduces the amount of additional damage needed for future breakdown nucleation.
Such surface deterioration in conditioned surfaces is not observed, and in line with all of the above, it is assumed that gradual evolution can not be used to describe intrinsic breakdown nucleation.  

On the other hand, these observations are consistent with an abrupt breakdown nucleation process.
Such abrupt processes related to plastic response were described by a critical transition in the controlling parameter. 
As most experimental work was done on OFE Cu, where plasticity is controlled by the highly mobile dislocations, a theory based on the collective motion of these defects was offered to explain the mechanism leading to the observed critical transition. 

Various efforts were made to explore the relation between applied fields and dislocation activity, 
which depends on the activation of preexisting sources, commonly sessile dislocation arrangements. 
This link may be due to the activation of the dislocation source by temperature, external stress, and variations in internal stresses.
A significant effort to elucidate the mechanisms of dislocation dynamics within the copper surface exposed to a high electric field was made in several studies using atomistic simulations coupled with the electrostatic field effects \cite{pohjonen_dislocation_2011, pohjonen_analytical_2012,pohjonen_dislocation_2013,vigonski_molecular_2015,vigonski_verification_2015,bagchi_atomistic_2022}. The incorporation of the latter in a fully consistent model describing atom-atom interactions allowed for valuable insights on possible mechanisms of dislocation dynamics within the metal surface under external electric fields. 
For example, high tensile stresses exerted by the electric field on the surface were shown to lead to a deterministic formation of surface asperities by dislocation motion to the surface, and new dislocation sources at preexisting defects can be activated, providing the self-reinforced growth mechanism for surface asperities \cite{pohjonen_dislocation_2011,pohjonen_analytical_2012,pohjonen_dislocation_2013}. 
However, the estimation of the stresses required in these small-scale simulations to trigger the dislocation nucleation processes remained significantly larger than the Maxwell stresses that a reasonable value of an applied electric field can exert on the exposed metal surface  
\cite{zadin_electrostatic-elastoplastic_2014,vigonski_molecular_2015,vigonski_verification_2015,bagchi_atomistic_2022} .

It is worthwhile noting that even without forming a large-scale structure on top of the surface, dislocations moving below the surface may lead to significant modifications of the electronic response to an externally applied field. Such effects were offered in the past as an explanation for the increase in corrosion rates of metal subjected to cyclic loading \cite{li_effects_2002,li_variations_2005}.
However, as of today, a direct link between sub-threshold deterministic structure formation and possibly observable variation in field emission related to breakdown nucleation has not been demonstrated.

\subsection{\label{ss:critical} Breakdown as a dynamical plastic process.}
At stresses lower than the threshold for deterministic immediate activation of plastic processes, stochastic activation of such processes occurs through non-linear interactions, which can lead to a critical transition in the density of dislocations. Such phenomena have been shown to control threshold plastic response in various scenarios \cite{ngan_transition_2010, friedman_statistics_2012, papanikolaou_quasi-periodic_2012} and even in brittle to ductile transitions \cite{khantha_brittle--ductile_1994}.
The intermittent nature of plastic processes and the fact that such intermittent response
can be connected with a critical transition in dislocation mobility was reproduced in a mean-field model that uses the mobile dislocation density as the leading order parameter.  \cite{nix_micro-pillar_2011,ryu_stochastic_2015} 
The mobile dislocation density fluctuation (MDDF) model, is a mean-field kinetic model, which uses the number of moving mobile dislocations, and was constructed to reproduce the expected dynamics of driven surfaces under the stresses caused by an external field. \cite{engelberg_theory_2019}
The model describes the kinetics of the creation and depletion of mobile dislocations in a single slip plane, neglecting interactions between slip planes and the spatial variation of the mobile dislocation density within one plane. In this model, mobile dislocations nucleate at existing sources, and their depletion is due to collisions with obstacles. The problem is formulated in terms of a birth-death master equation for the mobile dislocation population.  
The resulting model is unique in that, for the first time, it treats a breakdown in metals as a critical transition, due to the stochastic evolution of dislocations under an external field. In contrast with gradual (or linear) evolution models, the transition does not require the appearance of sub-breakdown surface features.

\begin{figure}
  \includegraphics[width=1\linewidth]{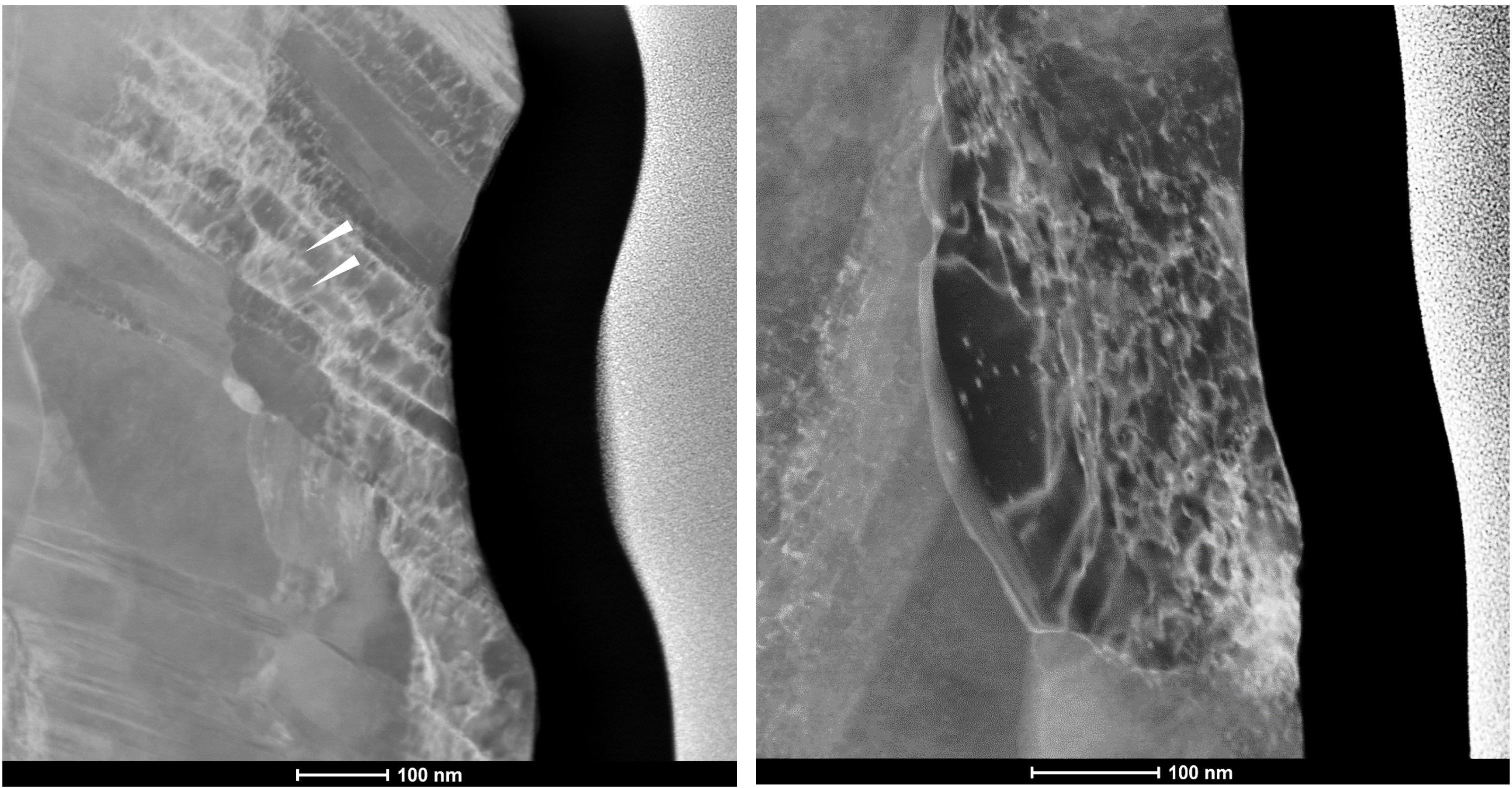} 
  \caption
  {
    \label{fig_TEM_cond}
    Cross-sectional TEM of a hard copper cathode. The left panel is from a reference region not exposed to a high field, and the right is from the same surface but on a region exposed to a high field and fully conditioned. The effects of the high field on the top 200 nm are evident.
  }
\end{figure}

It is noteworthy that such a model predicts that conditioning would involve evolution in the dislocation-related structures at the upper layer of the electrode. 
Indeed, microscopy of cross-sectional samples from conditioned electrodes, which were made from fine-grained hard OFE copper, showed the formation of dislocation-denuded zones with conditioning as demonstrated in Fig.~\Ref{fig_TEM_cond} \cite{jacewicz_surface_2024}.

\begin{figure}
  \includegraphics[width=1\linewidth]{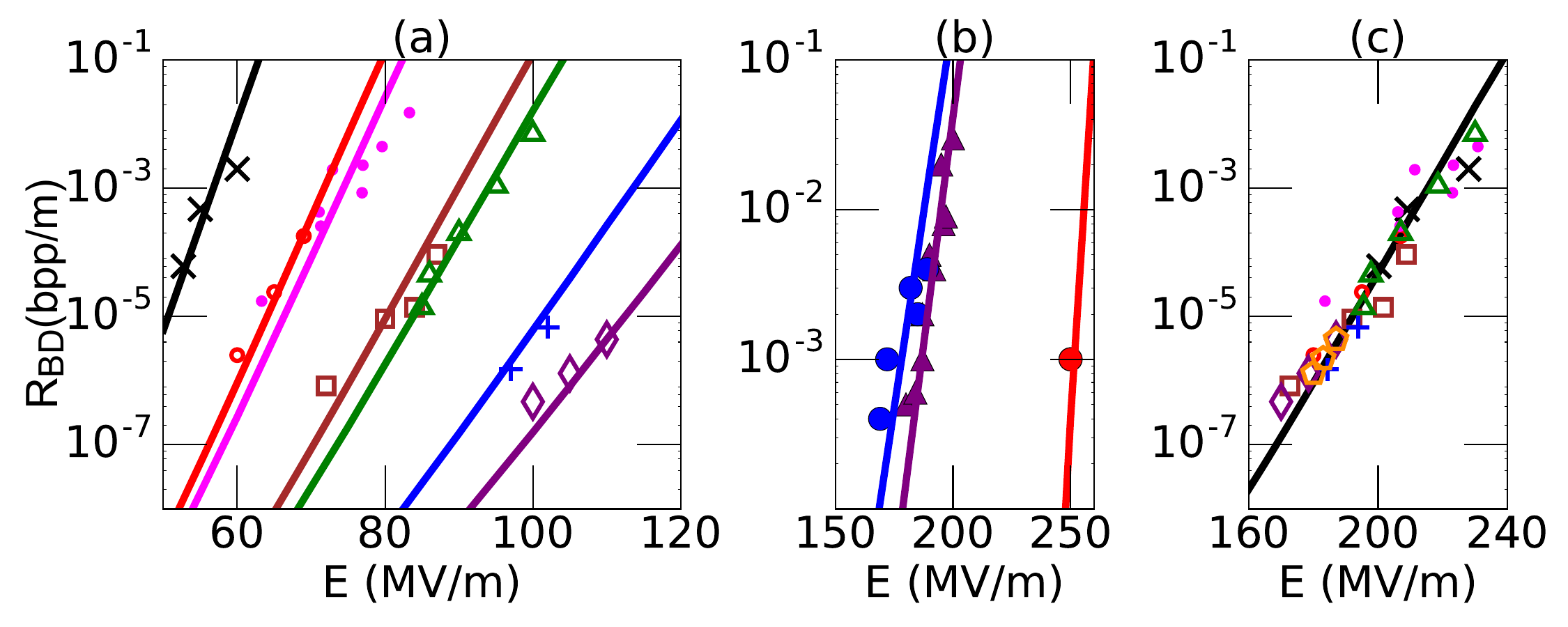}
  \caption
  {
    \label{fig_fits}
    Experimental \BDR~ in units of $(bpp/m)$ with fitted theoretical lines
    using Eq.~\eqref{eq:tau}:
    (a) \BDR~ versus $E$ for various Cu accelerating structures
    \cite{grudiev_new_2009}.
    (b) \BDR~ variation with $E$ at room temperature (two lines on the left)
    and at 45 K (line on the right) \cite{cahill_rf_2018}.
    (c) \BDR~ versus $E$ from sets (a) and (b) normalized so that all measurements are fitted with the same field enhancement factor (from \cite{engelberg_theory_2019}).
    }
\end{figure}

The MDDF leads to a non-trivial dependence of the mean time between breakdowns $\tau=1/R_{BD}$ and the applied field $E$. However, in two regimes, a simple approximated solution can be derived. For high fields, in the strongly driven state above the fluctuations controlled regime \cite{engelberg_theory_2019}, the mean time between breakdowns is dominated by an exponent of $E^2$ leading to $\tau \simeq exp(\alpha E^2)$, as was previously suggested \cite{nordlund_defect_2012} where $\alpha$ is a material and sample-dependent constant related to the plastic response of the surface to the applied field.   
However, for lower fields,  where breakdowns are fluctuation-driven, the full kinetic expression is approximated using a linear dependence of $\ln \tau$ on $E$,
\begin{equation} 
    \tau \simeq C \exp[\gamma (1 - E/E_0)] \label{eq:tau}.
\end{equation} \cite{engelberg_theory_2019}. 
Note that for $E$, the local field at the surface, the applied field is multiplied by a local field enhancement factor of the order of $1$, $E_0$ represents the threshold for the specific process and $C$ is a pre-exponential factor depending on current emitting area 
These dependencies reproduce observed behavior in field and temperature dependencies of the measured saturation breakdown rate over a wide range of drive conditions, demonstrated in Fig.~\ref{fig_fits}  \cite{engelberg_stochastic_2018,engelberg_theory_2019}.

In the described studies, the dislocations are modeled to respond to stress in the material. The applied field modifies the stress generated in the material without a direct interaction between the electrons and the dislocation within the material.
This assumption relies on studies that showed that only at very high current densities, direct momentum transfer from the electrons to the dislocation provides a significant contribution relative to the elastic interaction \cite{sprecher_overview_1986,liang_microstructure_2018}

\subsection{\label{ss:preBD} Pre-breakdown fluctuations}
A general characteristic of critical systems is that close to the critical transition the system state fluctuates. 
As was observed in other systems where plastic behavior was controlled by a critical transition, fluctuations in plastic activity and specifically dislocation motion precede the critical transition \cite{shekh_alshabab_criticality_2024, friedman_statistics_2012, zaiser_fluctuation_2005}.

In the case of pre-breakdown induced plastic activity, it is suggested that the system state is characterized by a surface property that affects the emitted pre-BD current and evolves through the motion of mobile dislocations. 
However, direct monitoring of the plastic response within the exposed surface in situ is still unavailable, and surface evolution is tractable only through indirect measurements.

Along these lines, in the MDDF model it was suggested that a critical transition in the density of the mobile dislocations within a metal can lead to a rapid increase in local current \cite{engelberg_theory_2019,engelberg_stochastic_2018}. 
While this model is in line with previous suggestions that the electric field is enhanced at the breakdown nucleation site due to localized plastic activity    \cite{vigonski_molecular_2015,bagchi_atomistic_2022}, it does not detail the specifics of the plastic process. 
However, it can be speculated that independent of the specific mechanism, an increase in the pre-BD dislocation fluctuations, which are an essential part of the MDDF, would lead to a rise in the field emission current.
Therefore, although no observable plastic evolution of the surface needs to be observed, an increase in field emission fluctuations should be measured.

In the model, subcritical fluctuations of the mobile dislocation density might be observed as fluctuations of the field emitted current during routine operation \cite{engelberg_stochastic_2018,engelberg_theory_2019}. 
It is important to note that the rate of these fluctuations differs significantly from the breakdown rate and is not linearly related to it.  
Theoretical estimates of these pre-BD fluctuations were derived based on estimates of surface annihilation rates for mobile dislocations as predicted by the MDDF \cite{engelberg_dark_2020}.

For a study of these fluctuations, measurements of the changes in the field emission current, rather than its average value, were conducted in CERN and compared to theoretical estimates of these fluctuations as derived from the full stochastic description of the MDDF. 
Measurements were done in a dedicated setup, and a sub-BD event identification served to identify the time distribution of the specific isolated dark-current spikes. 

\begin{figure}
  \includegraphics[width=0.9\linewidth]{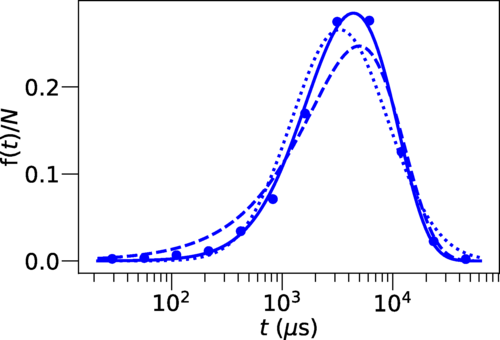} 
  \caption
  {
    \label{fig_fluc_dist}
    Representative histogram of the time interval between current spikes for a single electric field. The circles represent experimental results (with markers larger than the error bars), and the lines represent best fits to hypoexponential (solid), exponential (dashed), and log-normal (dotted) distributions \cite{engelberg_dark_2020}. 
   }
\end{figure}

\begin{figure}
  \centering
  \includegraphics[width=1\linewidth, trim= 0 0 0 70]{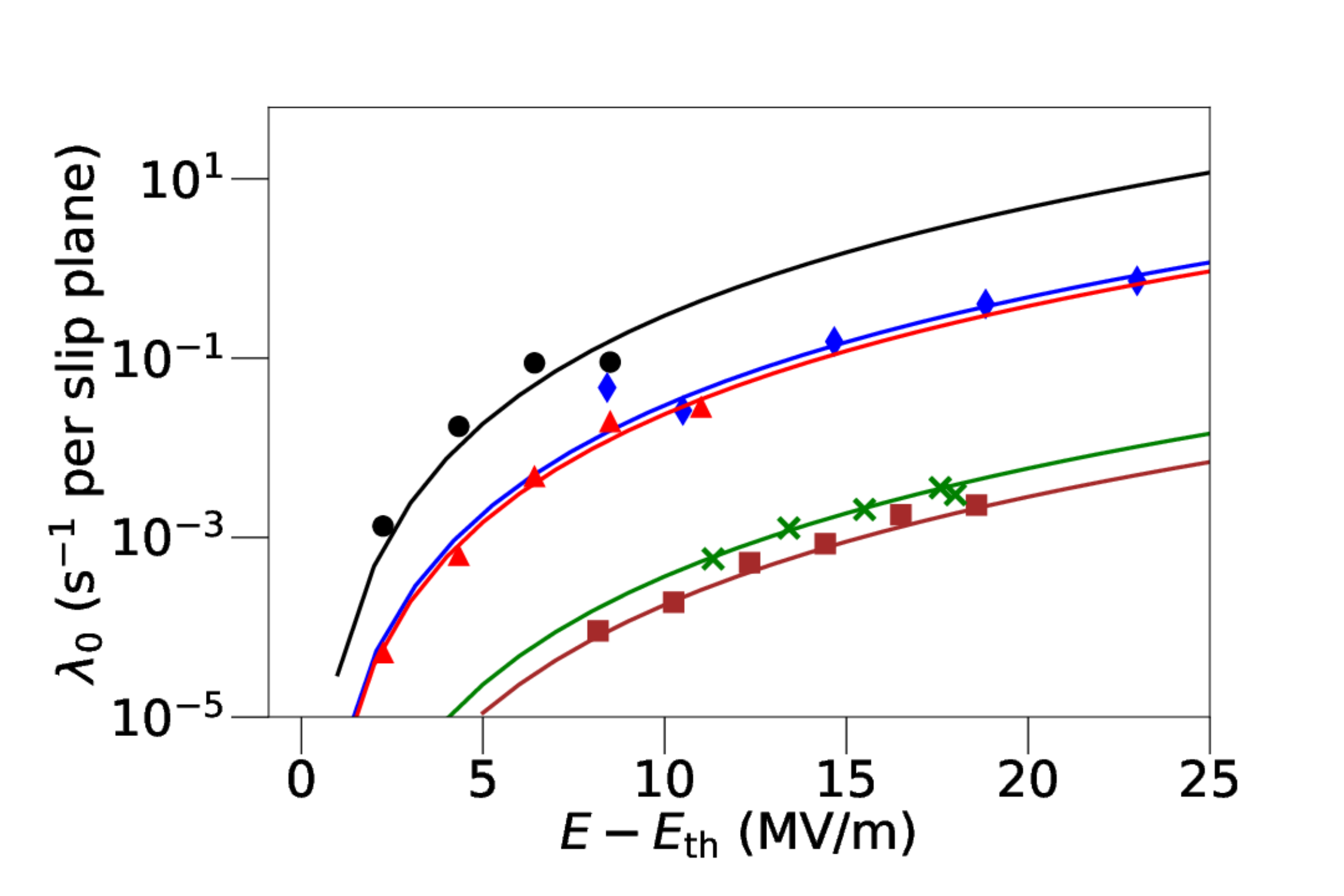}
  \caption
  {
    \label{fig_lambda0} 
    Rate of current spikes as a function of the surface electric field,
    experimental (markers) and theoretical (lines) from \cite{engelberg_dark_2020}.
  }
\end{figure}

Theoretical estimates of the dark current spikes were based on the assumption that these are correlated with multiplication events of the mobile dislocation population. 
The assumption that all transitions within the model are individual Poissonian processes led to a theoretical description of the time distribution between current spikes, which follows a two-parameter hypoexponential distribution with parameter values derived directly from the model \cite{engelberg_dark_2020}.
The model reproduced well both the distribution and the average fluctuation rate, as demonstrated in Fig.~\ref{fig_fluc_dist} and in Fig.~\ref{fig_lambda0} correspondingly.

Aside from providing support for the validity of an MDDF type model, an interesting prospect deriving from these observations is that this correlation may be utilized to reduce the breakdown rate in high-voltage setups and improve operational procedures in such systems, by using rates of observed dark current spikes to determine the level of conditioning of electrodes, or by monitoring spikes as observed in situ before breakdown.

\begin{figure}
  \includegraphics[width = 8.5cm]{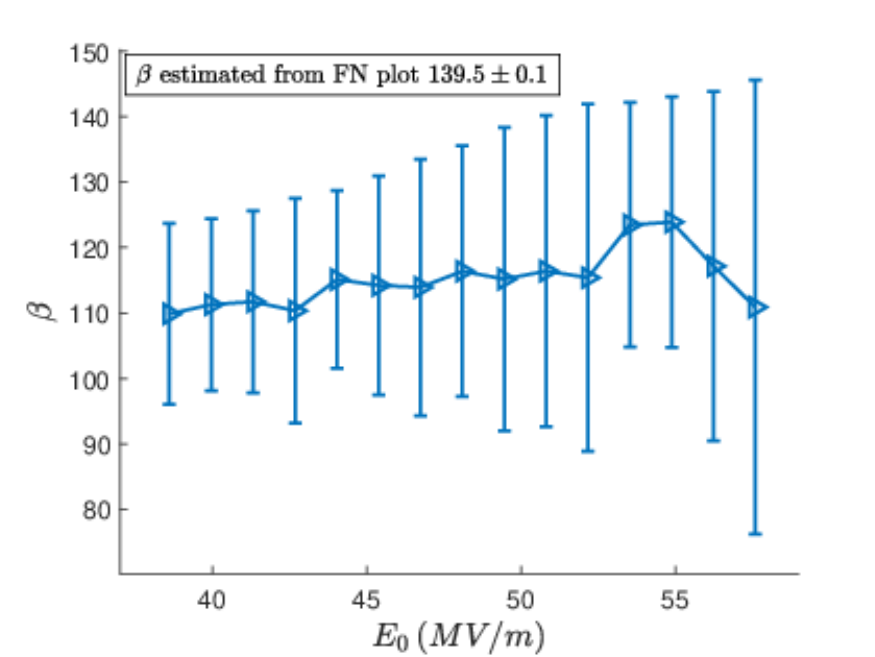} \\[-2ex]
  \caption
  {
    \label{fig_beta_var}
    Example of field-specific $\beta$ estimation at 30\,K with the field variation method. The estimated $\beta$ value with the average FN slope method is $139.5\pm0.1$ \cite{lachmann_statistical_2021}
  }
\end{figure}

Aside from the field-emitted current spikes distribution,  the fluctuations in the average field-emitted currents were used in \cite{lachmann_statistical_2021} to create a field-specific evaluation of the surface field enhancement factor $\beta$. 
The results of this analysis are presented in Fig.~\ref{fig_beta_var}, demonstrating insignificant variation of the field enhancement factor with the field. 
While these measurements are scarce and are yet to be performed up to the BD threshold itself \cite{lachmann_statistical_2021},  they do provide an additional indirect indication that field emission is not controlled by a gradual process, strengthening the hypothesis of a critical transition.

\subsection{\label{ss:dis pulse length}Pulse length dependence}

\begin{figure}
  \centering
  \includegraphics[width = 8.5cm]{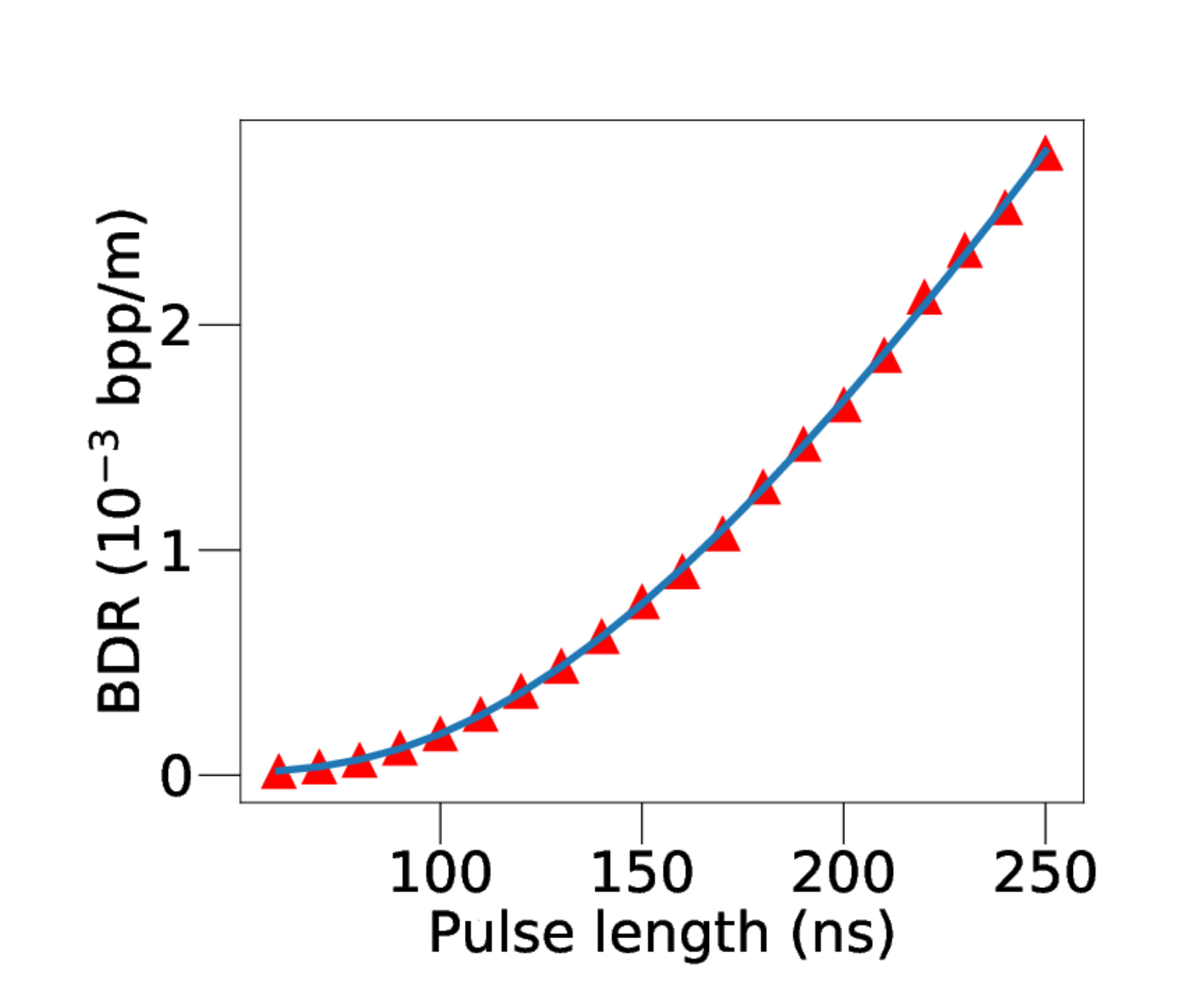}
  \caption
  {
    \label{fig_bdr}
    BDR as a function of the pulse length, $t_\text{pulse}$,
    for the nominal set of parameters and an electric field of 250 MV/m,
    found from the numerical simulations (triangles).
    The line is a fit to Eq.~\eqref{eq_bdr_for_pulse_length} (taken from \cite{engelberg_theory_2019})
  }
\end{figure}

In most applications, the electromagnetic field driving the breakdown is a pulsed rf signal.
Thus, the driving forces act for a limited pulse length, $t_\text{pulse}$,
which allows for exploration of the kinetics of the breakdown nucleation driving force.
Assuming that the plastic activity within the sample follows stochastic evolution  leads to a strong dependence of the \BDR ~ on $t_\text{pulse}$,
which can be empirically shown to satisfy
\begin{equation}
  R_{BD} = R_0 + m(t_\text{pulse} - t_0)e^{-\delta / t}, \label{eq_bdr_for_pulse_length}
\end{equation}
see Fig.~\ref{fig_bdr} \cite{engelberg_theory_2019}.
Here, \BDR~ is the breakdown rate and $R_0$, $t_0$, $m$, and $\delta$ are constants depending on the field.
Such behavior was measured in connection with the CLIC project where pulse lengths $t_\text{pulse}$ between 50 and 400~ns were examined,
with a duty cycle of 50~Hz \cite{wuensch_experience_2014}. Measurements of the dependence of breakdown rate on pulse length are described in subsection \ref{ss:behave-dependencies}.

In various scenarios, the BDR was shown experimentally,
to have an exponential or power-law dependence on
$t_\text{pulse}$ \cite{dobert_high_2004, grudiev_new_2009, degiovanni_comparison_2016}.
However, the validity of using the existing data to determine the dependence is limited, as it consists of either a small sample \cite{dobert_high_2004} or of measurements taken during the conditioning process, when the BDR is still dominated by extrinsic processes \cite{degiovanni_comparison_2016}.
At the very least, the BDR is expected to saturate for a continuous-wave rf signal. 
Therefore, the exponential or power-law dependence holds only for a limited range of pulse lengths.

\begin{figure}
  \centering
  \includegraphics[width = 8.5cm]{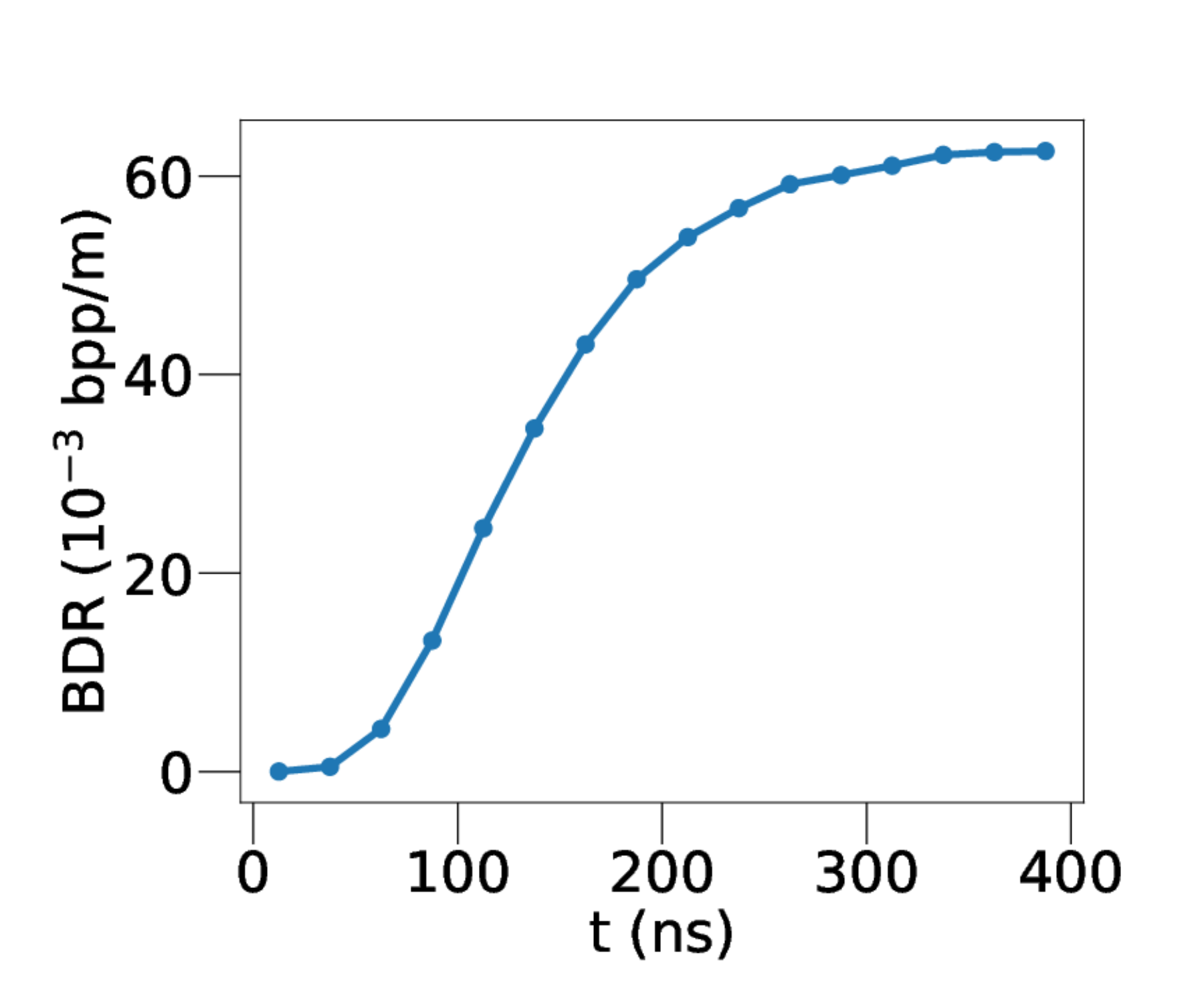}
  \caption
  {
    \label{fig_bd_distribution_within_pulse}
    The probability of a breakdown occurring as a function of the time within the pulse, found by simulation, for the nominal set of parameters, an electric field of 250 MV/m, and a total pulse duration of 400 ns.
    The probability distribution is presented as a histogram of sixteen bins, each bin 25 ns wide,
    and normalized by the total BDR at 250 MV/m and a pulse length of 400 ns. The total number of breakdowns is $10^6$.
  }
\end{figure}

Under the assumption of inter-pulse independence, as discussed in section \ref{s:behave} with relevant data shown in 
Fig.~\ref{fig:BD in pulse dist}, the distribution of breakdowns in time within each pulse should be
an increasing function due to the finite evolution of the time to transition $t_\text{tr}$. 
Measurements of the distribution of time of breakdowns within a pulse are described in Sec. \ref{ss:behave-Statistics}.
Figure~\ref{fig_bd_distribution_within_pulse} shows this
distribution for $E$ = 250 MV/m and a pulse duration of 400 ns.
For a given time interval $(t, t + dt)$ within a pulse,
an event will mature if it began
within the time interval $(t - t_\text{tr}, t + dt - t_\text{tr})$.
Given that a breakdown occurred,
the probability that it occurred within an interval $dt$ is,
therefore, $(dt/t_\text{pulse})\int_0^t P(t_\text{tr})dt_\text{tr}$,
where $P(t_\text{tr})$ is linearly proportional to the
probability distribution function.
This integral, however, is simply the CDF of $t_\text{tr}$, which is derived from the MDDF model.
This non-Poissonian distribution becomes predominantly Poissonian for times
that are significantly greater than $t_\text{tr}$.

If, however,
the interval between pulses is smaller than the typical relaxation time,
then the breakdown probability should not depend on the pulse duration alone,
but rather on the combined effect of exposure to the field
and the relaxation achieved between pulses.
In this case,
the variation in the BDR within the pulse can be small
and characterized by a constant probability,
similarly to the slow variation observed for $t > 500$ ns in Fig.~\ref{fig_bd_distribution_within_pulse}.
Indeed, in \cite{wuensch_statistics_2017} it was shown that the breakdown distribution
does not vary significantly within the pulse.
However, an increase in the breakdown probability was observed for one of the
structures in \cite{wuensch_statistics_2017}, and it was not clear whether full conditioning was achieved. 
Therefore, experiments involving variation of the duty cycle,
as well as additional data regarding the pulse-length dependence
during and after conditioning,
can help determine the exact nature of the pulse-length
dependence of the BDR and help determine the memory effect between and within pulses.

The success of dislocation-based models in describing distributions of BD, and pre-BD phenomena, together with the consistent microscopic observations, provides significant support to the hypothesis that indeed dislocation-based dynamics control the early stages prior to plasma onset.

\subsection{\label{ss:others}Other indications of plastic activity}
Significant dislocation activity can lead to various signals, including electric signals, due to effects on resistivity and acoustic emission due to the collision of dislocations with obstacles. 
In these cases, macroscopic signals help quantify variations in the global density of dislocations or significant increases in the number of moving dislocations.
Such a measurement, independent of microscopic observations, may help corroborate the assumption that dislocation motion is related to the observed processes. 
For plastically induced cyclic loading, it was shown previously that moving dislocations produce a measurable acoustic emission signal \cite{lhote_dislocation_2019}. However, efforts to measure the acoustic emission for surfaces exposed to electric fields have not yet produced measurable signals, due to the inaccessibility of the emitting area to direct acoustic monitoring.
Additional examples of such measurements include the effects of dislocation density on the local variation of resistivity in the top layer, which is expected to affect the resistivity of the uppermost grains and yield a measurable signal, as observed in direct plastic deformation \cite{miyajima_change_2010}.
For both methods, initial trials were performed, and for the resistivity system, a detailed design was simulated \cite{coman_situ_2023}, but as of now, the measurability of pre-BD signals remains an unsolved issue.

\section{\label{s:surf} Surface dynamics}

It is customary to associate the vacuum breakdown activity with the various surface irregularities of different origin that may give rise to the field emission currents eventually developing into full vacuum arcing events \cite{juttner_vacuum_1988, juttner_erosion_1979, behrisch_surface_1986, anders_review_2014, latham_high_1981,latham_high_1995}. Mainly these irregularities are associated with surface contamination or post-arcing sharp feature at rims of cathode spots. Alternative hypothesis of a potential whisker growth to serve as an initial field emitting tip, has also been proposed \cite{lee_spontaneous_1998}. To date, however, there is no consistent theory which can show the relevance of these processes to the pre-breakdown condition on metal surfaces exposed to high electric fields. To understand better the physics of metal surface evolution in the presence of a high electric field, it is important to take a deeper look into materials properties and the mechanisms that may govern modification of these properties under applied electric fields.

Surfaces subject to high electric fields undergo several dynamic processes that can create and/or amplify various surface irregularities. As was shown in Section \ref{s:dis}, some few atomic layer protrusions, step edges, or even larger irregularities may develop on the surface as a result of under-the-surface dislocation dynamics triggered by the stress that is a consequence of the application of an external electric field. 

Strictly speaking, the interatomic forces that hold the material together at the surface act to smooth out any irregularities on the surface following the principle of minimization of surface energy \cite{cahn_contribution_1984}. This implies that any atomically sharp surface feature experiences a strong thermodynamic force towards a blunter and, hence, more energetically stable shape. This effect was observed in kinetic Monte Carlo atomistic simulations in Ref. \cite{jansson_long-term_2016}. However, the features resulting from the dislocation activity under the surface can develop further as a result of electrostatic forces acting on partially charged surface atoms by the applied electric field. Although the electric field itself does not penetrate below the first atomic layer, these forces translate into tensile stress throughout the metal bulk because of the interatomic interactions. Because of the nature of these interactions, the force on surface atoms is always directed out from the surface, imposing tensile stresses within the material. These stresses can trigger dislocation dynamics that leads to the growth of asperities on the surface of the metal exposed to a high electric field \cite{pohjonen_dislocation_2013}.

There is significant experimental evidence that in the presence of an external electric field, the diffusion of surface atoms becomes biased towards field gradients (i.e. inhomogeneity of the electric field around surface irregularities) \cite{tsong_measurements_1972,tsong_direct_1975,kellogg_measurement_1977}. This effect was shown to lead to the formation of surface features; for example, nanoprotrusions were seen to grow in several experimental studies \cite{binh_electron_1992,nagaoka_field_2001,fujita_mechanism_2007,yanagisawa_laser-induced_2016, meng_situ_2024}. 
 
\begin{figure}
    \centering
    \includegraphics[width=0.9\columnwidth]{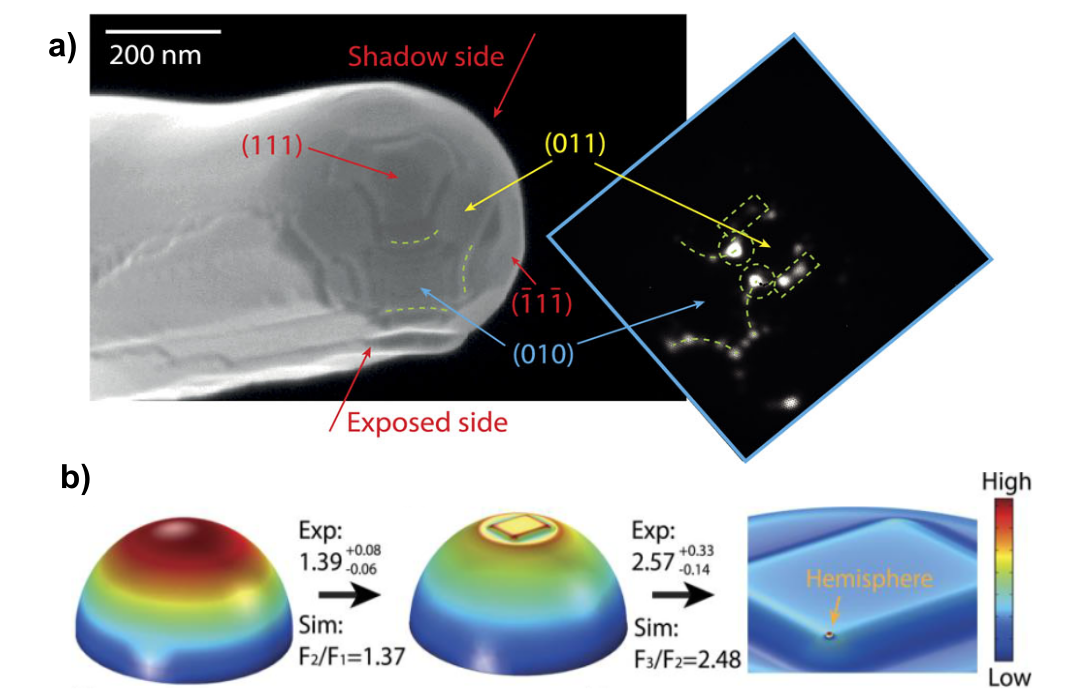}
    \caption{a) SEM image of the apex of the tungsten tip, which was exposed simultaneously to an electric field and femtosecond laser irradiation for nearly 5 hours. The inset is the electron emission pattern with the bright spots corresponding to the locations of the developed field-emitting protrusions. The identical facets are illustrated with Miller indices in both SEM and field-emission pattern images. b) Finite-element method calculations of the field enhancement are due to faceting (middle image) and the formation of a small tip at the corner of the faceted plateau (far right image). The color bar shows the value of the electric field. The highest values from left to right are 2 GV/m, 2.74 GV/m, and 7.34 GV/m.  The figure is generated from the images in Ref. \cite{yanagisawa_laser-induced_2016}.}
    \label{fig:laser_Wtip}
\end{figure}

Figure~\ref{fig:laser_Wtip}a illustrates the faceted surface of a W tip exposed for 5 hours to the simultaneous effect of an electric field and local heating by a laser \cite{yanagisawa_laser-induced_2016}. The inset shows the field emission map after 2.5 hours from the beginning of the experiment. Here, the bright spots correspond to the localized field emission currents as a result of the appearance of field-emitting spots along the edges of the surface facets. Weak emission is still visible from the ridges, which was stronger at the beginning of the experiment but started fading away when more localized field-emitting spots started to grow. Eventually, several initial spots merged into a single one at the sharpest corner (similar to that shown in Fig.~\ref{fig:laser_Wtip}b) with a much stronger current at a lower applied voltage. We note that the W tip was carefully relaxed prior to the experiment to ensure the absence of any irregularity on the tip surface, which could trigger the development of the observed faceted structures with the subsequent formation of field-emitting spots. 

The significant reduction of the applied voltage (up to three times) observed in this experiment, which still led to a strong field emission signal, cannot be explained by any other phenomenon.  
The corresponding finite-element method (FEM) calculations, see Fig.~\ref{fig:laser_Wtip}b, confirmed that similar field enhancement (a.k.a. the applied voltage reduction) can be obtained if one assumes the formation of a protrusion at the corner of a surface facet. The latter may form as a result of thermal stresses during rapid heating and cooling of the surface by femtosecond laser irradiation. Here, the color bar indicates the electrostatic field value with the highest value changing from left to right from 2 V/nm to 7.34 V/nm, measured at the small corner protrusion.

This experiment indicates that, indeed, a metal surface under an electric field can develop tip-like structures that cause local field enhancement. 
Although rapid cyclic thermal stress may trigger the formation of ridge-like structures at the facet boundaries, further development of these surface features into a nanoprotrusion is more gradual and can be associated with possible biased diffusion of surface atoms under the electric field gradient, driving the atoms toward regions with locally enhanced electric fields.  

\subsection{Theoretical modeling of surface charge effects on atomic dynamics}
\label{sec:helmod}

Metal surfaces exposed to an external electrostatic field become charged but the field does not penetrate deeply beneath the surface, only within 1-2 monoatomic layers \cite{jackson_classical_1975}, because free electrons re-distribute within the metal bulk to screen the external electric field at the surface. Hence, it is sufficient to understand the nature of the surface charge in terms of excess or depletion of electron density at the surface atoms depending on the direction of the applied electric field. Generally speaking, the redistribution of electron density is not uniform and depends on the positions of the atoms with respect to the positions of the surrounding surface atoms. This results in a non-uniformity of the partial charges that can be assigned to individual surface atoms. The partial charge associated with an atom responds to the applied electric field, adding electrostatic forces to the atomic dynamics that may modify the evolution of the surface morphology under the applied field. 

\begin{figure}
    \centering
    a)\includegraphics[width=0.8\columnwidth]{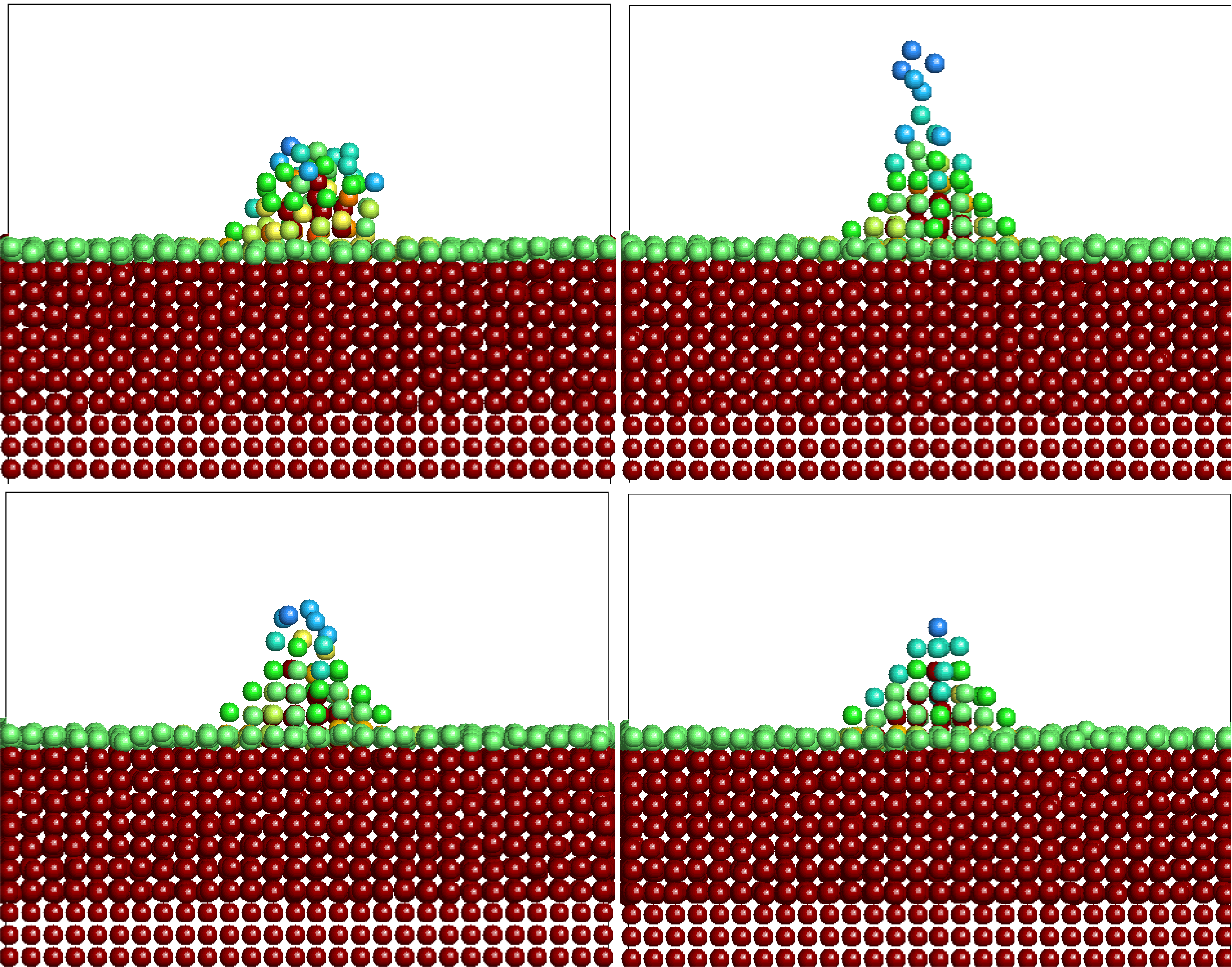}
    b)\includegraphics[width=0.8\columnwidth]{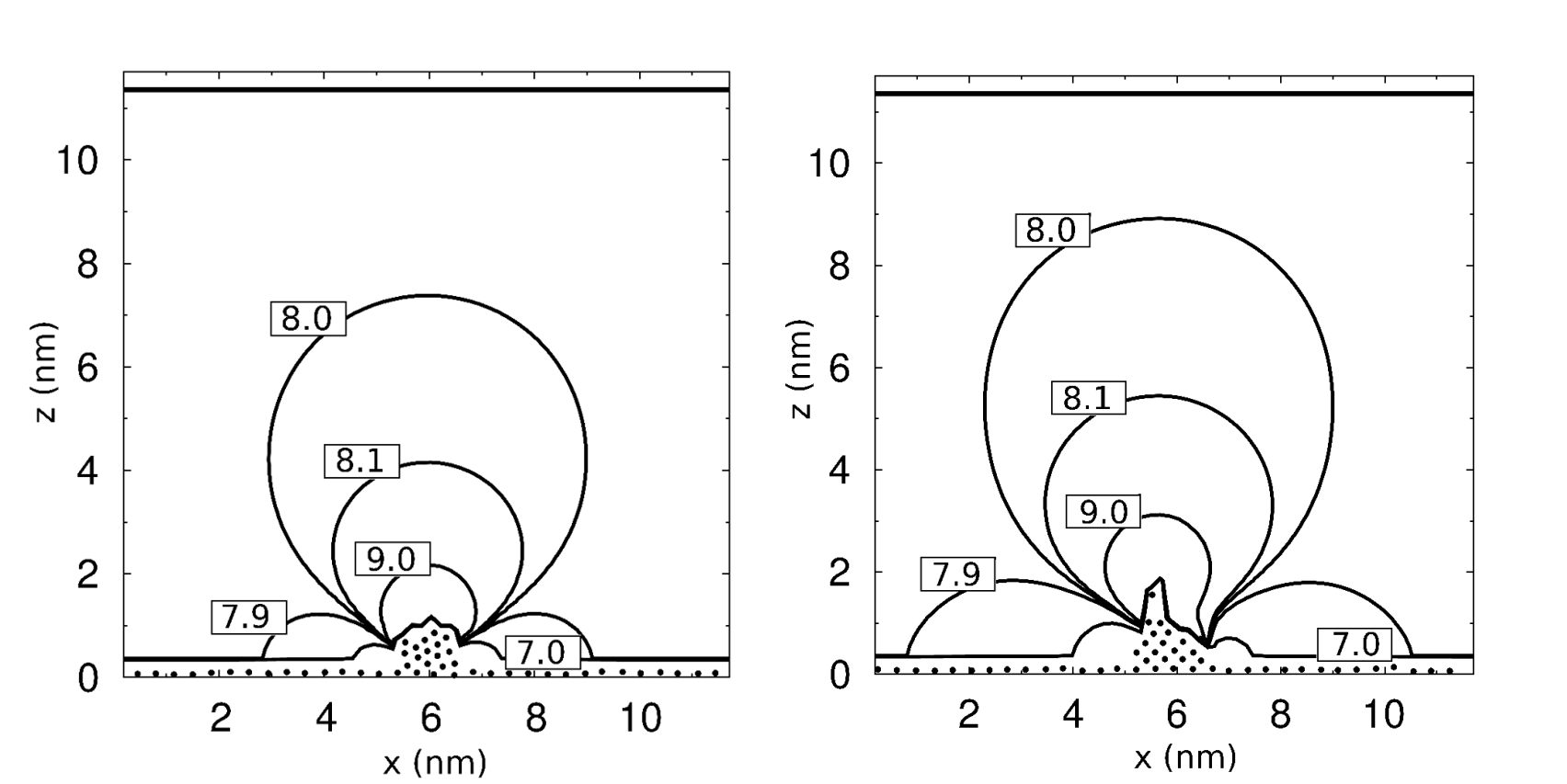}
    \caption{ (a) shows four snapshots of atomistic simulations of evolution of a surface tip under the applied electric field of 8 V/nm at the temperature of 600K. The balls representing Cu atoms are colored according to the partial charge from 0e (dark red) to 0.48e (light blue) which is induced on surface atoms by the applied electrostatic field. (b) shows the electrostatic field dynamically updated during the simulation following the change in surface morphology. The contour lines in the cross-sectional images show the  equipotential surfaces of the field with the values given in the boxes with an accuracy of 0.1 V/nm. Adopted from Ref. \cite{djurabekova_atomistic_2011}. }
    \label{fig:helmod_dynamic}
\end{figure}

The first approach to tackle this problem was developed by \cite{djurabekova_atomistic_2011}. They proposed an algorithm for assigning partial charges (i.e. a fraction of an electron) using the Gaussian pillbox to estimate the charge induced by the local field according to Gauss's law at every atom \cite{jackson_classical_1975}. In this way, each individual surface atom is assigned a partial charge depending on its position at the surface and the local atomic environment. In this work, the distribution of the electrostatic field in vacuum is calculated by solving the Laplace equation for the applied field imposed as a boundary condition far above the surface. The solution is updated on the fly following the dynamic change in the surface morphology. A snapshot series in Fig.~\ref{fig:helmod_dynamic}(a) illustrates atomistic simulations of the resulting dynamic evolution of a nanometer-size tip on the Cu surface under an electric field. 

Slow diffusion processes are very challenging for molecular dynamics simulations, which can only cover very short time spans of a few nanoseconds \cite{thompson_lammps_2022}. To allow any dynamics to be observable within this short time span, elevated temperatures ($\sim$600 K) and electric fields ($\sim$ 8-10 GV/m) are needed. 
The atoms are colored according to the partial charge induced on the surface atoms by the applied electric field. The latter is enhanced around the tip as can be seen in Fig.~\ref{fig:helmod_dynamic}(b), where the cross-sectional views of the same surface tip in Fig.~\ref{fig:helmod_dynamic}(a), including the equipotential contour lines to indicate the field enhancement evolution following the shape modifications of the surface feature. As one can see, the partial charge in these simulations is sufficient for field-assisted evaporation of the top atoms already within the short time span of the molecular dynamics simulations. The dependence of the partial charge value of a surface atom on its immediate environment has been verified by employing density functional theory (DFT) calculations 
for a single and a couple of adjacent Cu adatoms on a Cu \hkl{100} surface \cite{djurabekova_local_2013}. In the context of the theoretical model, the adatom is an atom that lies above the top layer of a crystal and has fewer atomic bonds to the surface atoms than the surface atoms themselves within a complete layer. As such, adatoms are exposed to vacuum and, hence, to an applied electrostatic field stronger than regular surface atoms.

\subsection{Work function  variation due to surface irregularities}
\label{sec:WorkFunction}

Geometric field enhancements resulting from sharp surface features with a high aspect ratio and small radius curvature at the top are named the most frequently among the reasons for strongly enhanced field emission currents measured in various experiments from seemingly flat surfaces \cite{feng_model_2005, kildemo_breakdown_2004, pogorelov_field_2009, navitski_field_2013, nagaoka_field_2001}. The field enhancement is used as a fitting parameter to match the applied voltage and the measured current to some version of the Fowler-Nordheim equation (here we give the simplistic original one) \cite{fowler_electron_1997}: 
\begin{equation}
\label{eq_FN}
 j = \frac{AF_L^2}{\phi}e^{-\frac{B\phi^{3/2}}{F_L}}   
\end{equation}
where $j$ is the current density and $\phi$ is the work function. $F_L$ is the local electric field that is estimated as $F_L = \beta\times F_0$, where $F_0$ is the applied electric field, and $\beta$ is the enhancement factor obtained from the fit to the experimental data in order of 20-80 \cite{kildemo_breakdown_2004}. It is apparent from this formula that it is often convenient to plot the experimental data as $\log j/F^2$ as a function of $1/F$, as illustrated, for example, in Fig.~\ref{fig:FE different materials}. 

Since high sharp features have not yet been reported on regular surfaces, there is also a suggestion that a strongly increased field emission current density may be caused by local variations in the work function ($\phi_{\text{Cu}} \approx 4.6$ eV \cite{anderson_work_1949}) due to surface roughness and various surface defects \cite{takeuchi_local_2001}. This is a plausible assumption since, according to the Fowler-Nordheim theory, the reduction of the work function causes the same effect as the enhancement of the local electric field, with the two being practically indistinguishable in a current-voltage measurement. Moreover, $\phi$  is well known to be sensitive to the crystal orientation of the surface \cite{smoluchowski_anisotropy_1941}, as well as to defects. The difference between ridged and closed surfaces in FCC structures may reach a few tenths of eV \cite{fall_work-function_2000, birhanu_copper_2018}. Even the vicinal surface can change the work function by a couple of percent \cite{godowski_work_2013}.
Moreover, photoemission spectroscopy reported a local drop of nearly 0.1 eV of the work function at the step edges on atomically smooth surfaces \hkl{111} relative to the measured overall value \cite{takeuchi_local_2001}. However, strong plastic deformations were suggested to also alter the value of the work function \cite{li_effects_2002, li__effects_2004}. 

In computational modeling, this effect can be accurately assessed using density functional theory calculations, where the electron density distribution around the surface atoms is calculated explicitly. Because of the limited length scale in these calculations, the analysis of only small defective structures on a metal surface is feasible. However, some information about the variation of the work function around surface defects can be deduced from these calculations. For instance, despite insignificant dimensions, a single adatom on a flat surface can already contribute to the surface roughness, causing a local disturbance to the electric field distribution around it. This disturbance causes a local decrease in the dipole layer, which in turn is expected to modify the work function. Earlier DFT calculations showed that a drop in work functions can be expected due to the presence of a single adatom on the copper surface \hkl{100}, and it was found to be as large as 5.6\% in the very close vicinity of the defect that smeared at 1. 5\% while averaging the potential energy curve on a larger surface area \cite{djurabekova_local_2013}. Later, similar calculations, but for the Cu surface \hkl{111} \cite{toijala_ab_2019}, confirmed a similar reduction of the work function in the presence of an adatom, a monoatomic step. A further reduction of up to 10\% was found assuming the growth of a pyramid on the surface. However, the calculations suggested that no further decrease in the work function can be expected since the dominant reason for this decrease is the maximal redistribution of electron densities around the top surface atom in a position which is the most exposed to the vacuum. Any other position of the atoms on the surface will result in less exposure (i.e., more neighboring atoms on the surface) and hence in lower effect on the work function.

The presence of geometric asperities on the surface was determined to induce a minor decrease in work function, contributing multiplicative factors of 1.14 and 1.24\cite{toijala_ab_2019} to the apparent field enhancement in the case of a single adatom and a pyramid. This illustrates that geometric protrusion can cause simultaneous reduction of the work function, affecting field enhancement factors measured in experiments \cite{kildemo_breakdown_2004}. A similar conclusion was derived in the experimental study \cite{chen_surface-emission_2012}, where a synergistic effect between reduction in the work function and reasonable geometric field enhancement could explain the high values of measured field emission currents. However, it is clear that the possible reduction of the work function alone cannot explain the high values of field enhancement factors measured experimentally, still necessitating the assumption of a purely geometric field enhancement to explain the experiment.

\subsection{Role of electron density distribution on surface evolution under external electric fields}\label{subsec:Polarization}

\begin{figure}
    \centering
    \includegraphics[width=0.8\linewidth]{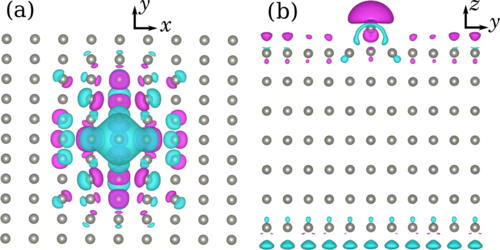}
    \caption{Charge redistribution around the surface atoms (a) in the absence and (b) in the presence of the positive external electric field with a strength of 1 GV/m (anode). The cyan and magenta colors (online) show the increased and decreased electron density that differ from the density in the bulk by 1\% and 0.1\% for (a) and (b) cases respectively. The calculations are done for W surface with \hkl{110} orientation. Reprint from Ref.\cite{kyritsakis_atomistic_2019}.}
    \label{fig:polarization}
\end{figure}

Surface asperities, even single adatoms, i.e. the atoms that randomly appeared on the top of the surface, modify the electron density landscape on the surface. The distribution of non-uniform electron density due to the presence of an adatom on the tungsten (W) surface can be seen in Fig.~\ref{fig:polarization}a. Here, the results of quantum-mechanical density functional theory (DFT) calculations \cite{kyritsakis_atomistic_2019} are shown for a W atomic structure with a W adatom in the middle of it, viewed from the top. The position of the adatom with respect to the surface atoms is seen in Fig.~\ref{fig:polarization}b, where the side view of the same geometry is shown. Although the nature of the chemical bonds is the same for all atoms in this system, the smaller number of neighbors of the adatom reduces its binding to the surface. The position of the adatom also promotes accumulation of the electron density right above it (cyan cloud in the middle). This electron density redistribution occurs already in the absence of an external electric field. The excess of electron density around the adatom creates a localized alternation of the electron density around the surface atoms beneath the adatom, which can be seen in Fig.~\ref{fig:polarization}a as magenta and cyan colors that, respectively, show depletion and excess of electron density . As discussed in Section \ref{sec:WorkFunction}, this effect is responsible for a minor local reduction in the work function \cite{neugebauer_adsorbate-substrate_1992,toijala_ab_2019}.  

Figure~\ref{fig:polarization}b shows that under a positive external electric field (anode), the surface is positively charged. The color around the surface atoms is now magenta, while the bottom of the atomic structure used in the DFT calculations is colored in cyan, indicating the negative charge of this surface. (Note that in static calculations no electric current is included.) Again we see that the adatom is responsible for the non-uniform charge distribution. Although in this case, all surface atoms become charged, the charge associated with the adatom is much higher. This non-uniformity is expected to play an important role in the evolution of the surface morphology for metals exposed to high electric fields, since the interaction of the atoms with the electric field will be different in different surface locations.

First, since the interaction of the charged surface with the field generates tensile stresses also under the surface \cite{djurabekova_atomistic_2011}, the nonuniform charge around the surface asperities will give rise to the stress concentration underneath these regions where the charge is higher. These forces follow the dynamic changes in the electric field distribution that act more strongly on the atoms with higher partial charge. This process alone can trigger dramatic changes on the surface, in particular, when an under-surface defect is present, such as a void or cavity \cite{pohjonen_dislocation_2013}. Moreover, these forces contribute to field-assisted evaporation of atoms from sharp surface features at elevated temperatures \cite{djurabekova_atomistic_2011}. 

\begin{figure}
    \centering
    \includegraphics[width=1\linewidth]{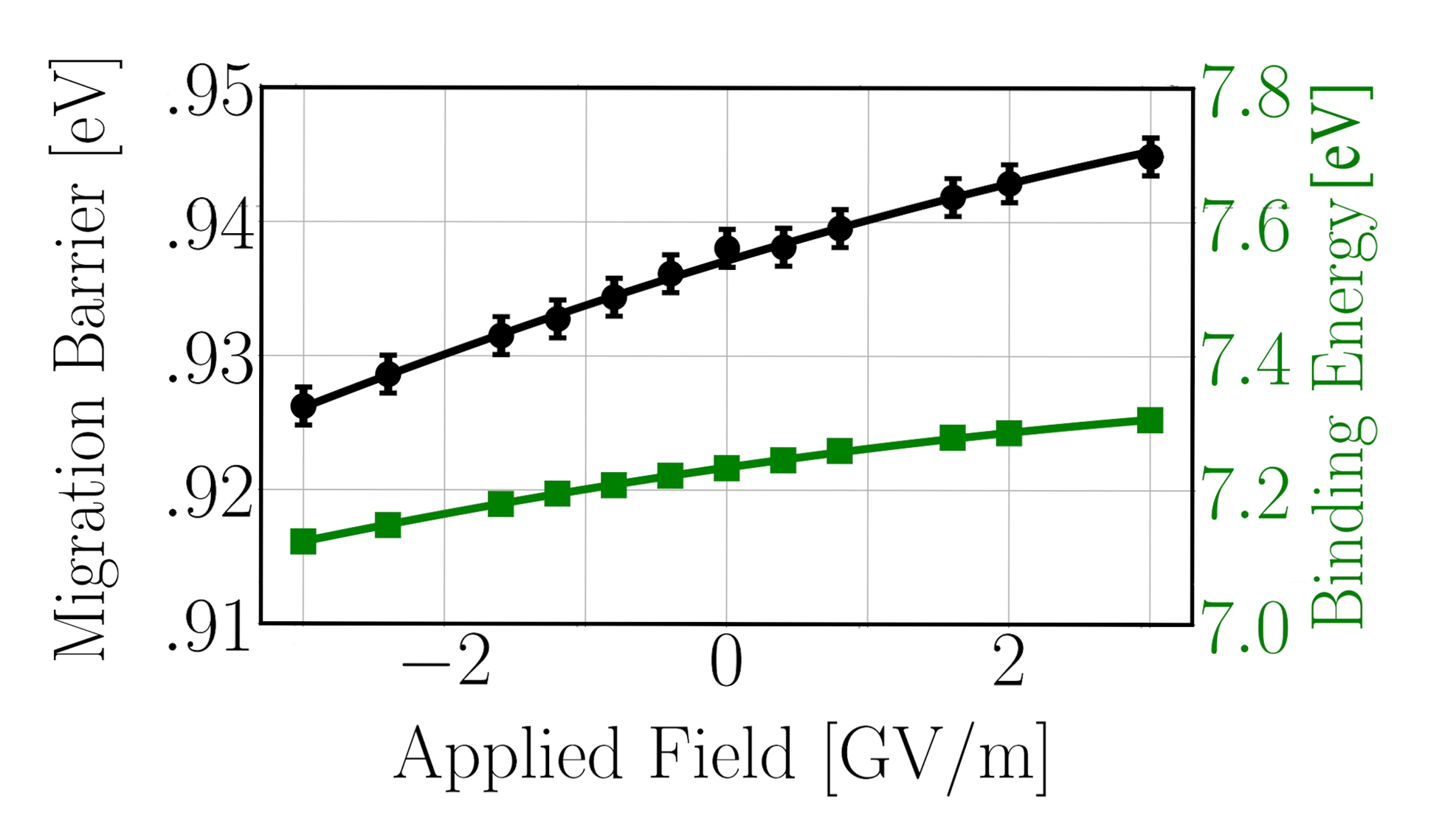}
    \caption{Migration energy barriers (black circles, the left-side $y$ axis) as well as the binding energies (green squares, the right-side $y$ axis) of a self-adatom on the \hkl{110} W surface for the constant applied electric field condition. The markers show results from the dft calculations and the lines are theoretical predictions based on the model of surface atoms polarizability. Reprint from Ref. \cite{kyritsakis_atomistic_2019}.}
    \label{fig:W_Em_EF}
\end{figure}

On the other hand, the non-uniform charge distribution on surface atoms may lead to the evolution of surface morphology, which proceeds via surface diffusion processes. These thermally activated hoppings of surface atoms and adatoms from one lattice site to another \cite{andersen_practical_2019} will be affected by an instantaneous change of the charge accumulated on the atoms that perform hoppings due to the interaction of the induced partial charge on the surface atoms with the applied electric field. These energy barriers are known as migration barriers that define the hopping rates. To incorporate this effect within computational models, one can also consider the interaction of a charged metal surface with the applied electric field as follows.

A piece of metal is known to develop a macroscopic permanent dipole when exposed to an external field. The permanent dipole moment is defined by the accumulated charge on the metal surfaces that compensates for the electric field within the metal itself. Microscopically, this process can be seen as charge accumulation on surface atoms, which is, in turn, not uniform and depends on the configuration of surface atoms. Depending on the direction of the applied electric field, the atoms facing the vacuum will have excess or depletion of electron density above them  \cite{neugebauer_adsorbate-substrate_1992, djurabekova_local_2013, muller_migration_2006}, and the stronger the atoms stand out from the surface, the higher the partial charge around them.

Field enhancements around surface asperities imply appearance of field gradients which can be used for modelling the migration of partially charged atoms on the surface. Because the increase of the electric field increases the partial charge on the surface atom, it can be expected that such a partial charge associated with the atom will promote migration of this atom towards stronger fields. The first idea of associating the behavior of partially charged surface atoms with the surface dipoles was proposed in Ref. \cite{tsong_direct_1975}, where the adatoms were suggested to migrate preferentially toward the surface electric field gradients, similarly to the way isolated dipoles move in the presence of field gradients. This analogy was proposed to explain experiments in which the biased diffusion toward sharp features or edges of sharp tips was observed 
\cite{utsugi_field_1962, kellogg_electric_1993}. Better physical models to explain biased diffusion in the presence of electric field gradients were based on DFT calculations \cite{feibelman_surface-diffusion_2001}. However, these calculations are too slow to assess the dynamic effect of the biased diffusion under applied electric fields. To enable more systematic observation with statistical analysis of the effect, it is necessary to quantify the effect to enable its incorporation into large-scale models or analytical theories. 

Recently, a consistent theory of the polarization of a metal surface and its localized effect on surface atoms and adatoms was proposed in Ref. \cite{kyritsakis_atomistic_2019}. In this work, the polarization of the macroscopic surface is considered as a whole, while significant charge redistribution around a surface feature (an adatom in Fig.~\ref{fig:polarization}a) is taken into account as an additional dipole moment $\mu$ associated with it. This additional dipole moment exists even in the absence of an external electric field. It has also been seen that the additional dipole moment depends on the local atomic environment and the position of the atom with respect to the lattice sites. 

The dipole moment itself and the rate with which it can change depend on the strength of the external electric field. The interaction of this dipole with the external electric field $F$ introduces an additional energy component into the system, which is an electrostatic energy $\mu F$. Figure~\ref{fig:W_Em_EF} demonstrates this effect and how the energy $\mu F$ modifies the binding energies (green) and migration barriers (black) as a function of the field strength. 
These DFT calculations clearly show the reduction in the binding energy to the surface (right-hand $y$-axis) and the corresponding decrease in the migration barrier (left-hand $y$-axis) for negative fields. However, some increase in these values for positive electric fields is associated with the compensation of the additional permanent dipole moment generated at the adatom in the absence of the field. At low positive fields, the charge is pressed inwards, reducing the dipole moment and, hence, decreasing the effect of the electric field on the height of the barrier.


Moreover, this consideration can take into account the effect of the field gradient, which can be introduced as an additional term in the migration barrier calculations as follows \cite{kyritsakis_atomistic_2019}:
\begin{equation}\label{eq:barrier}
    E_m = E_0 - (\Delta \mu) F - \frac{1}{2} (\Delta \alpha) F^2 - (\mu + \alpha F) \Delta F
\end{equation}
In Eq.~\eqref{eq:barrier} $E_0$ is the migration barrier without a field, $\Delta \mu$ and $\Delta \alpha$ refer to the difference in the effective dipole moment and the effective polarizability between the saddle point (the highest energy point between the lattice sites) of a migration process and the lattice site itself, and $F$ is the applied field. Having these considerations in mind, one can attempt to simulate surface diffusion processes, explicitly including the effects of electrostatic fields in the computational models.



\subsection{Modeling atomistic surface diffusion under high electric fields}

As discussed in Section \ref{subsec:Polarization}, the partial charge induced by an external electric field on surface atoms affects the thermodynamic equilibrium due to additional forces of an electrostatic nature. In turn, this affects the thermally activated dynamics of atoms, inducing biases towards stronger fields in the random migration of surface atoms, which is known as surface diffusion.

A kinetic Monte Carlo (kMC) method that was parameterized to specifically describe the hopping events for surface atoms was successfully used to simulate surface diffusion \cite{jansson_long-term_2016, jansson_tungsten_2020}. The parameterization of such a model is based on calculations of migration barriers that are functions of the local atomic environment. Enabling the effects of an electric field in these simulations is not straightforward since the continuous electron density distributions must fit within the discrete realm of the kMC method. Applying the methodology described in Section \ref{subsec:Polarization}, one can use a single formula to modify the set of existing barriers introducing the effect of an electric field. 



\begin{figure}
    \centering
    \includegraphics[width=0.9\linewidth]{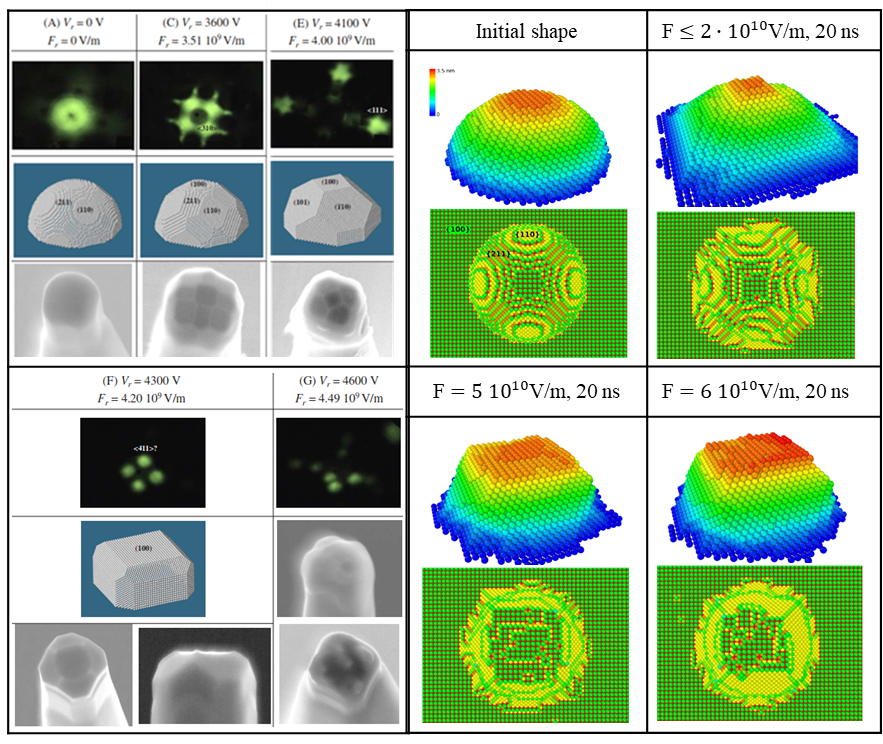}
    \caption{Comparison of the kMC simulation results with the experiment of the top shapes of the W tip under an applied electric field. The field emission experiments (left panel) are performed for the $\mu$m size tip at 2300 K for a few seconds, while the simulations (right panel) were performed for a few nm size tip at 3000 K for tens of ns. Despite scale differences, surface modifications are similar in experiment and simulation. The top and the bottom panels show the evolution of the W tip at low and strong electric fields, respectively. Experimental images comprise the three rows (top to bottom): field emission maps with increase of the voltage from left to right; tip surface reconstruction showing facets to match the FE maps above; the scanning electron microscopy (SEM) images of the experimental tip. Adopted from Ref.\cite{fujita_mechanism_2007}. The kMC results present the 3d image of the tip above and the corresponding top view below for different applied fields from 0 GV/m to 60 GV/m. Adopted from Ref. \cite{jansson_growth_2020}. Green-colored atoms belong in the \hkl{100} oriented facets, and yellow and orange color is used for atoms in the \hkl{110} and in \hkl{211} oriented facets, respectively. The color code in the 3d images shows the $z$ coordinate. See the main text for more discussion.}
    \label{fig:Wtipchange}
\end{figure}

Strictly speaking, the polarization characteristics around the surface atoms positioned in different atomic environments should differ, since the electron densities are also distributed differently around the atoms differently bound at the surface. Implementation of the environment-dependent electric field effects is a challenging task as the functional form becomes very complex and requires multiple calculations of these parameters for various atomic configurations at the surface. However, the use of fixed values for both characteristics in the KMC model of surface diffusion on the W tip under an electric field \cite{jansson_growth_2020} already gave encouraging results which are shown in Fig.~\ref{fig:Wtipchange}. In the figure, the left panels show the experimentally observed surface modification of the W tip during field emission (FE) experiments at elevated temperatures. Three rows in the experimental images (from the top down) show the FE mapping, the hard-sphere model tip reconstruction to match the corresponding FE maps above, and the corresponding scanning electron microscopy (SEM) images of the actual tip. Although there is a clear difference in length and time scales between the experiment and the simulation ($\mu$m vs nm and s vs ns, respectively) the trend of surface modification is very similar. The kMC snapshots in the right panel of Fig.~\ref{fig:Wtipchange} show the evolution of a W tip. The top row shows the initial tip and the tip held at low electric fields, which do not induce sufficient bias in surface diffusion to influence the surface dynamics. The result is similar to the experiment on the left, where heating the tip leads to its facetting with a large fraction of the lowest energy \hkl{110} surfaces. The increase in field promotes the growth of the \hkl{100} surface at the top, since this allows maximal growth of the total length of sharp edges, where an electric field gradient appears giving rise to the biased diffusion. When the field becomes too large, the order gets confused, as any new defect on the surface will create the spot of the field gradient, reducing the order of the surface; see the last bottom image in the experimental and simulation panels.

\begin{figure}
    \centering
    \includegraphics[width=0.85\linewidth]{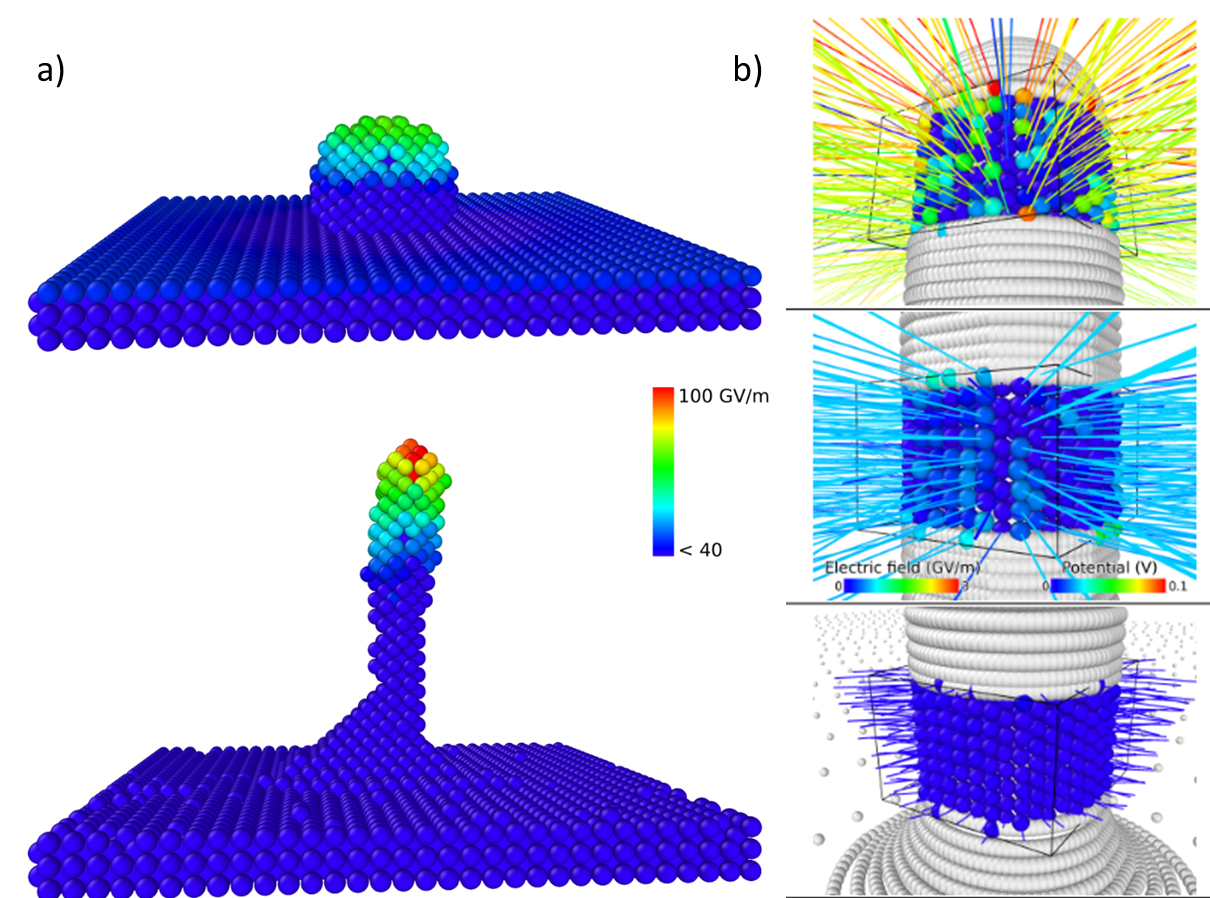}
    \caption{Effect of biased diffusion on sharpening of surface tips. a) shows the kMC result of sharpening a W tip from its initial (top) configuration at the applied electric field of 40 GV/m. b) is the CVHD simulation of a biased surface diffusion on a Cu tip. The simulations are done separately for different parts: at the bottom, at the middle part, and near the top of the tip. The arrows indicate the electric field distribution which is gradually increasing from the bottom to the top inducing the electric field gradient. }
    \label{fig:biased_diffusion}
\end{figure}

 Generally, sharp surface features with high aspect ratio are often assumed to explain localized FE currents from seemingly flat metal surfaces \cite{kildemo_breakdown_2004}. However, such tips on a nanoscale were shown not to survive on a metal surface without an external electric field \cite{jansson_stability_2015} due to the principle of minimization of surface energy. When an electric field is applied, the dynamics of surface atoms changes due to the redistribution of charge densities around them, as discussed in Section \ref{subsec:Polarization}. The stronger the atom stands out from the surface, the stronger the field enhancement and the higher the partial charge accumulates on it. The simulations where Eq.~\eqref{eq:barrier}  was used to estimate the migration barriers\cite{jansson_growth_2020} are shown in Fig.~\ref{fig:biased_diffusion}a. The top snapshot in Fig.~\ref{fig:biased_diffusion}a demonstrates an initial small surface feature, which develops into a sharp needle-shaped feature (bottom panel in Fig.~\ref{fig:biased_diffusion}b) due to the biased surface diffusion under an applied electric field.

The bias in surface diffusion caused by the presence of electric field gradients was also confirmed in the atomistic dynamic simulation augmented by collective variable hyperdynamics, which allowed a significant increase in the simulated time \cite{kimari_biased_2022}. The snapshots in Fig.~\ref{fig:biased_diffusion}b show the calculations performed separately for different parts along the height of a surface tip. The three images stacked in Fig.~\ref{fig:biased_diffusion}b from bottom to top illustrate the atomic dynamics as a function of the surface electric field, which is shown by arrows colored according to the strength of the field. The color of the atoms indicates the magnitude of the partial charge: dark blue (close-to-zero charge) to bright red (0.5e). As one can see, at the bottom of the tip the field is low and no diffusion of surface atoms is observed. This changes significantly at the top of the tip where the field is the highest, and the atoms with the high partial charge move toward the stronger field more easily. Quantitatively, the biased effect in this work was assessed as an increase in drift velocity towards the top of the tip, in particular on the closed-packed \hkl{111} surface. In these simulations, only the partial charge induced by the field near the surface atoms was estimated (see Section \ref{sec:helmod}) and included in simulations of surface dynamics instead of considering dipole moments, which was discussed in Section \ref{subsec:Polarization}. Following the drift velocity, the authors confirmed the effect of biased diffusion on metal surfaces under external high electric fields, particularly on close-packed surfaces where the energy barriers compare closer with the changes introduced by the interaction of surface charges with the external electric field.

\section{\label{s:emit} Electron emission and atom evaporation} 

\subsection{Thermal effects in electron emitting tips}

The observed high values of field enhancement factors \cite{kildemo_breakdown_2004} imply the existence of field emitting tips, which either exist prior to high electric field application or grow on the surface as a result of field-induced surface dynamics. It is natural to consider the intensively emitting tips as a potential source of neutrals, which, in combination with the field-emitted electrons, initiate the plasma that forms the conducting channels and eventually leads to the discharge. The heat generated in the tip by the strong field emission currents affects both the electron emission and the atomic dynamics in sharp tips.  Hence, it is necessary to calculate the temperature distribution in a tip of a given geometry. This problem is already quite complex and multi-physics in nature since it requires calculating the electric field distribution around the tip, the field emitted current on its surface, the internal distribution of the current and heat, and the solution of the heat equation for the temperature distribution. 
The problem can be approached numerically in various levels of approximation. The numerical solution of the Laplace equation for the field distribution can be combined with the field emission and Ohm's laws to estimate the generated heat in a protrusion of given geometry \cite{jensen_electron_2008,keser_heating_2013, mofakhami_unveiling_2021}. This approach provides an assessment of the Joule heating effect. However, any possible shape modification of the protrusion remains inaccessible. 

An additional complication in this problem is that the tip shape might change dynamically under field-induced stress, especially at elevated temperatures that might even melt the tip locally. Such tip changes will consequently change the electric field distribution, requiring a concurrently coupled model to follow. This behavior can be followed using atomistic simulations by means of the molecular dynamics method, concurrently coupled with the electrodynamic calculations of electric field effects \cite{djurabekova_atomistic_2011, parviainen_molecular_2011, veske_electrodynamics_2016}. In these simulations, shape modification due to electric field-induced tensile stress \cite{veske_electrodynamics_2016} or tip melting due to current heating \cite{parviainen_molecular_2011} is used as feedback to recalculate the field distribution in a self-consistent fashion. These early simulations already revealed field-assisted evaporation of atoms from the sharpest features at the top of the tip, which may occur even before the phase transition to a molten state. 

A dominating contribution of Joule heating can be assumed only for field emitters with an apex curvature radius on the order of a $\mu$m and higher, since tips of this size will heat up more efficiently through the processes that take place in the bulk. For nanometer-sharp surface features with a very large surface-to-volume ratio, the Nottingham effect \cite{nottingham_thermionic_1936, nottingham_remarks_1941} becomes a leading process in heat exchange between the tip and the electrons leaving it \cite{mofakhami_unveiling_2021, behboodi_nottingham_2023}. Furthermore, the increase in temperature within the electron-emitting tip results in the transition from cold field emission to thermionic emission, as well as the intermediate thermal-field regime, which needs to be described accurately in a single model.

Finally, an important effect that needs to be considered is the space charge. Typically, the emission current densities involved in such phenomena fall in the range where the space charge built by the emitted electrons becomes dense enough to suppress the electric field that extracts the electrons. Moreover, this space charge depends very strongly on the emission at any given time in the evolution of the tip, affecting and being affected by its entire characteristics (shape, temperature distribution, distribution of emitted particles).

It is evident from the above that building a model that accurately captures the evolution of intensively emitting nanotips at elevated temperatures is extremely challenging and requires developing novel computational techniques. Such techniques evolved gradually during the past decade, starting with the development of the first hybrid electrodynamics-molecular dynamics model HELMOD \cite{djurabekova_atomistic_2011}. The first attempt to introduce electron emission and thermal effects in HELMOD was performed by \cite{parviainen_electronic_2011}. \cite{eimre_application_2015} introduced a more accurate emission model that covers all emission regimes based on the General Thermal Field (GTF) equations \cite{jensen_reformulated_2019}, including the Nottingham effect, but only for static tips. The first model that accurately included all emission and thermal effects on a dynamically evolving tip was proposed by \cite{kyritsakis_thermal_2018}. It revealed a thermal runaway mechanism, thanks to the introduction of a more versatile FEM-based technique for electrodynamics \cite{veske_dynamic_2018} that allows for a multiscale extension of the simulation domain, as well as a generalized numerical tool for electron emission calculations that accurately covers all emission regimes \cite{kyritsakis_general_2017}. This model was later enhanced with Particle-In-Cell capabilities by \cite{veske_dynamic_2020}, allowing the inclusion of space-charge effects.

\subsection{Thermal runaway and atom emission as a precursor of vacuum breakdown}

\begin{figure}
    \centering
    \includegraphics[width=0.95\linewidth]{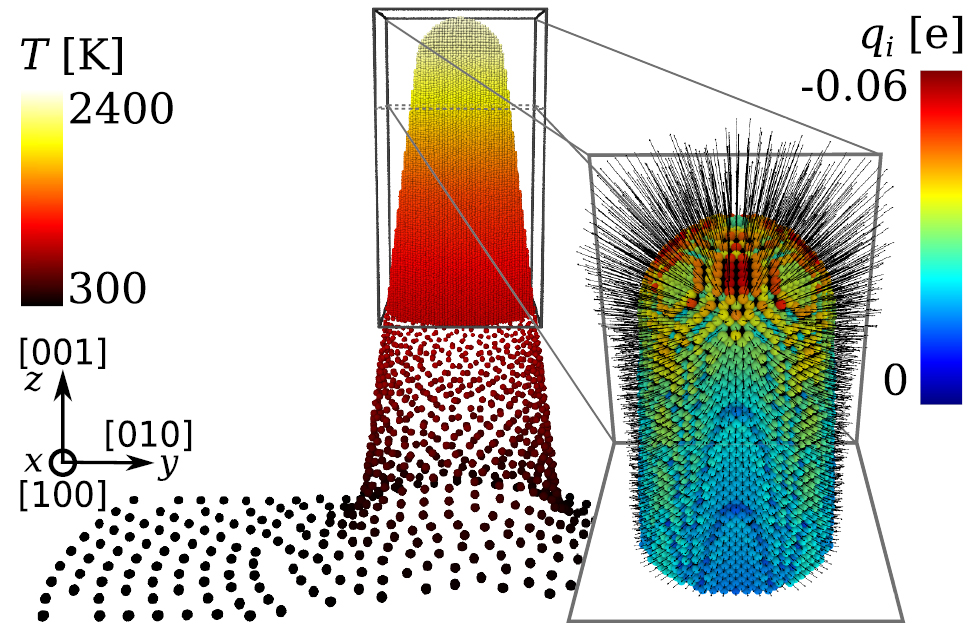}
    \caption{Schematics of the model adopted in Ref. \cite{veske_dynamic_2018, kyritsakis_thermal_2018} to describe the electron emitting tip in the pre-breakdown regime. Dots at the upper part of the tip are the atoms which were explicitly simulated using molecular dynamic (MD) method. The atoms are colored according to the temperature evolved in the tip as a result of Joule and Nottingham heating. Rarefied dots in the lower part of the tip are the finite element method nodes indicating the extended surface of the tip where calculations of the heat exchange is done in the continuum limit. Inset shows the partial charge of the surface atoms. Black arrows show the vectors of the electric field whose length represents the strength of the field. Reprint from \cite{kyritsakis_atomistic_2019}.}
    \label{fig:largetip}
\end{figure}

The full dynamics of the tip evolution due to the pre-breakdown strong electron emission currents can be described by multi-physics multiscale models \cite{veske_dynamic_2018, kyritsakis_thermal_2018, veske_dynamic_2020}. In this model, the atomic dynamics is calculated explicitly within the top part of the tip, while both the electric field, current density, and temperature distributions are calculated within a finite element mesh that extends to the entire height of the tip and the flat surface surrounding it, see Fig.~\ref{fig:largetip}. 

\begin{figure}
    \centering
    \includegraphics[width=\linewidth]{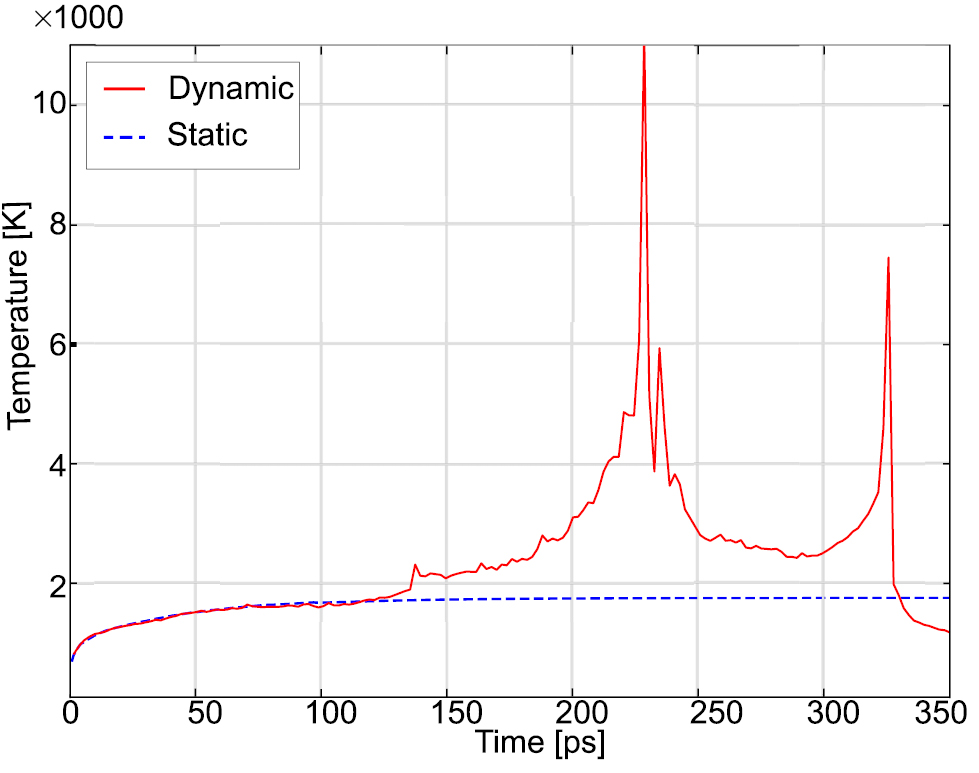}
    \caption{Temporal evolution of temperature at the top of a nanotip with the aspect ratio of $\simeq$ 30 under the applied electric field of $\sim$ 0.6 V/nm. Red line shows the temperature in the dynamically evolving tip while the blue dashed line indicate the temperature evolution in the assumption of a static tip. }
    \label{fig:runaway_temp}
\end{figure}

The developed concurrent multiscale model allows us to peek into the dynamics of the atoms under an externally applied electric field, taking into account both the field-induced stresses and the emission-induced thermal effects. These dynamics play a paramount role in determining the behavior of an intense field-emitting nanotip, as they contribute the major positive feedback in the thermal runaway mechanism \cite{kyritsakis_thermal_2018}. Figure~\ref{fig:runaway_temp} demonstrates this, comparing the evolution of the maximum temperature in a static (blue dashed line) to a dynamic (red solid line) simulated tip. Although the two cases exhibit the same temperature evolution in the initial stages, the dynamic rearrangement of atoms after the emission-induced melting results in a thermal runaway that causes the temperature spikes of Fig.~\ref{fig:runaway_temp}. Under tensile stress due to the interaction of the electric field with the charged surface, the tip apex sharpens after melting. The local field and emitted current density increase within the sharpened apex, increasing the generated heat and forcing the atoms to break the atomic bonds and evaporate while promoting further sharpening in a positive feedback loop. In the assumption of a static tip, where no shape modification occurs, the temperature does not rise above a given saturated value (blue dashed line in Fig.~\ref{fig:runaway_temp}). Underestimation of temperature dynamics within the current-emitting tip can compromise the analysis of processes that take place in the pre-breakdown stage of vacuum arc onset. 

\begin{figure}
    \centering
    \includegraphics[width=\linewidth]{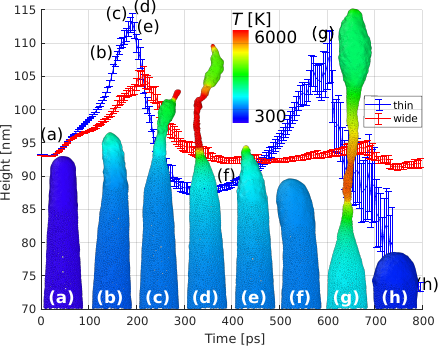}
    \caption{Time evolution of the averaged nanotip heights (left) and some characteristic excerpts from the simulation (right). Labels (a)–(h) on the left correspond to the frames on the right side. Error bars show the variation of the data between parallel runs in a form of standard error of the mean.}
    \label{fig:runaway}
\end{figure}

Figure~\ref{fig:runaway} shows the dynamic evolution of the shape of the tip. The plot shows the maximum height of the tip for two different initial tip geometries. The insets show the tip shape and temperature distribution (color coding) in specific frames (designated on the plot). The initial aspect ratio of both tip geometries was about 30, but two different shank openings at the bottom of the tip were simulated, with "thin" and "wide" referring to a characteristic shank radius of 17 nm and 54 nm respectively, compared to 3 nm at the apex of the tip. One can see that (c) and (d) correspond to the spike in height, while the shape of the tip is highly changed showing narrow necking with a strong temperature increase (scale in the color bar). This necking and the corresponding extremely highly local temperatures cause the detachment of entire metal droplets, resulting in the height rapidly decreasing, and the tip becoming blunt again. However, as it is still molten and hot, the thermal runaway mechanism restarts, leading to the next spike in height, in the case of the thin tip. In contrast, this does not occur in the wide geometry due to the more efficient heat conduction into the bulk with the broader opening at the bottom. Still, even in the wide tip case, some repeated dynamic still continues.

By counting the number of atoms detached from the top of a tip per unit time, one can estimate the evaporation rate of neutrals with respect to the emission rate of electrons given by the emission currents. These estimations were consistent with the values predicted in Ref. \cite{timko_field_2015} (see next section for details), where it was shown that the plasma discharge takes place if $\sim$ 15 neutral atoms are injected into the system for 1000 electrons. Hence these studies suggest that the thermal runaway process in a metal tip acts a precursor to a full plasma onset.

Finally, we should note that the results shown in Fig.~\ref{fig:runaway_temp} are taken from \cite{kyritsakis_thermal_2018}, while the ones of Fig.~\ref{fig:runaway} are from \cite{veske_dynamic_2018}. The latter was run with a more advanced version of FEMOCS, which included PIC-MCC simulations that accurately calculate the space charge. In contrast, the earlier ones of \cite{kyritsakis_thermal_2018} had an approximate correction for the space charge suppression of the emission. However, there is no significant difference between the results of the two methods regarding the main conclusions they lead to, as described above. 

\section{\label{s:plasma} Plasma ignition and evolution}
One of the processes in vacuum arc initiation that is paramount to understanding is the initiation of plasma. The plasma ignition dynamics are extremely hard to study experimentally, as the phenomenon's spatial and temporal scales are extremely small (nm-ns). However, advanced modeling has given significant insights into vacuum-arc plasma-onset physics, employing mainly particle-in-cell plasma simulations. 

The main challenge posed for these simulations is the complex interaction of the plasma with the metal surface, which is the main determining factor of plasma evolution. The following sections give an overview of the numerical simulation models developed for arc ignition and the main results they offered.

\subsection{\label{ss:arcpic} Arc ignition models based on flat surface morphology}

\begin{figure}
    \centering
    \includegraphics[width=0.85\linewidth]{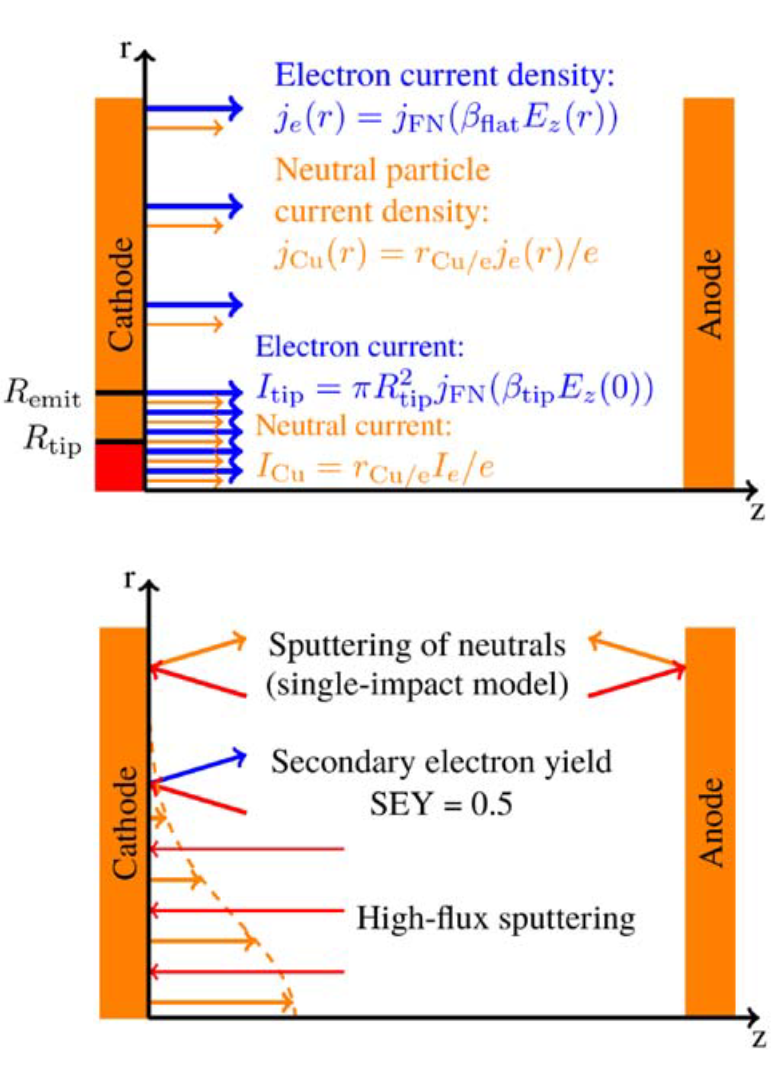}
    \caption{Schematic representation of the processes considered in 2D ArcPIC. The figure is generated from the images in \cite{timko_field_2015}.}
    \label{fig:ArcPIC_schematic}
\end{figure}

\SC{Description of the arc ignition processes, from electron emission to plasma initiation, has been discussed semi-quantitatively by several authors in the past, see for example~\cite{chatterton_theoretical_1966, williams_field-emitted_1972,mesyats_ecton_1995}. The advent of numerical simulations allowed for the research of a quantitative relationship between the field-emitted current and the amount of neutral atoms emitted required to establish a runaway plasma. However, these early attempts were based on a static effusion model of the injected neutrals~\cite{knobloch_explosive_1998} or simply a static background of neutrals~\cite{mahalingam_modeling_2008}.} The first attempt to simulate the vacuum arc plasma initiation process with a self-consistent relationship between electron emission and neutral injection was made by Timko et al. \cite{timko_one-dimensional_2011, timko_modelling_2011} with the one-dimensional (1D) Particle-In-Cell with Monte Carlo Collisions (PIC-MCC) model ``ArcPIC''. This model treated the cathode surface as a source of neutrals and electrons. The injection flux of electrons is defined by the cathode electric field via the Fowler-Nordheim equation to model field emission, assuming a certain field enhancement factor $\beta$, which is removed after the evaporation of a certain amount of neutrals. The injection flux of neutrals is assumed to be proportional to the flux of electrons with a given flux ratio $r_{Cu/e}$. This model was later expanded to a 2D axisymmetric geometry \cite{timko_field_2015}, assuming that most of the cathode surface $z=0$ has a small distributed field enhancement $\beta_{flat}$ and in the center of the system $r=0$ there is a microscopic field enhancement feature with a field enhancement factor $\beta_{tip}$. A schematic representation of this model is shown in Fig.~\ref{fig:ArcPIC_schematic}.

\begin{figure}
  \centering
  \includegraphics[width=1.0\linewidth]{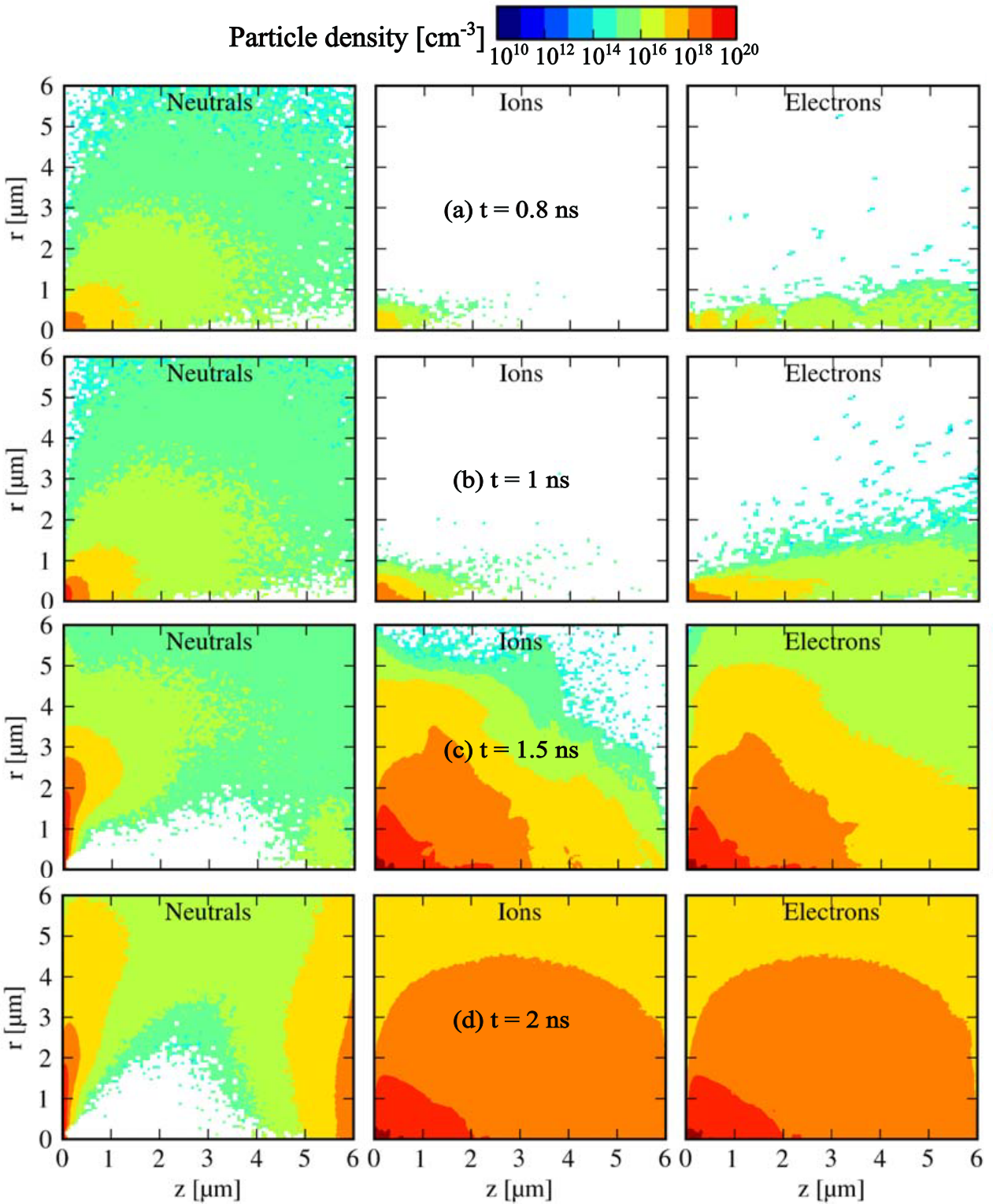}
  \caption{Evolution of particle densities in the ArcPIC simulation in four different time frames. Based on \cite{timko_field_2015}.}
   \label{fig:arcpic_results}
\end{figure}

Figure~\ref{fig:arcpic_results} shows the evolution of particle densities as simulated by the above ArcPIC model. We see that in the beginning of the simulation a strong electron beam appears near the center of the simulation ($r=0$), accompanied by the evaporation of a strong neutral cloud. At about 1 ns, a significant number of ions appear near the tip, as a result of electron impact ionization. This ion cloud gradually condenses and expands, eventually filling up the entire simulation box, developing into a full dense plasma. This plasma is accompanied by a potential sheath forming a strong electric field near the cathode. This field is necessary to maintain both a field-emitted electron flow into the plasma and the bombardment of the cathode with sufficiently energetic ions that sputter neutrals and maintain a neutral flow.

 Despite its simplifications, ArcPIC gave deep insights into the plasma initiation process. It was shown that a full plasma could develop under the above assumptions, and the threshold values of the key parameters were determined to $r_{Cu/e}=0.015$ and  $\beta_{tip}=30$, at a macroscopic applied field of 290 MV/m.

The injection of neutrals at this rate $r_{Cu/e}$ was shown to be a crucial assumption for the plasma onset. The microscopic mechanism behind such a neutral injection was yet unknown at the time. However, there were strong indications that field emitter self-heating and consequent evaporation could possibly explain it, instigating the research on the dynamics of intensively field-emitting nanotips presented in section \ref{s:emit}.

\subsection{ \label{ss:femocsPlasma} Plasma ignition from intensively emitting protrusions}

The quasi-planar models presented in section \ref{ss:arcpic} (ArcPIC), in combination with the thermal runaway mechanisms revealed by the concurrent multi-physics approach presented in section \ref{s:emit} offered significant qualitative insights into the plasma initiation mechanisms. Yet, they do not yet offer a precise quantitative description of the plasma initiation process that can make predictions comparable to experimental data. This is mainly because the small-scale FEMOCS simulations of section \ref{s:emit} are disconnected from the large-scale ArcPIC model. There is an indirect connection in that FEMOCS results justify the neutral injection assumptions of ArcPIC, but this does not capture the interplay of the plasma and tip runaway processes during tip thermal runaway.
The concurrent coupling of PIC to the hybrid ED-MD model has been proposed and developed to capture this process. 

The first step in this direction was performed by \cite{veske_dynamic_2020}, which incorporated PIC-MCC routines in FEMOCS. A significant advancement of this compared to ArcPIC is that the FEMOCS PIC implementation is based on finite elements rather than ArcPIC's finite differences. This is crucial as it allows unstructured meshes to capture vastly different spatial scales in the same simulations. Combined with a variable weight superparticle approach, this can yield a model that fully couples the surface and plasma dynamics. 

\begin{figure}
    \centering
    \includegraphics[width=0.99\columnwidth]{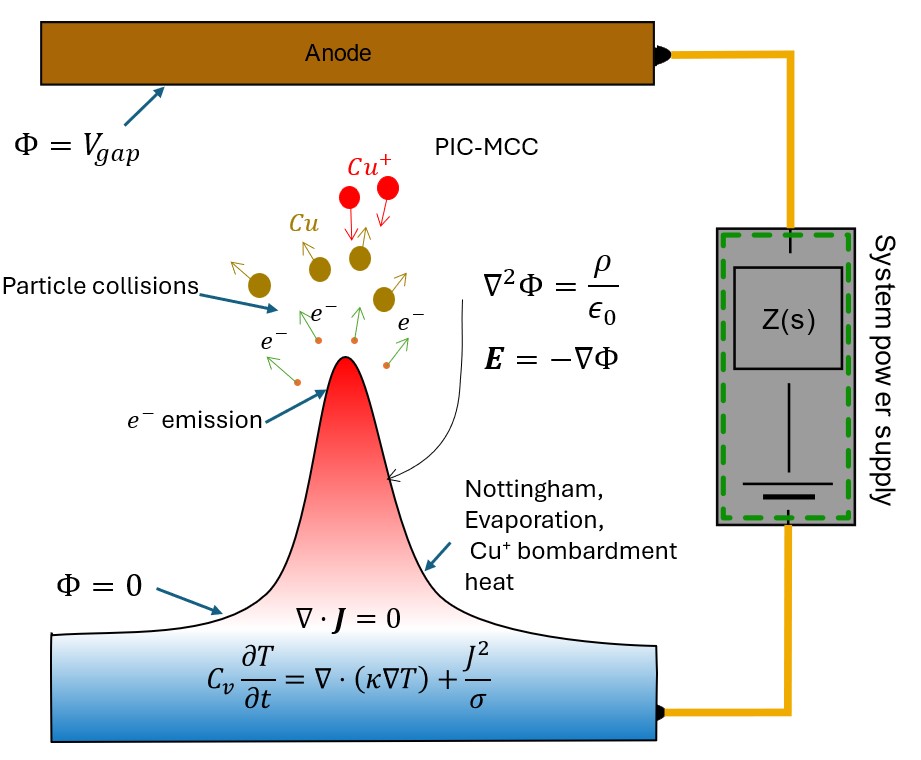}
    \caption{Schematic representation of the processes considered in 2D static FEMOCS.}
    \label{fig:femocs2D_schematic}
\end{figure}

However, in this first step, the PIC capabilities were used only to calculate the electron cloud space charge, and no atomic (neutral or ion) species were introduced. The next step that tackled this issue was performed recently by Koitermaa et al. \cite{koitermaa_simulating_2024}, who added and further advanced the plasma simulation techniques of ArcPIC. Figure~\ref{fig:femocs2D_schematic} gives an overview of the physical processes considered in this static version of FEMOCS.

This model investigated the physical processes beyond the dynamic evolution of the tip that are most significant for the development of plasma, focusing on the plasma ignition itself, i.e., the mechanisms that produce and maintain the supply of vapor and the ones that ionize it. To this end, the full tip geometry is simulated, yet in a static manner. The evaporation rates on the surface of the tip are calculated based on the local temperatures and the corresponding PIC superparticles being injected accordingly. For the ionization, in addition to the standard impact ionization processes considered in ArcPIC, FEMOCS includes direct field ionization via tunneling, as well as double and triple-ionized species. Finally, unlike ArcPIC, FEMOCS considers a normal sputtering yield on the surface without introducing the empirical enhanced sputtering due to ion impact heating. This ion bombardment heating is formally taken into account in FEMOCS in the heating problem, as the energy of the impacting ions is introduced as a surface heat component in the heating problem. Finally, a similar circuit model to ArcPIC was considered, yet the implementation is much more general to potentially capture any linear coupling between the macroscopic electromagnetic power sources and the localized microscopic breakdown site (see Sec. \ref{s:circ} for details).

Simulating a standard hemisphere on a cone tip geometry with the above model predicts that local heating at the tip can lead to plasma ignition, even for a static tip. The tip gets into a thermal runaway even when its shape does not change. The heating feedback loop is closed in the early stages by the increase of the electrical and thermal resistivity of the tip and later by the bombardment heating and the formation of an early stage of a plasma sheath that increases the local field and draws increased field emission current.

\begin{figure}
    \centering
    \includegraphics[width=0.99\columnwidth]{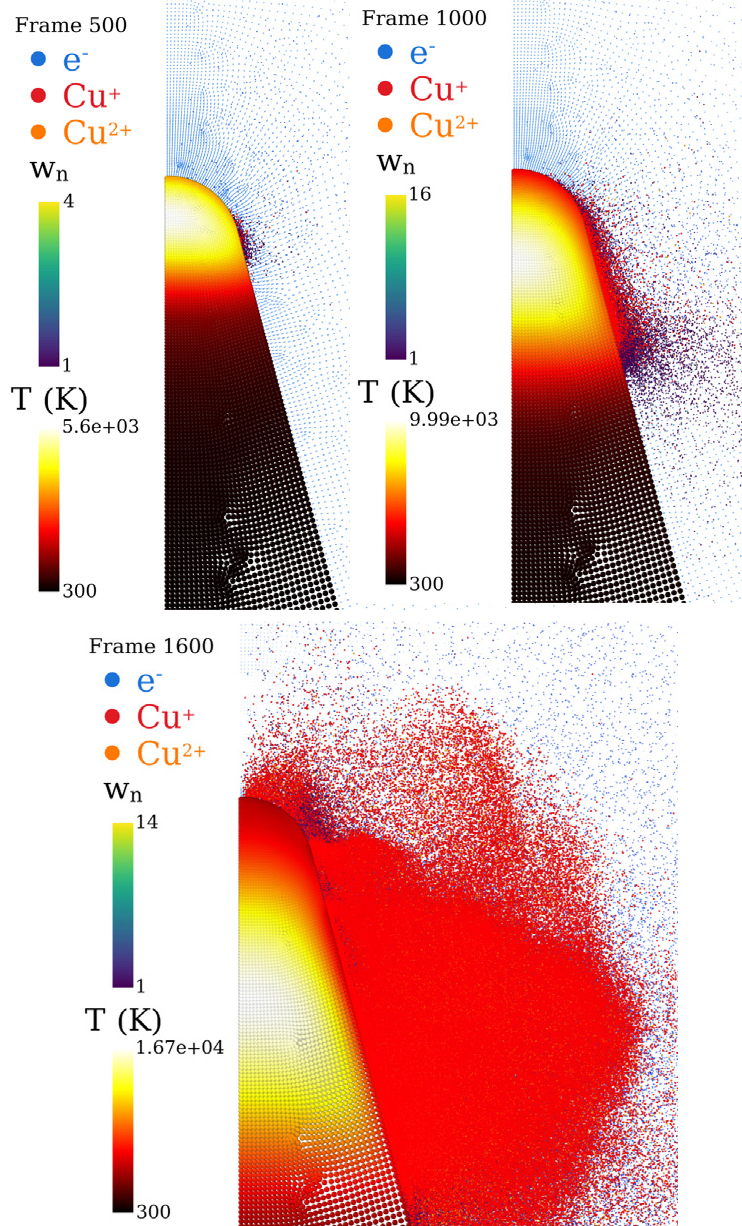}
    \caption{Visualization of the evolution of the nanotip simulated in \cite{koitermaa_simulating_2024} (only top segment shown). The evolution is shown in three different frames corresponding to 125~ps (frame 500), 250~ps (frame 1000) and 400~ps (frame 1600) ps simulation time. The figures combine the temperature distribution in the bulk (mesh points), the electric field in the vacuum (blue arrows), and the positions of the various superparticle species. Adopted from \cite{koitermaa_simulating_2024}.}
    \label{fig:plasma_ignition}
\end{figure}

\begin{figure}
    \centering
    \includegraphics[width=0.99\columnwidth]{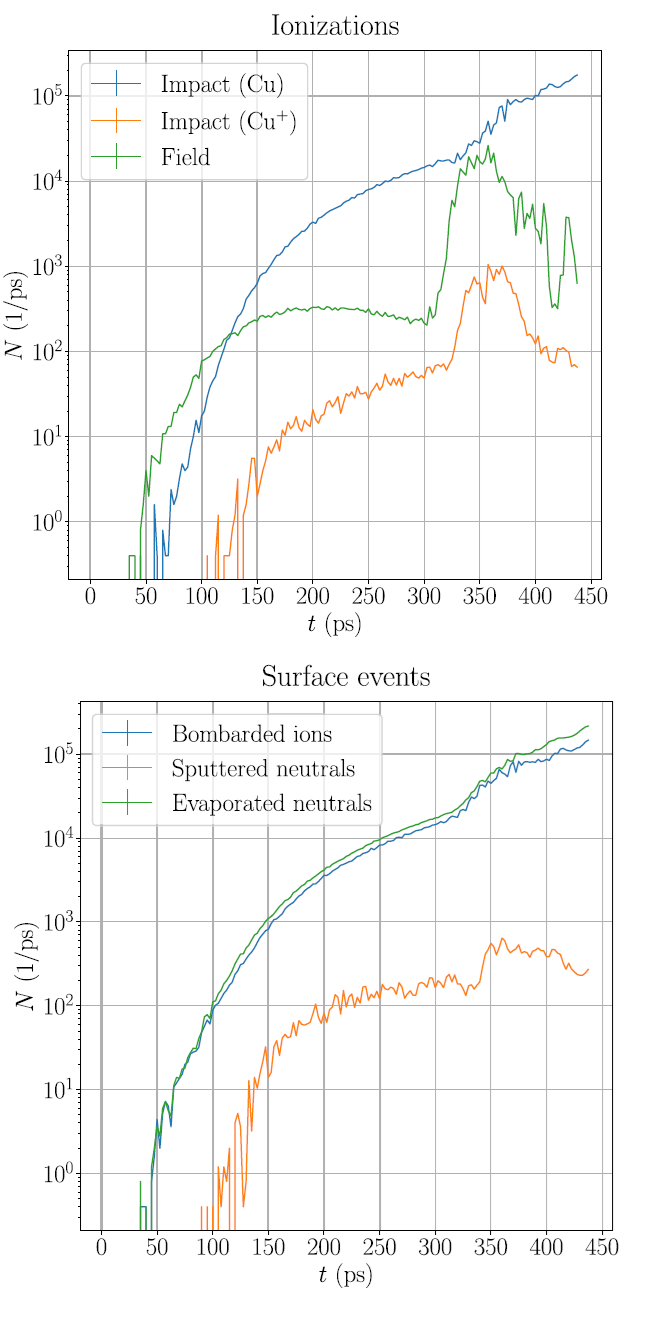}
    \caption{(a) Evolution of ionization rates for all three ionization processes included in the plasma initiation simulations; (b) evolution of the corresponding surface process rates.}
    \label{fig:event_rates}
\end{figure}

Figure~\ref{fig:plasma_ignition} shows the evolution of the temperature distribution on the tip along with the distribution of all superparticle species in the vacuum region. It is evident that a significant plasma starts on the side of the tip and then gradually expands around the surface area of the tip.
The most significant insights of these simulations are shown in Fig.~\ref{fig:event_rates}, where we plot the evolution of the various processes occurring on the tip. At the beginning of the plasma process, we see that the dominant ionization mechanism is field ionization, which was not considered in previous models. The fact that FEMOCS resolves the local field distribution allows us to properly take into account this effect, as field ionization is significant only in the vicinity of the tip apex where the field strength is in the range of ~10 GV/m. Yet, in later stages, when the plasma develops significantly and the field in the plasma core drops, the impact ionization becomes dominant, with the field ionization oscillating.

Finally, in the bottom graph of Fig.~\ref{fig:event_rates}, we see that the main source of neutrals in the system is thermal evaporation, with sputtering playing a minor role. The evaporation rate curve follows very closely and slightly exceeds the one by ion bombardment, which is expected considering the particle balance in the plasma system, with only a small excess of injected neutrals contributing to the growth of the plasma. 

\begin{figure}
    \centering
    \includegraphics[width=0.99\columnwidth]{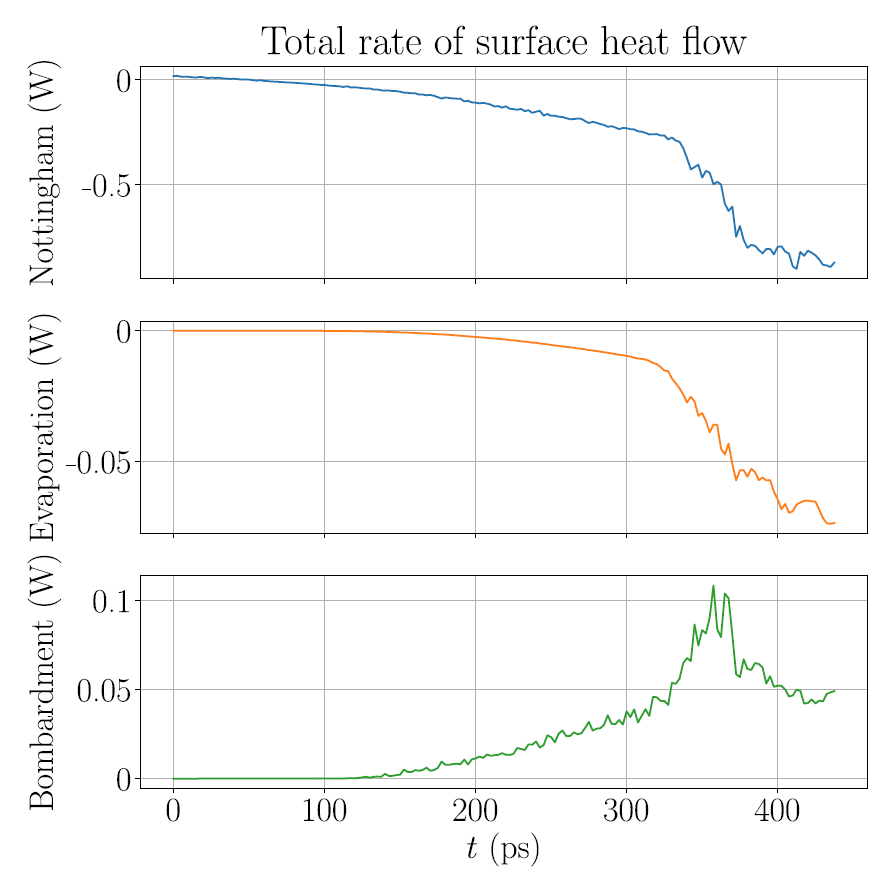}
    \caption{Evolution of the heat deposited on the tip included in the PIC-MCC plasma initiation simulations, separated into the three different heat sources (Nottingham heating-cooling, evaporative cooling, bombardment heating).}
    \label{fig:heat_sources}
\end{figure}

Another important aspect of these results is the heat balance. Since the main source of neutrals in the system is thermal evaporation, high temperatures need to be maintained to sustain the plasma growth. Figure~\ref{fig:heat_sources} shows the evolution of various heat sources in the system. We see that after the initial heating of the tip, the main positive heat source is Joule heating, with surface components being mainly negative. Although surface heat components (Nottingham heating-cooling, evaporative cooling, and bombardment heating) become significant in later stages, the Joule heating remains higher, thus maintaining a constant temperature increase, further increasing plasma growth. Among the surface heat sources, Nottingham cooling remains the most significant throughout the simulation, with bombardment heat and evaporating cooling playing a minor role.

Finally, we should note that we expect the behavior of the Joule heating to be substantially different in a more realistic dynamic tip simulation. As the tip evaporates away and flattens, the total current has more space to be emitted, thus reducing the current density and the Joule heat, resulting in much more moderate temperatures. However, we expect the heat balance to remain the most important factor determining plasma growth. 

The results of this section, in combination with the ones shown in section \ref{s:emit}, provide a comprehensive understanding of the most plausible mechanisms proposed to explain the transition from intense field emission at sharp surface protrusions to the development of a fully formed plasma. These mechanisms cannot be observed directly in experiments due to the extremely low spatial and temporal scales, and the simulation results cannot yet be directly and quantitatively compared to experimental measurements. However, they are compatible with all experimental findings and provide invaluable insights into the physics of vacuum breakdown formation, which can enable mitigation techniques. 

One of the most promising insights these simulation models offer is related to the dependence of breakdown occurrence on coupling the breakdown site to external electromagnetic power sources. Significant attention is currently being focused on studying microscopic power coupling via a complex linear external circuit. Preliminary results from both ArcPIC and FEMOCS have already found a strong dependence of the breakdown occurrence on the external circuit. Obtaining more detailed insights into this process can produce the first quantitatively comparable results to experiments and propose effective breakdown mitigation techniques. A more detailed discussion of these aspects can be found in the next section.

\section{\label{s:crater} Post vacuum arc surface damage}

\subsection{Formation of post vacuum arc surface damage} 

\begin{figure}
    \centering
    \includegraphics[width=0.95\linewidth]{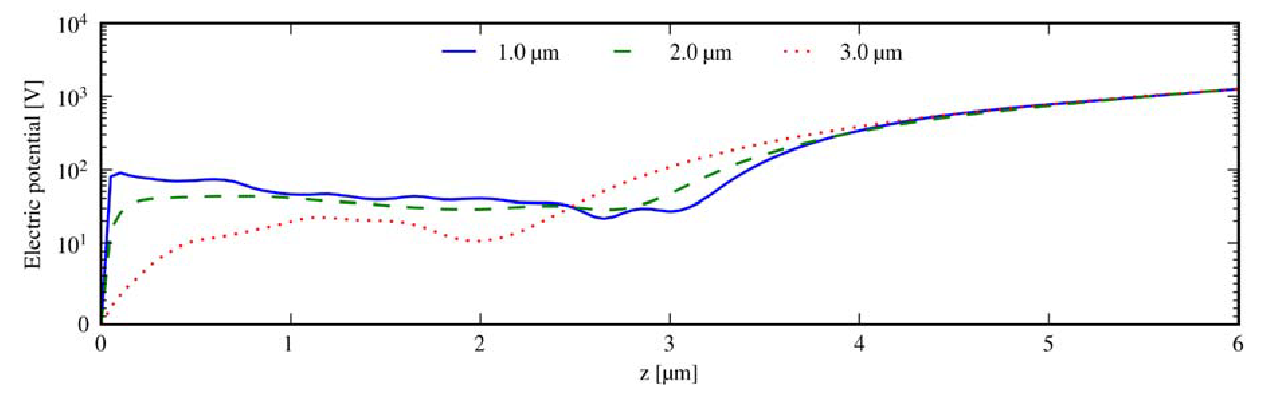}
    \caption{The electric potential as a function of distance from the cathode averaged within cylinders of three different radii after 1.5 ns after ignition of plasma. Note the overall logarithmic scale of the $y$ axis is linear from 0V to 10V. }
    \label{fig:sheathpot}
\end{figure}

Plasma built up that fills the gap between the electrodes inevitably causes intensive heat exchange between the plasma and the metal surfaces of the electrodes. Heat exchange is mediated by plasma particles, that is, ions, electrons and electromagnetic radiation, since the temperature of the plasma can easily reach a few eV \cite{anders_review_2014}. The mass difference between electrons and ions causes reorganization of plasma particles in such a way that positively charged ions accumulate above the negatively charged surface, forming a so-called sheath \cite{langmuir_positive_1923} with a burning voltage of $\sim$ 20 eV which drops over a distance of the order of a Debye length. An example of a potential drop near the surface, that is, the sheath potential, can be seen in Fig.~\ref{fig:sheathpot}. The figure shows that this drop is not uniform and changes away from the center following the density distribution within the plasma itself.

\begin{figure}
    \centering
    \includegraphics[width=0.9\linewidth]{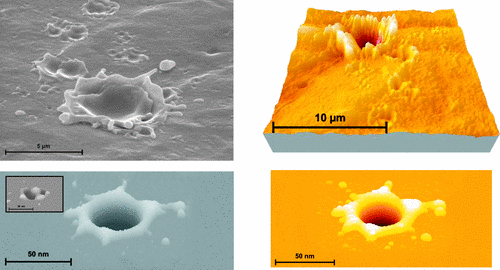}
    \caption{Comparison of the crater shapes obtained in simulations and measured in experiments. The top images are the SEM image (left) and AFM image (right) of a side crater with a well-defined shape. The bottom images are the same image obtained in the MD simulation with the total deposited energy of 4.34 keV/ nm$^2$. The colors of the images were selected to match the colors of the corresponding experimental ones. The inset in the bottom left part shows a complex double crater from the MD simulation. Scale bars emphasize the difference in the length scale between experimentally measured and simulated craters. Reprint from  Ref. \cite{timko_mechanism_2010}.  
    }
    \label{fig:crater_images}
\end{figure}

The ions escaping the plasma are accelerated via the electrostatic potential towards the surface, causing a strong channel for the heat exchange along with ballistic collisions with surface atoms that also contribute to the surface morphology as a result of the plasma impact. In early simulations of crater formation by plasma ions \cite{timko_mechanism_2010,djurabekova_crater_2012}, the ideal circuit with the one-dimensional geometry of the plasma was assumed as the first-level approximation. In this approximation, the applied voltage to the electrodes constituted the voltage drop in the sheath without loss. Although these simulations are of rather academic interest, they already emphasize the ion flux effect on the crater shape. From the visual agreement between the simulation and the experimental images shown in Fig.~\ref{fig:crater_images}, one can see the signature features of the violent response of the surface material to the plasma impact. Although the experimental images are selected to show the side craters rather than the main cathode spot, one clearly sees that even in these regions, the crater rims exhibit distinct shapes with well-visible fingers extended outside the rim. 

In the simulations where the same amount of energy as deposited in the plasma impact simulation was instantaneously deposited as thermal energy ($\sim$ 10 eV/atom) within a cylinder of the same radius as the plasma impact spot, the formed crater was fully symmetric with smooth rounded rims and no apparent asperities. The result indicates a very fast but still uniform phase transition from solid to molten, and to vapor. No overlapping impacts of plasma ions took place in these simulations, hence, no thermal instabilities responsible for asymmetric shape of rims in the simulations shown in Fig. ~\ref{fig:crater_images}. 


\begin{figure}
    \centering
    \includegraphics[width=0.9\linewidth]{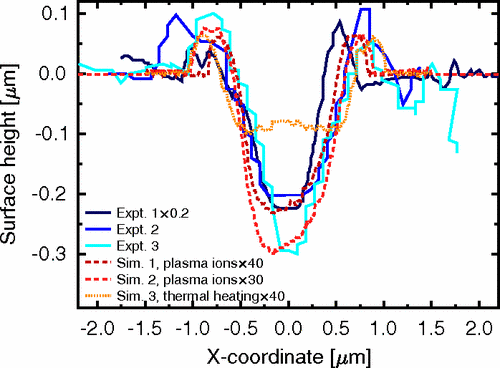}
    \caption{Comparison of crater shape profiles obtained from experiments and simulations. The experimental shapes were obtained from AFM measurements and the simulated ones were from a simulation of a spherical AFM tip approaching the atomic coordinates obtained from the MD simulations. The multiplication factor of 30–40 was used to enable the comparison. The absence of impacting plasma ions with directional momenta towards the surface in the simulations with the thermally deposited energy reduces the crater's depth, making it different from the experimental ones. Reprint from Ref. \cite{timko_mechanism_2010}.}
    \label{fig:crater_scales}
\end{figure}

Fig.~\ref{fig:crater_scales} shows a remarkable agreement between the crater profiles measured experimentally and in the simulations of plasma impacts. Although the profiles were scaled 30-40 times to enable comparison, the aspect ratio (crater depth to crater diameter) is extremely well reproduced in simulations. For comparison, one crater profile obtained in the simulations of thermally deposited energy is added to emphasize the necessity of considering the directional momenta of plasma ions towards the cathode surface. This single profile is much shallower than those obtained in the MD simulations of plasma ion showers, which results from the lack of plasma pressure on the cathode spot due to energetic plasma ions accelerated toward the surface. 

\begin{figure}
    \centering
    
    \includegraphics[width=0.9\linewidth]{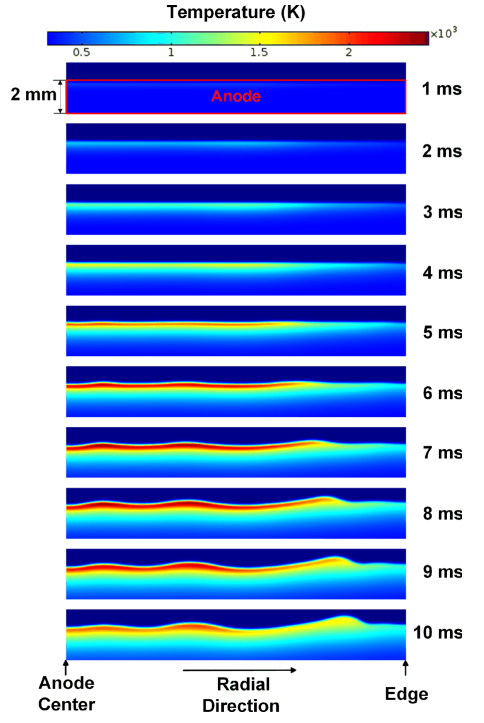}\\
    \caption{Evolution of the anode surface during the 20 kA arcing. The model includes the heat source from ion and electron currents at the anode surface during the arc phase.  Reprint from Ref. \cite{tian_modelling_2016}.}
    \label{fig:macro_spot}
\end{figure}

There were several attempts to simulate the formation of cathode spots in larger-scale simulations using the finite-element solvers of continuum equations. The macroscale of a typical cathode spot created by plasma on electrode surfaces motivates the description of plasma-surface interactions using continuous two-dimensional thermoplastic equations solved numerically  \cite{lasagni_fem_2004,mesyats_simulation_2014,mesyats_hydrodynamics_2015}. In these simulations, plasma is not seen as a shower of individual ions, but as a continuous pressure exerted on the surface within a radius roughly corresponding to the assumed lateral spread of plasma \cite{tian_modelling_2016, cunha_detailed_2017, cunha_numerical_2023, cro_phenomenological_2024}. These simulations can provide insights into the macroscopic behavior of metal surfaces simultaneously exposed to the pressure of plasma ion impacts and the heat exchange due to ion and electron currents, an example is shown in Fig.~\ref{fig:macro_spot}, where the surface of the anode was modified due to the plasma impact. The processes at the cathode are more violent and more accurate physical models where many ongoing processes can be included explicitly are needed. The present models are based on manually built-in assumptions, hence, they are limited in providing the first-principles insights into the plasma-surface interaction processes. 

\section{\label{s:circ} Circuit and macroscopic geometrical dependence of field holding strength}
\subsection{\label{ss:circuit intro}Introduction}

All the physical processes of breakdown ignition described in the preceding sections - dislocation dynamics, field-biased surface diffusion, electron field emission and atomic thermal/field evaporation - are driven by the applied surface electric field and result in a very small nucleation site. The central role of the surface electric field and locality of breakdown initiation might lead one to expect that the limiting high-field performance of a device would be determined solely by the maximum surface electric field in the device. And indeed over many decades, and for many devices, the maximum surface electric field has been used as a design constraint as given by the so-called Kilpatrick limit \cite{kilpatrick_criterion_1957}.   

However recent studies have show clearly that the high-field limit is not given by a unique value of maximum surface electric field, but that the value is instead is dependent on the characteristics of the system as a whole. The ensemble of results from the linear-collider high-gradient testing programs show a systematic variation of maximum surface electric field among same frequency and technology structures tested with common protocols but of different rf designs. Most generally, those with lower group velocity and lower localized power flow can support higher surface electric fields. Another example where the maximum surface electric field limit is dependent on global system characteristics is in the gap dependence in the pulsed-dc systems described in section \ref{s:exp}. Here measurements show that a larger gap can only support a lower surface electric field compared to a smaller gap with a proportionality of $d^{0.7}$.  

In general terms, the observed dependence of peak surface electric field on device geometry is a consequence of the strength of the coupling of electromagnetic energy of a system to the power absorbed by a developing breakdown arc. As a breakdown arc forms, it absorbs energy through the acceleration of emitted electrons, resulting in a reduction of the fields around the breakdown site, slowing or stopping the evolution of the arc. Nearby stored energy and incoming power from the power source re-supply the volume, either enough for the breakdown to fully form or, if insufficient, then the arc extinguishes and is not observed in the macroscopic scale of the device. The coupling of the stored energy and supply power to a volume affected by an arc depends on the geometry of the system and the characteristics of the powering circuit. This coupling dependence means that the limiting performance of a system is given by a complex function of the electric and magnetic field properties of the system as a whole. 

A sequence of power-flow based high-field limits are presented in this section, in the order that they were developed. The earlier, simpler analyses help provide the logical basis and motivations for the subsequent more complex analyses. These models elucidate the mechanisms of breakdown initiation, but also have significant practical importance. An accurate, predictive, and quantitative high-field limit allows the global optimization of a facility that uses a high-field system. For example, a linear collider can be optimized for cost and energy consumption by varying the geometry of an accelerating structure to give the highest gradient and efficiency while simultaneously considering the effects on the beam. This dual consideration is particularly important because geometrical variations that result in higher gradients typically result in a degraded beam transport.  

\subsection{\label{ss:phenomenological}Phenomenological power-flow based models Implications for breakdown evolution}

The main quantitative high-field limits that capture the geometric dependence of high-field systems are now described. A series of limits have been developed with a progressively better description of observed rf test results and based on an increasingly sophisticated underlying physical models. The high-field limits are described in chronological order in order to make clear the limitations of each of the models and highlight how subsequent models addressed these limitations. This provides an effective basis for explaining the underlying physics of breakdown initiation. On the practical side, the successive limits have resulted in a progressively improved ability to predict gradient, and so to better perform global optimizations for example of accelerator systems. 

One of the earliest attempts to relate the field holding limit to the macroscopic characteristics of a high-field rf system, in this case an accelerating structure, is described in \cite{adolphsen_processing_2001}. The focus of this work was to explain the difference in the level of surface damage caused by breakdowns between the more highly damaged higher, and less damaged lower-group velocity ends of a tapered traveling wave accelerating structures. Such structures were tested in the context of the NLC/JLC linear collider studies. The authors argued that the difference in damage was caused by a relative difference in the impedance match between a breakdown and the rf fields at that area of the structure. The difference in damage level was observed despite the near flat longitudinal profile of the peak surface electric-field on successive iris of these structures. The paper argues that a breakdown has a low impedance and it acts like a short circuit. This is then compared to the the impedance of the structure at the high group velocity end, where it is low, and the low group velocity end, where it is high.  This results in a better impedance match, and consequently higher transfer of rf power to an arc at the high-group velocity end compared to the low group velocity end. The group argued that the higher power coupling resulted in more damage to the structure surface. \cite{adolphsen_processing_2001} quantifies this relationship by: 
\begin{equation}
\label{eq:vg adolphsen}
    P_{abs}\simeq\frac{v_g^2}{(R/Q)^2}\frac{\sin(\varphi)}{\varphi\sin(\varphi)+2v_g\cos(\varphi)}G^2
\end{equation}
where \(P_{abs}\) is the power absorbed by the arc, \(G\) is the local accelerating gradient, \(v_g\) is the group velocity and \(R/Q\) relates the accelerating gradient to the stored energy per unit cell. Equation~\eqref{eq:vg adolphsen} is a proportionality for the absorbed power so does not give a specific limit that can be used in the design of an accelerating structure.

A subsequent attempt to describe the high field limit, also focusing on rf accelerating-type structures, is described in \cite{wuensch_scaling_2006}. The limit is given by an empirically derived value of the expression:  
\begin{equation}
\label{eq:PoverC}
    \frac{P}{C}=Constant
\end{equation}
where \(P\) is the power flow,  \(C\) is the innermost circumference of the structure and \(Constant\) is an empirically derived constant.

This limit is based on the general argument that the growth of an arc requires a certain level of power flow density in the volume occupied by the arc, and this density can be approximated by the quantity \(P/C\). This general argument was combined with trial comparisons of various functional dependencies to observed data from a series of about sixteen rf structures. The structures were made from roughly similar technology, from precision machined copper parts assembled by brazing or bonding and run in similar conditions. The quantity \(P/C\) was most consistent with the general argument and with the available data. By basing the value of the limit on a set of tested structures, \(P/C\) can then be used to predict the achievable gradient in a new design, but using similar technology and operated similarly. 

The power flow density is approximated by the input power divided to the circumference of the inner surface rather than the cross section because power flow in the disk loaded waveguide geometries typically used in accelerating structures is confined to the near the surface of the inner radius of the iris. The height of the power flow region is similar to that of a typical plasma spot thus the the quantity \(P/C\) describes this power density reasonably well.

The $P/C$ limit predicts that, at fixed frequency, larger aperture, and consequently higher group velocity structures have a lower surface electric field limit. As $C$ increases, $v_g$ increases, as well as $R/Q$ resulting in a lower surface electric field limit. 

The \(P/C\) limit is consistent with the observation that an arc's impedance evolves to match the power source impedance. The adaptability of arc impedance to drive impedance is seen in dc arcs where the impedance of an arc adjusts itself so that the burning voltage goes to a material-specific value that depends on its ionization levels, around 20 V in the case of copper \cite{anders_cathodic_2008}. This focus on power density rather than impedance matching is one major difference compared to using Eq.~\eqref{eq:vg adolphsen} as a high-field limit. 

In \cite{wuensch_scaling_2006} the high gradient limit \(P/C\) was applied to all structures high-field tested in the linear collider studies up to that date for which relevant data was available. A number of observations emerge in this comparison with the available data. 

One is that \(P/C\) predicts very well the high-field limit for traveling wave structures of a given frequency. This is true over a wide range of structure types, from accelerating structures with a group velocity of around 10 \% to power-generating structures with a group velocity of around 50 \%. It is likely that Eq.~\eqref{eq:vg adolphsen} would fit the same constant frequency traveling wave subsets of data equally well since a larger aperture structure has a higher group velocity for the same phase advance per cell.

On the other hand, \(P/C\) clearly does not apply to standing wave structures. There is no net power flow in such structures which implies an infinite field holding capability. Equation~\eqref{eq:vg adolphsen} fails in the same way.

In addition, the frequency dependence of \(P/C\) implies that the achievable gradient for a scaled geometry should go up as frequency goes down, which is in contradiction to the observation that the gradient is approximately constant for scaled structures. Indeed, some early publications have concluded that the gradient goes up with frequency. This is addressed in \cite{wuensch_scaling_2006} and resolved by explicitly adding a proportionality to frequency dependence. This is equivalent to the statement that an achievable gradient is independent of frequency for a scaled geometry. It is also the flat frequency dependence predicted by Eq.~(\ref{eq:vg adolphsen}).

Finally, neither Eq.~\eqref{eq:PoverC} nor Eq.~\eqref{eq:vg adolphsen} capture the pulse length dependence observed in short, sub-$\mu$s pulses. We have seen in section \ref{ss:behave-dependencies} that a pulse length dependence is only observed in pulse lengths up to around 1 $\mu$s  after which field holding remains constant. The models that will be derived below also do not capture the pulse length dependence. This is likely because the mechanism that gives the pulse length dependence does not come from circuit and power flow-related effects. The rf and dc voltage pulses described in this report have a minimum duration of the order of 100 ns. On the other hand the turn-on time for breakdowns is much shorter, of the order of ns as seen in the pulse shape data shown in section \ref{s:behave} and the simulation results shown in section \ref{s:plasma}. The pulse length dependence is more likely to come from dislocation processes as described in section \ref{s:dis}. 

The next generation high-field limit, referred to as $S_c$, extends the concept that fields are limited by the level of the local power flow at the position of the igniting arc. The quantity was introduced in  \cite{grudiev_new_2009}. The publication begins a greatly expanded compilation of 30 GHz and X-band testing results compared to \cite{wuensch_scaling_2006}. and also includes more details of the tests including measured breakdown rates as well as field quantities such as peak $E_{surf}$, peak $S_c$ and others re-simulated from structure geometries collected from the groups who built the structures and carried out the tests. Although many differences between different structure tests remain, this broader and deeper compilation of experimental results provides an improved basis for evaluating the applicability of high-gradient limits.  

The quantity $S_c$ is based on the Poynting vector, so is explicitly related to the local power flow inside an rf structure. It is similar to $P/C$ in the sense that it is related to power flow, but different in that it is calculated locally. $S_c$ is a scalar quantity that combines the real part of the Poynting vector with fraction of its imaginary part. The real part of the Poytning vector is the net power flow, while the imaginary part is the power flow that occurs within an rf cycle as energy is transferred back and forth between high electric and magnetic field regions of an rf structure. $S_c$ captures the fact that both flows of energy contribute to maintaining fields in the local region of an arc. The quantity is: 

\begin{equation}
\label{eq:Sc}
    S_c=|\Re(\vec{S})|+\frac{1}{6}|\Im(\vec{S})|
\end{equation}

\begin{figure}
    \centering
    \includegraphics[width=0.8\linewidth]{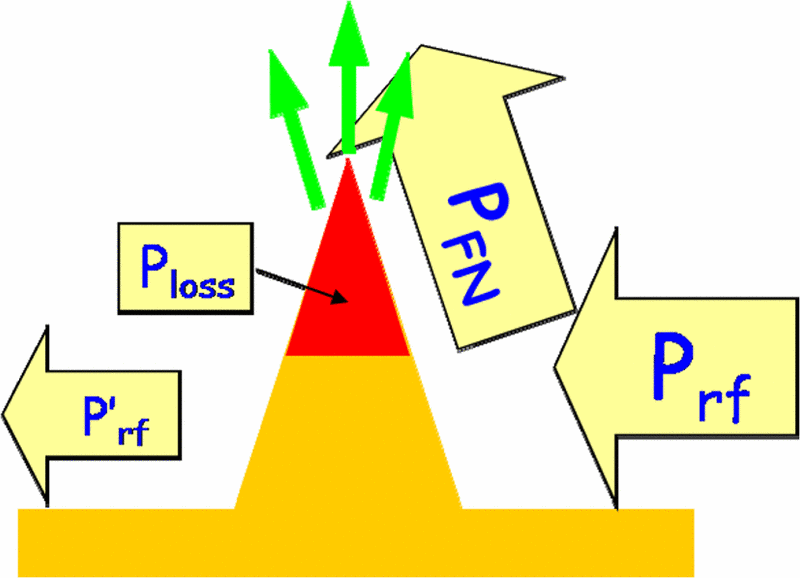}
    \caption{Power flows near tip}
    \label{fig:power near tip}
\end{figure}

The power flowing at a specific point on the surface can power the processes driving the evolution of an arc and is found to be comparable to the power necessary to heat a field-emitting tip. The value of the fraction $1/6$ is determined by a phenomenological fit and is explained by the relative coupling of the real and imaginary power flow to the tip. A schematic image of power flow near a field-emitting tip is shown in Fig.~\ref{fig:power near tip}. 

\begin{figure}
    \centering
    \includegraphics[width=1\linewidth]{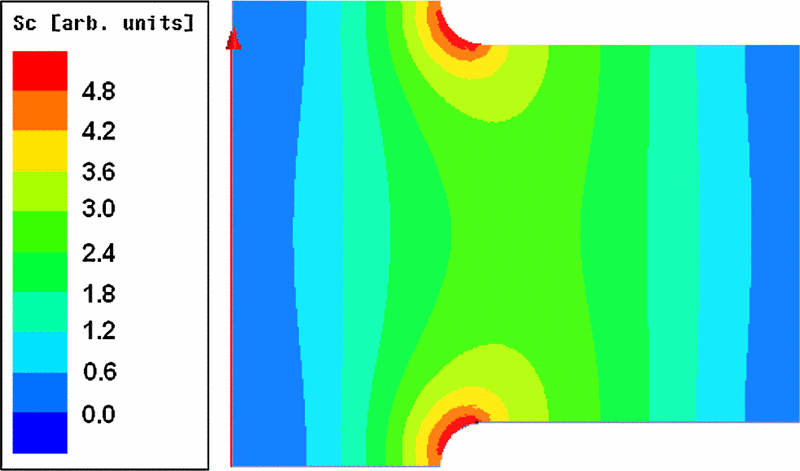}
    \caption{Cross section of one cell of a travelling wave accelerating structure showing the distribution of Sc. The left most line of the geometry is the central axis of a circularly symmetrical structure, the rightmost is the outer cavity wall. Taken from \cite{grudiev_new_2009}}
    \label{fig:Sc inside a TW cell}
\end{figure}

The inclusion of imaginary power flow is crucial for the applicability of $S_c$ to standing wave structures. Standing wave structures do not have a real power flow (strictly speaking they do have a power flow related to replenishing power loss due to finite quality factors, but this is extremely low). As we have seen this means that the $v_g$ and $S_c$ limits would imply infinite field levels. Standing wave structures do however have imaginary power flow, the power flow that occurs within each cycle of rf fields, which is available to power a breakdown. This power is captured in the second term in section Eq.~\eqref{eq:Sc}. The applicability of $S_c$ to standing wave structures resolves a major deficiency of the previous limits.

Because $S_c$ is calculated locally, it predicts naturally that there should be no frequency dependence of high-gradient limit of structures with a scaled geometry. Such a frequency dependence is implicitly validated by the quality of the fit of $S_c$ to both X-band and 30 GHz data in the compiled data, even if there isn't an example of a geometry exactly scaled between X-band and 30 GHz. This is consistent with the the $v_g$ based limit and comes without a somewhat arbitrary addition of $f$ to $P/c$.

Also because $S_c$ is calculated locally, it is used to optimize the details of the internal shape of an rf device. The high-field limits $v_g$ and $P/C$ take into account only the global features of an accelerating structure. They do not for example take into account a sharp edge of feature in a geometry, features that have been identified as causing locally-enhanced breakdown density. $S_c$ is also used to optimize shapes, for example of the cross section of the coupling iris shown in Fig.~\ref{fig:Sc inside a TW cell} in order to avoid local 'hot-spots'. 

It has also been suggested that gradient is limited by surface magnetic field, based on results of a particular set of structures with different irises \cite{dolgashev_geometric_2010}. However, there does not appear to be a numerical limit generally applicable across a broad range of devices. Also in the particular structures of \cite{dolgashev_geometric_2010}, other quantities such as $S_c$ are often also constant. Also looking at tests of structures beyond the group, for example tests of higher-order-mode damped and undamped CLIC structures show that there is of the order of $5 \% $ difference in achievable accelerating gradient and similar values of peak surface electric field $E_s$ and $S_c$ while there is a difference of more than a factor of two in surface magnetic field. Magnetic field is indeed relevant for the pulsed surface heating limit, but it is not a breakdown limit.

\subsection{\label{ss:powerCoupling} Power coupling between the macroscopic system and the microscopic arc development}

We have seen that successive generations of high-field limits have been capable of describing an increasing range of rf structures; from describing performance based on global characteristics only, to capturing the effect of local features and providing an increasingly detailed physical picture of the breakdown initiation process. 

None of the three rf limits, $v_g$-based, $P/C$, and $S_c$, however, are able to describe a dc system, like that described in Sec. \ref{ss:exp-dc} or other high-voltage dc systems, because there is no power flow ($v_g=0$). There is a power flow during charging and discharging of the electrode capacitance in the dc system, but the value is extremely low compared to typical $S_c$ values in rf structures. In addition, breakdowns are distributed in time throughout the high voltage pulse, not only during the rise and fall. On the other hand, a clear dependency of the surface electric field limit is observed for the pulsed dc system in analogy to the geometrical dependency observed in rf systems and described by $v_g$-based, $P/C$, and $S_c$. The dependency observed in the dc system is that the maximum voltage is proportional to the gap between the electrodes to the power of 0.7, with the consequence that the peak surface electric field goes down as the electrode gap increases. This 0.7 dependence can be seen in varying gap data taken with the pulsed dc system as shown in Fig.~\ref{fig:gap dependence DC}.

\begin{figure}
    \centering
    \includegraphics[width=0.75\linewidth]{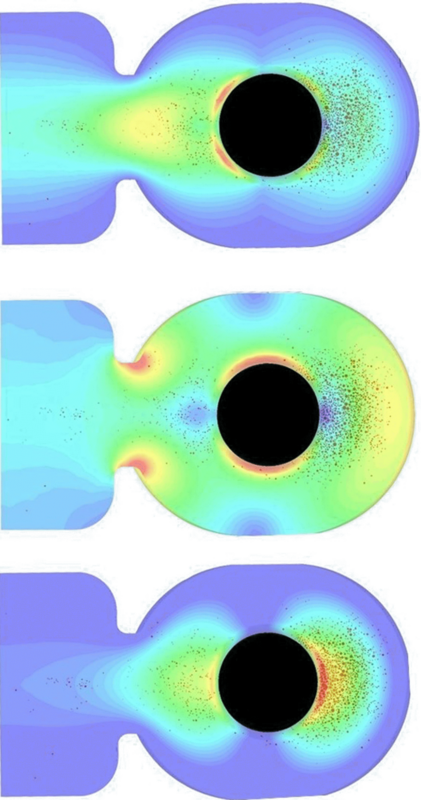}
    \caption{S$_c$, B and E distributions of the input coupler of a deflecting (crab) cavity with superimposed breakdown positions taken form post-test optical imaging. From Ben's paper. From \cite{woolley_high-gradient_2020}}
    \label{fig:crab cavity}
\end{figure}

A specific discrepancy with an experiment of $S_c$ emerged from an rf structure which was tested after the publication of $S_c$ \cite{grudiev_new_2009}. The structure was a so-called crab cavity, also know as a deflecting structure. The fundamental difference to previously tested high-gradient structures is that the crab cavity has an angularly varying field pattern. It operates in the TM$_{110}$ mode in contrast to the circularly symmetric, non-angularly varying pattern of the previously tested and discussed structures which all operate in the TM$_{010}$ mode. The field pattern is shown in Fig.~\ref{fig:crab cavity}. The structure was high gradient tested and cut open after test. The breakdown locations from the entirety of the test were determined by optical imaging. Breakdowns are seen as dots in Fig.~\ref{fig:crab cavity}. One can clearly see that the breakdown density is highest where the surface electric field is highest, not concentrated at the location of maximum $S_c$ nor the location of maximum surface magnetic field. Surprisingly, the structure operated at the typical value of $4  \textrm{W}/\mu \textrm{m}^2$, but the lack of correlation between regions of high $S_c$ and high breakdown density indicates that $S_c$ is not a sufficient quantity as a general high gradient limit. Neither $v_g$-based nor $P/C$ limits make any statement about where inside a structure breakdowns are likely to occur. 

In addition to the discrepancy of $S_c$ with the experimental result of the crab cavity, a further theoretical concern became increasingly clear which also applied to the $v_g$-based and $P/C$ high-field limits. This is that all three quantities are functions of single-frequency, steady-state solutions of the electromagnetic fields whereas breakdown is a very fast, ns timescale, phenomenon. Since the high-field limits are meant to capture the threshold of breakdown, a more comprehensive limit must capture the fast time response of the electromagnetic fields induced by the rapidly evolving breakdown. The focus of this review is on the initiation of breakdown, so the earliest moments of the breakdown process become critical. The consequence is that a more broadband analysis of the high-field devices is necessary.

The generalization from earlier high-field limits is how much power is available to flow into the small volume containing the breakdown in the short, ns time-scale when the breakdown is evolving from quiescence, and whether this is sufficient or insufficient to drive the breakdown evolution process. This generalization is crucial to covering dc high-field systems since it moves from steady-state power flow to the transient flow of power into the breakdown area. In the case of both dc and rf, the power is available from locally stored energy on short timescales (the speed of light in free space is 30 cm/ns to give an approximate scale) and from the power circuit on longer timescales.

Another way of describing this concern is posing the question will the growing electron emission from the breakdown site load the local fields enough that the field drops below the values needed to sustain the physical processes described in section \ref{s:plasma}. To this end, a next-generation high-field limit has been developed. It seeks to consider all data and results as previous limits and, in addition, incorporate the newest data and considerations just described. Furthermore, it is based on a rigorous formulation of the power coupling dynamics that incorporates the transient electromagnetic phenomena that occur during breakdown. The fundamental aspects of this formulation were introduced by \cite{paszkiewicz_studies_2020}, but here we shall present it in a slightly different and more general manner. 

The main goal of this formulation is to estimate the dependence of the transient response of the local field $E(t)$ at the breakdown site to the arc current. This relationship then defines the capacity of the electromagnetic system to feed sufficient power into the arc within its development time scale and consequently determines whether an arc shall develop or not. 

Due to the linearity of the Maxwell equations, $E(t)$ can be considered as a superposition of the unloaded field $E_0(t)$ originating from the external power sources and a loading field $E_L(t)$ caused by the moving charges of the arc itself. In principle, $E_L(t)$ depends on the entire distribution of the emitted current density $\mathbf{J}(\mathbf{r})$ in a system, i.e. the entire trajectories of electrons and ions emitted by the arc. However, we can approximate that $E_L(t)$ at the breakdown site depends solely on the current density distribution in its close vicinity, as the contribution of farther moving charges decays rapidly due to two effects: the moving charge dispersion that reduces the current density and the dropping contribution of charges with increasing distance, in view of the generalized Coulomb's law (Jefimeko's electromagnetic field equations \cite{griffiths_introduction_2014}).

Under this approximation, the microscopic breakdown site is reduced into a Hertzian (point) dipole. This assumes that the dimensions of the space containing the moving charges that significantly contribute to $E_L$ are much smaller than the macroscopic dimensions of the high-field structure and the wavelengths of the electromagnetic fields they produce. Under this assumption, which is true in most relevant cases of high-field systems, the entire spatial distribution of the current density induced by the arc is concentrated at a single point, mathematically described by a Dirac delta function or a Hertzian dipole. Such an electromagnetic source is characterized solely by the spatial integral of the current density distribution
\begin{equation}
    \int \mathbf{J}(\mathbf{r}) d^3r = \mathcal{J} =  
  IL
\end{equation}
which is the product of the Hertzian dipole antenna current $I$ and its infinitesimal length $L$. For a developing arc, this quantity is highly time-dependent. Thus, as an electromagnetic power source, the developing arc is characterized by the function $\mathcal{J}(t)$. 

This point dipole model can be used to derive the dependence of $E_L(t)$ on $\mathcal{J}(t)$. An important aspect of this dependence is that for any high-field structure, either rf or dc, it is a linear time-invariant functional that depends solely on the macroscopic characteristics of the high-field structure and not on the microscopic dynamics of the breakdown itself. This is a direct consequence of the linear time-invariant (LTI) nature of the Maxwell equations. Note, however, that the dependence of the arc current $\mathcal{J}(t)$ on $E(t)$ is, in general, non-linear and time-variant, as it stems from the complex dynamics of the developing arc.

Given the LTI dependence of $E_L(t)$ on $\mathcal{J}(t)$, it is convenient to describe them in the frequency domain via their Fourier (or equivalently Laplace) transforms $\hat{\mathcal{J}}(\omega), \hat{E}(\omega)$, and  $\hat{E}_L(\omega)$. In the frequency domain, we can generally express an LTI dependence as a multiplication, i.e. 
\begin{equation}
\label{eq:impedance_field}
    \hat{E}(\omega) = \hat{E}_0(\omega) - \hat{E}_L(\omega) = \hat{E}_0(\omega) - \hat{\mathcal{J}}(\omega) \hat{\zeta}(\omega),
\end{equation}
 where $\hat{\zeta}(\omega)$ is the local impedance function of the given breakdown site. The latter is a characteristic property of each point in the system and its inverse Laplace transform $\zeta(t)$, known as the impulse response, represents the evolution of the electric field at a given point after an impulse excitation by a unit Hertzian dipole at that point. 
 
 $\zeta(t)$ has units of $\Omega \textrm{m}^{-2}$, whose physical meaning becomes apparent by multiplying Eq.~\eqref{eq:impedance_field}  by $L$, and defining $V(t) = E(t)L$, $V_0(t) = E_0(t)L$ and $Z(t) = \zeta(t)L^2$. Thus we get the dependence that motivates the circuit structure shown in Fig.~\ref{fig:femocs2D_schematic}. This analysis reduces the entire electromagnetic coupling of the macroscopic system to the arc into a simple local impedance function. The $L^2$ term scales the impedance with respect to the length of the arc domain, where the presence of moving charges is considered. 

The impedance function can be accurately calculated numerically for any geometry using numerical methods by standard electromagnetic computational software. This calculation is performed by placing a unitary short dipole source antenna at the point of interest, treating it as a port, and calculating its scattering parameters for a wide range of frequencies (for details, see ref. \cite{paszkiewicz_studies_2020}). The resulting values give $\hat{\zeta}(\omega)$ for any range of frequencies domain, which can then be reversely Fourier transformed into the time domain to obtain the impulse response $\zeta(t)$. The resulting values can be convoluted with $\mathcal{J}(t)$ (e.g., as obtained by an ArcPIC or FEMOCS simulation) to obtain the time evolution of the loaded field at the arc site
\begin{equation} 
\label{eq:convolution}
    E(t) = E_0(t) -  \int_0^t \mathcal{J}(\tau) \zeta(t-\tau)d \tau \textrm{.}
\end{equation}
Then, the value of $E(t)$ is fed back as a time-dependent boundary condition in the PIC simulation. 

The above formulation is applicable to any high-field system and geometry, dc or rf. It provides a general and rigorous formalism to connect the macroscopic experimental findings of breakdown rate dependence to the microscopic arc ignition simulation models presented in sections \ref{s:emit} and \ref{s:plasma}. Furthermore, Eq.~\eqref{eq:convolution} is a generalized version of the external coupling circuit model already implemented in ArcPIC and FEMOCS. In fact, the RC circuit implementation of \cite{timko_field_2015, koitermaa_simulating_2024} is a special case of Eq.~\eqref{eq:convolution}.

A systematic simulation of the full coupling of this model with PIC is an ongoing work. However, simple analytic approximations for the impedance functions can be considered for various systems, providing valuable insights and an intuitive understanding of the processes that impede plasma development and result in the empirically observed trends described above. Furthermore, by considering such a simplified approach, as proposed by \cite{paszkiewicz_studies_2020}, a direct comparison with empirical data is possible, yielding a very good prediction of the empirically observed limits.

This is based on approximating the complex breakdown dependence of $\mathcal{J}(t)$ on $E(t)$ with a simple time-independent relationship $\mathcal{J}(t) = f(E(t))$. Such a relationship can be the Child-Langmuir law $\mathcal{J} \propto E^{3/2}$, which gives an upper limit for the current density of an electron beam. Yet, the qualitative comparison presented below is not specific to the choice of this function. 

\begin{figure}
    \centering
    \includegraphics[width=1\linewidth]{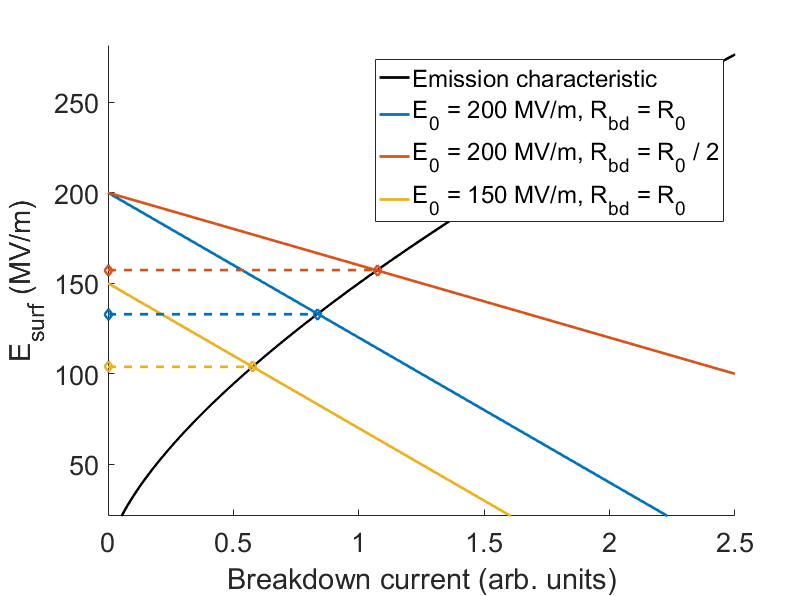}
    \caption{Graphical representation of the concept of breakdown impedance. The colored curves represent the surface field as a function of the loading current drawn by the breakdown, for different values of the unloaded field $E_0$ and breakdown impedance $R_{bd}$. The black curve represents the increasing dependence of the loading breakdown current to the surface field. The crossing point of these curves represents the "operating point" of a breakdown. The lower this point is, the lower the chance that a breadown will fully develop.}
    \label{fig:Rbd_graph}
\end{figure}

A second simplification that can be done for rf accelerating structures, which gives quite a good agreement with empirical data, is to look at the value of $\mathcal{R}\{Z(\omega)\} = R_{bd}$ at the single operating frequency of the accelerating structure. At this frequency, the system becomes equivalent to a series resistor circuit, yielding a simple picture of the breakdown impedance, graphically depicted in Fig.~\ref{fig:Rbd_graph}. In this figure, the straight lines represent the linear dependence of $E$ on $I$ for three cases, with the slope of the line defining the breakdown impedance $R_{bd}$. The black curve gives the BD emission characteristic. The crossing point corresponds to the operating field during the breakdown, which \cite{paszkiewicz_studies_2020} proposed as a new empirical limit. In other words, locations in a structure with a higher operating field $E^*$ are expected to have a higher breakdown rate as they maintain a higher flow of electromagnetic power into the breakdown. 

The value of $R_{bd}$ needs to be calculated numerically, but simple models can yield analytical formulae. This formula has been derived \cite{paszkiewicz_studies_2020}. 
By modeling an accelerating structure as a series of coupled LCR circuits, each representing a cell of the rf structure. This model yielded
\begin{equation}
    R_{bd} \propto \frac{1}{v_g} \frac{R}{Q} \left( \frac{E_0}{E_{acc}} \right)^2 \textrm{,}
\end{equation}
which was also confirmed by numerical simulations. The dependence on the group velocity, as well as the ratio of surface to accelerating fields, appears consistent with the early empirical predictors of breakdown performance, such as the group velocity or $P/C$, as the field ratio scales with the circumference.

Beyond this simplification, the value of $E^*$ has been calculated numerically for multiple systems, yielding very promising results that agree with experimental observations. These results can be found in ref. \cite{paszkiewicz_studies_2020}. Here, we present the results for two specific systems, namely the dc large electrode system and the crab cavity. 

\begin{figure}
    \centering
    \includegraphics[width=1\linewidth]{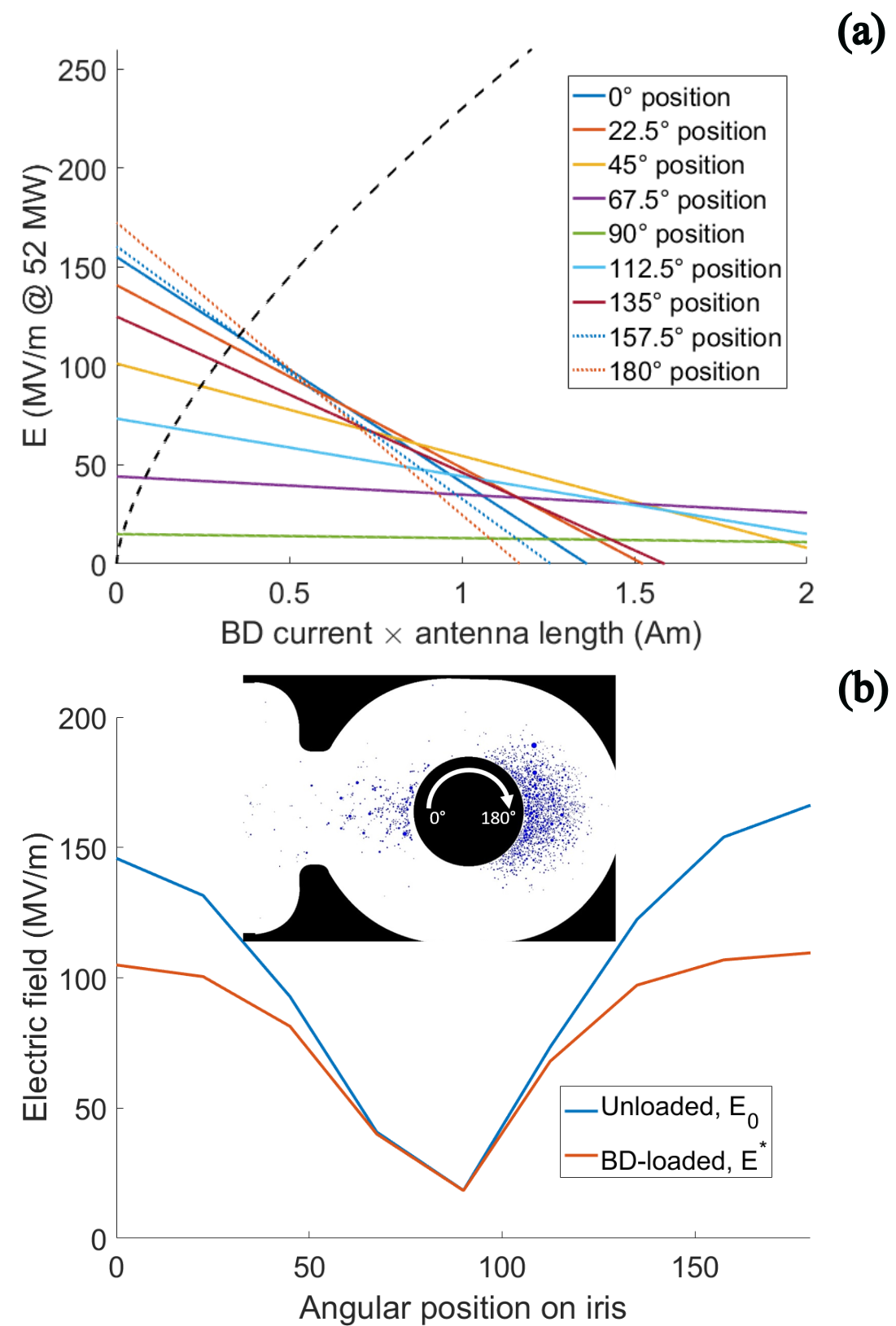}
    \caption{(a) Field as a function of the antenna input current $\mathcal{J}$ for various positions along the iris of the crab cavity, calculated numerically at a cavity input power of 52 MV/m. The unloaded field $E_0$ corresponds to the value at $\mathcal{J}=0$, the slope of the curves corresponds to the impedance $R_{bd}$, and the loaded field $E^*$ to the crossing point with the black dashed curve $I\propto E^{3/2}$. (b) The angular distribution of the loaded and unloaded fields on the iris of the cavity. The inset shows the crab cavity cell geometry along with the experimentally observed breakdown distribution (blue dots).}
    \label{fig:crab_Estar}
\end{figure}

Figure~\ref{fig:crab_Estar} demonstrates the success of this approach for the crab cavity, in contrast with the $S_c$ picture of Fig.~\ref{fig:crab cavity}. The distribution of the loaded field $E^*$ follows the same trend as the unloaded electric field $E$, thus restoring the prediction that the high-field performance is determined by the power flow rather than the surface electric field value.

Another system where previous high field limits ($S_c, v_g$, etc.) failed to predict the breakdown distribution is the pulsed dc system, as group velocity and $S_c$ are not definable in dc in general. In addition the dc systems show a $V \propto d^{0.7}$ gap dependence as discussed in Sec. \ref{ss:behave-dependencies}. However, the above approach gave excellent predictions. One of the approximations that is necessary to abandon when studying the dc system is the single frequency. In the dc, there is no reason to choose a certain operating frequency. For that reason, instead of $R_{bd}$, the indicating limit is given by the RMS value of $Z(\omega)$ over a wide range of frequencies. Using this RMS average as a metric, both the positional dependence of breakdown density inside the surface of pulsed dc systems and the gap scaling have been successfully predicted \cite{paszkiewicz_studies_2020}.

\section{\label{s:conclusions} Summary }
\subsection{\label{ss:currentUnderstanding}Current understanding}

In this review, we have summarized a body of experimental, theoretical, and simulation studies 
that address the fundamental processes of vacuum breakdown. Several notable results significantly advanced the overall understanding of vacuum breakdown. Enabling this advance is a body of consistent and comparable experimental work that allows for a quantitative comparison and combination of data. The experimental work covers high-field and breakdown phenomena over a wider range of conditions and system types than have been before. The work contributes both to fundamental understanding and practical considerations for real-world devices.

One of the most striking theoretical insights is the identification of the central role in the breakdown initiation of the plastic response, i.e., the dislocation dynamics of the electrode material to the stresses caused by surface electromagnetic fields. This insight explains the ordering of field-holding strength of different materials through the different material crystal structures and corresponding energies required for dislocation motion. A formalism based on dislocation dynamics provides a predictive model for the observed dependence of breakdown rate on the applied electric field. This strong dependence has been observed and measured in radio frequency and pulsed dc electrodes. The same formalism also accurately predicts the observed dependence of field-holding on temperature, including the striking increase in field-holding as devices are cooled to cryogenic temperatures. In addition, the dislocation model provides explanations for details of experimentally observed breakdown behavior, for example, the pulse-to-pulse memory effect, which is deduced from the uniform distribution of breakdown timings within pulses, despite a strong dependence of breakdown rate on pulse length. 

Analysis of systematic and detailed data from the conditioning and long-term operation of many tens of  radio frequency structures and dc electrodes reveals that the underlying conditioning process correlates with the number of pulses, rather than the number of breakdowns. This observation indicates the existence of intrinsic features that are conditioned under applied-field induced cyclic tensile stress. The conditioning of intrinsic features occurs at the highest fields while conditioning of extrinsic features dominate the conditioning of lower electric field level systems, the main domain of previous studies. The way in which the field holding capability of a device evolves in response to the application of fields has practical operational importance but also gives strong insight into the breakdown initiation process. The strong correlation to the number of pulses gives strong evidence that the evolution of the dislocation structure is responsible for the long-term conditioning of high-field devices. This insight into the pulse number dependence has allowed the optimization of conditioning strategies, aided by a Monte Carlo simulation tool that models the conditioning and breakdown process. 

Data from testing of multiple device types show a remarkable variation of achievable peak surface field, which turns out to be dependent on device geometry. This means that the macroscopic geometry influences the microscopic breakdown process. The key element in explaining this macroscopic-to-microscopic dependence is the power flow. Simply stated, a certain level of power inflow is necessary to support the breakdown evolution process, in particular while the plasma is forming and ever-increasing currents are emitted from the breakdown site. A series of increasingly comprehensive theoretical models based on the coupling of power flow to breakdowns have been developed. The earliest models considered global power flow, a later one local complex power flow, and finally a full dynamic beam-loading. These successive generations of models quantitatively predict an ever wider range of experiments.  

In addition to the dislocation dynamics under the metal surface, the dynamics of surface atoms play an important role in the initiation of the vacuum breakdown. It may lead to sharpening and further growth of surface irregularities whose appearance was triggered by an undersurface dislocation reaction. 
Surface dynamics is driven by biased diffusion of surface atoms that become partially charged and respond to the presence of a surface gradient.  This process is also thermally activated and, hence, slows down at cryogenic temperatures. The process is self-reinforcing; As features grow via biased diffusion, the field gradient increases, leading to stronger diffusion and growth, consistent with the observed critical BD process. 

A long-standing mystery of high-field systems is the nature of the features that cause emission enhancement in macroscopic systems. A large correction factor is needed to reconcile the current versus applied field measurements with electron field emission theory, as expressed by the Fowler-Nordheim equation. A widespread assumption is that there are microscopic surface features, commonly called tips, that cause a geometrical field enhancement, which reconciles experiment with theory. Field enhancement factors of 30 are typically needed for the fully conditioned structures described in this report, and factors of 100 to 200 are needed for the dc electrodes. Advanced modeling capabilities have been developed and applied to tips that would give such field enhancements. The models include the response of the tips to the applied field, including heating and cooling processes due to current flow and electron emission, mechanical stresses, and emission of copper atoms, and also include plasma formation and evolution. This modeled the evolution process, including all important physical processes of the breakdown from an initial condition through the observed transition of the vacuum to a conductive medium and collapse of applied fields. The tip simulations provide a simulation basis for the time-evolving impedance used in the power coupling modeling that gives geometrical dependence field holding. Simulations provide insight into different aspects of breakdown sites, including, for example, constraints on the nature of the tips and thermal coupling to the base. 

Reduction of work function due to the application of an electric field was also considered as a mechanism to enhance field emission. It was found to be unlikely to expect a significant effect as a result of electron density redistribution around the surface atoms exposed to the external electric field; however, some experimental work reports an observed reduction in work function. A variation of the work function value is observed both theoretically and experimentally for different surface facets; however, further theoretical studies are needed to investigate whether the bulk defects developed under the applied pulsed electric field produce a sufficient variation in the work function.

\subsection{\label{ss:prospects}Prospects and open questions}

This report describes a coherent theoretical foundation for a large body of mature experimental work. 
There are, inevitably, several open questions some of which are the subject of ongoing efforts, in particular in connecting remaining important theoretical constructs with direct experimental observations. 
In addition there are efforts underway to apply the insights and methodologies described here to a wider range devices, with a correspondingly wider range of operating conditions and parameters. This expansion will hopefully help these applications, but will also provide new data acquired in a way that can compared qualitatively with those presented here. This will provide further testing of ideas and models and help improve our understanding of breakdown. 

One of these ongoing research focus areas is to identify exactly what is changing as a device conditions, through direct observations of how a conditioned surface and subsurface differs from an unconditioned one. As we have seen, the dislocation model described in Sec. \ref{s:dis} predicts the field and temperature dependence of breakdown rate. An extension of the role of dislocations is that their structure or population changes as conditioning proceeds, in analogy to work hardening of metals. One of the primary experimental lines of this effort is to perform microscopic imaging of devices at various conditioning stages and, in complement, as a function of position in spatially varying surface fields. The primary experimental tool is electron microscopy, for example with FIB sample preparation followed by TEM analysis.

Another major open question is the nature of the enhanced emission sites, and how they populate high-field surfaces. Data from field emission measurements from all macroscopic devices require significant emission enhancement factors as we have seen in Secs. \ref{s:exp} and \ref{s:emit}. But at the time of this writing, no sufficiently sharp tip-like feature has ever been found in many hundreds of SEM images taken in the course of the experiments described in this report nor is anyone able to predict the emission enhancement factor independently of a field emission measurement. Imaging of of the spatial distribution of field emission is being carried out by upgrading the breakdown positioning cameras of pulsed dc systems described in \ref{s:exp} to higher sensitivity.  The upgraded system will allow the question of whether breakdowns actually preferentially occur where stable field emitters are observed, or not, to be answered. 

Examples of devices where the experimental methodology described here is now being applied include radio frequency quadrupoles, electrostatic separators and muon capture cavities, all devices used in accelerators but much larger than those described here and operating at lower peak fields. Additional effects reduce field holding; H$^-$ irradiation in radio frequency cavities \cite{serafim_h-_2024}, $\gamma$ irradiation in electrostatic separators and immersion in multi-tesla external magnetic fields for muon capture cavities. The expanded parameter space and the role of additional breakdown-inducing effects will help us refine our understanding. 

Finally, there is a clear role for an experimental set-up that combines a high-field device with in-vacuum diagnostic systems. This could be, for example, a pulsed dc system combined with in-vacuum manipulators to transfer electrodes to a scanning electron microscope, ideally combined with a focused ion beam system or a scanning tunneling microscope. Such a setup could help identify the origin of field emission sites or identify evidence of the link between plastic activity and field emission leading to BD processes. Direct observation of these effects is made complicated when samples are exposed to air between exposure to high-fields and analysis. Initial attempts at in-vacuum experiments have been made, but technical challenges have not allowed robust and conclusive results \cite{muranaka_-situ_2011}. We are hopeful that similar work, if pursued in the future, will provide a direct observational link.

\bibliography{references_fixed} 

\appendix

\section{Glossary} \label{sec:glossary}

\begin{table}[h!]
\centering
\begin{tabularx}{\linewidth}{|p{2cm}|X|p{1cm}|}
\hline
\textbf{Term} & \textbf{Definition} & \textbf{Abbr.} \\
\hline
Vacuum arc & A plasma appearing near the metal surface under vacuum conditions after high electric field exposure & --- \\
\hline
Vacuum Breakdown & Transition from an insulating to a conducting state via a localized plasma above the surface & VBD \\
\hline
Breakdown Rate & number of pulses with breakdown divided by the total number of pulses in a measurement window & BDR \\
\hline
Conditioning & controlled process of gradually increasing field and/or pulse length to improve field-holding capability & --- \\
\hline
Field-holding capability & ability to sustain a given surface electric field without breakdown & --- \\
\hline
Field (electron) emission & electron emission by quantum tunneling from a metal surface under strong electric fields & FE \\
\hline
Dark current & Current caused by field emitted electrons & --- \\
\hline
Maxwell stress & Tensile stress acting on a conductor’s surface due to the applied electric field & --- \\
\hline
Biased diffusion & Diffusion of surface atoms with uneven jump rate didstribution, resulting in atomic migration toward regions of higher electric field & --- \\
\hline
Vacuum bursts & sudden pressure spikes in a vacuum system & --- \\
\hline
Large Electrode System & A pulsed dc test system with two large, flat electrodes separated by a precision spacer & LES \\
\hline
Anode tip system & pulsed dc setup with a shaped anode tip facing a flat cathode at micrometer-scale gaps & --- \\
\hline
Faraday cup & current detector that collects charged particles along a beam axis & --- \\
\hline
Modified Poyinting vector & local power-flow metric characterizing power available to initiate an arc & $S_c$ \\
\hline
Group velocity ($v_g$) & speed at which rf power propagates through a traveling-wave structure & $v_g$ \\
\hline
Accelerating gradient & on-axis average field experienced by the beam in an accelerating structure & --- \\
\hline
Pulse length & duration of the applied rf (or dc) high-voltage pulse & $\tau$\\
\hline
Thermal runaway & positive-feedback heating of an emitting nano-tip leading to melting, evaporation, and enhanced emission & TR \\
\hline
Plasma sheath & near-surface region where potential drops by $\sim$20 V (``burning voltage'' for Cu arcs), accelerating ions into the cathode & --- \\
\hline
Breakdown impedance & effective local impedance coupling the microscopic breakdown current to the macroscopic electromagnetic environment & Z \\
\hline
\end{tabularx}
\caption{List of terms with definitions and abbreviations or symbols.}
\label{tab:glossary}
\end{table}

\end{document}